\documentclass[twocolumn,aps,showpacs,floatfix,superscriptaddress, prr]{revtex4-2}

\usepackage{graphicx}
\usepackage{amsmath}
\usepackage{amssymb}

\usepackage[normalem]{ulem}
\usepackage{soul}

\usepackage{xcolor}
\usepackage[utf8]{inputenc} 
\usepackage{parskip}
\usepackage{booktabs}
\usepackage[export]{adjustbox}

\begin{document}
 
\title{Entanglement and Dynamical Scaling Laws in Quantum Superabsorption}

\author{Juan David Álvarez-Cuartas}
\email{j.d.a.cuartas@fys.uio.no}
\affiliation{Departamento de Física and Centre for Bioinformatics and Photonics---CIBioFi, Universidad del Valle, Cali 760032, Colombia}
\affiliation{Department of Physics and Center for Computing in Science Education, University of Oslo, N-0316 Oslo, Norway}

\author{John H. Reina}
\email{john.reina@correounivalle.edu.co}
\affiliation{Departamento de Física and Centre for Bioinformatics and Photonics---CIBioFi, Universidad del Valle, Cali 760032, Colombia}
\affiliation{Department of Physics and Center for Computing in Science Education, University of Oslo, N-0316 Oslo, Norway}

\date{\today}

\begin{abstract}
Quantum batteries (QBs) exploit collective quantum resources to surpass the limits of classical  energy storage and power delivery. We analyze $N$-qubit cavity-coupled QBs governed by Dicke and Tavis--Cummings models under Gaussian driving and open-system dynamics. Finite-size scaling laws $\mathcal{O}(N)\!\sim\!N^{\alpha}$ demonstrate an optimal region of relaxation and dephasing where coherent driving stabilizes entanglement entropy growth for thermodynamic observables (maximum energy $E_{\mathrm{max}}$, charging time $\tau$, and maximum power $\bar{P}_{\mathrm{max}}$) and for qubit and cavity entanglement entropies. The Dicke model exhibits entropy-suppressed extensive behavior, while the Tavis--Cummings model achieves super-extensive scaling with $\alpha_{E_{\mathrm{max}}}\!\in\![1.08,1.26]$, $\alpha_{\tau}\!\approx\!-0.49$, $\alpha_{\bar{P}_{\mathrm{max}}}\!\in\![1.57,1.73]$, supported by qubit-cavity entanglement. We demonstrate that dissipation can act as a stabilizer source, yielding scaling benchmarks that are relevant to several experimental platforms. Our findings connect entanglement, dissipation-enhanced scaling laws and superabsorption, outlining a pathway towards scalable quantum batteries offering practical quantum advantage.
\end{abstract}

\maketitle

\section{Introduction}
\label{sec:introduction}

Superabsorption is the cooperative enhancement of light absorption in an ensemble of quantum emitters, arising from the same collective coherence that gives rise to superradiance, but operating in reverse. In a superabsorbing system, the rate at which excitations are absorbed increases faster than linearly with the number of emitters, leading to faster energy transfer than any incoherent sum of independent units~\cite{higgins2014superabsorption}. 
This phenomenon was theoretically predicted as the time-reversed analogue of Dicke’s superradiance~\cite{dicke1954original}, in which a collection of emitters coupled to a common electromagnetic mode undergoes collectively enhanced spontaneous emission, releasing radiation with an intensity that scales as \(N^{2}\). Both effects originate from the formation of collectively coupled Dicke states, whose shared coherence gives rise to non-classical scaling in the emission and absorption rates. Understanding how such cooperative effects persist or degrade in open quantum environments is a central question for quantum technology.

Quantum batteries 
are quantum devices designed to store and deliver energy by exploiting resources such as coherence and entanglement~\cite{horodecki2009quantum}. These uniquely quantum features raise the prospect of outperforming classical batteries in charging and power delivery. While classical thermodynamics establishes strict limits on extractable work from any system~\cite{carnot1824reflections,fermi2012thermodynamics,kardar2007statistical}, it does not constrain the time scales over which energy transfer occurs. This distinction has motivated intense interest in seeking a quantum advantage in out-of-equilibrium processes~\cite{vinjanampathy2016quantum,binder2018thermodynamics,herrera2023correlation}. Although entanglement is not strictly necessary for optimal work extraction~\cite{hovhannisyan2013entanglement,campaioli2017enhancing}, it can significantly impact the rate and efficiency of charging, making it a key resource in quantum superabsorption technology~\cite{campaioli2018quantum,Campaioli_2024_Review}.

Like conventional batteries, QBs are temporary storage units subject to physical constraints such as finite energy storage capacity and bounds on power density~\cite{julia2020bounds,yang2023battery}. Strategies to enhance their performance have therefore been explored, including collective charging schemes that boost capacity~\cite{campaioli2018quantum,ito2020collectively},  stabilization mechanisms that mitigate losses to the environment~\cite{liu2019loss, gherardini2020stabilizing, xu2021control} and the use of decoherence-free subspaces~\cite{reina2002decoherence,quach2020dark,JHR_DPhil}. Since the concept of a QB was introduced~\cite{alicki2013entanglement}, several key theoretical advances have followed. In particular, Campaioli \textit{et al.} showed that the choice of charging protocol can lead to superextensive behavior: when $N$ identical qubits are charged collectively, the total charging power can scale faster than $N$, resulting in a $\sqrt{N}$ speedup compared to parallel charging~\cite{campaioli2017enhancing}. Further progress came with the realization that the Dicke QB model could be implemented in solid-state architectures~\cite{ferraro2018,campaioli2018quantum}. Quach~\textit{et al.}~\cite{quach2022superabsorption} achieved superabsorption at room temperature in a molecular QB and compared their measurements with results from the Tavis--Cummings model, 
demonstrating a quantum charging advantage in a physical system.
\begin{figure*}[!ht]
    \centering
    \includegraphics[width=0.75\linewidth]{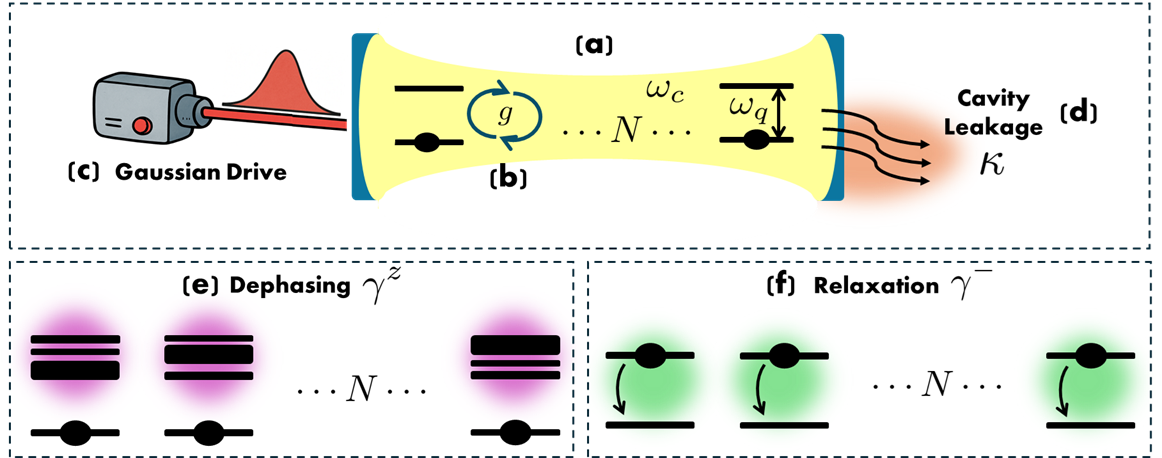}
    \caption{\textbf{Scheme of the cavity–driven QB and dissipation channels}. 
    (a) An ensemble of $N$ identical qubits, each with a transition frequency $\omega_q$, are (b) collectively coupled  to a  single cavity mode of frequency $\omega_c$ with a coupling strength $g$. (c) The battery is charged by a resonant Gaussian drive $\eta(t)$ that excites the cavity (see below). (d) The cavity mode loses photons through leakage at a rate $\kappa$. 
    (e) Each qubit undergoes pure dephasing at rate $\gamma^z$, which randomizes the relative phases without exchanging energy; (f) qubit relaxation occurs at a rate $\gamma^-$, corresponding to spontaneous decay from the excited state to the ground state.}
    \label{fig:system}
\end{figure*}

These findings highlight a key point: quantum batteries are inherently open systems. 
They interact with their environment through processes such as spontaneous emission, dephasing and photon leakage, which 
 weaken quantum correlations and reduce stored energy. Improved energy retention and faster charging have also been reported under non-Markovian dynamics, where system–environment correlations persist over time~\cite{kamin2020nonmarkovian,ghosh2021enhancedperformance}. However, in this work, we focus our attention on the Markovian regime, as discussed below.

A second subtle yet significant aspect relates to the nature of light–matter interactions in quantum batteries. Many studies use the rotating-wave approximation — which neglects counter-rotating terms — leading to the Tavis–Cummings model~\cite{tavis1968exact, agarwal2012tavis, fujii2017rwa}. While this approximation remains valid in the weak-coupling regime ($g \ll \omega_q,\omega_c$, where $g$ is the coupling strength and $\omega_q$ and $\omega_c$ are the qubit and the cavity transition frequencies, respectively), it breaks down in the ultrastrong regime~\cite{forndiazRMP2019usc}. The full Dicke model, which includes counter-rotating terms, predicts phenomena  such as vacuum-induced excitations, entangled ground states, and superradiant phase transitions~\cite{jaako2016ultrastrong}. However, the implications for charging dynamics in quantum batteries are not fully understood.

In this work, we present a comprehensive analysis of a QB comprising 
$N$ identical two-level systems (TLSs), which we refer to as ``qubits'',
that are collectively coupled to a single-mode cavity and driven by a Gaussian pulse. We compare two interaction regimes: the Tavis--Cummings model and the Dicke model, which includes counter-rotating processes. Dissipation is incorporated via three channels: (i) TLS relaxation  through spontaneous emission, 
(ii) TLS pure dephasing, 
and (iii) cavity photon loss. 
Through a systematic analysis of 
 these models, we quantify how the energy storage capacity, charging power, charging time and entanglement entropy scale with the size of the system. This clarifies the role of dissipation in the thermodynamic performance of quantum batteries and the corresponding scaling laws.

The remainder of this paper is organized as follows. Section~\ref{sec:theory} introduces the QB models, the Tavis--Cummings and Dicke Hamiltonians, the Gaussian driving protocol, and the Lindblad dissipation channels. Section~\ref{sec:numerical} describes the numerical implementation, which exploits permutation symmetry to access system sizes up to $N\!\sim\!50$. Section~\ref{sec:results} presents the results: quasi-ideal dynamics, open-system behavior under cavity loss, relaxation and dephasing, and finite-size scaling laws $\mathcal{O}(N)\!\sim\!N^\alpha$ for the thermodynamic observables. Section~\ref{sec:hardware} maps the operating regimes onto experimentally realistic parameters in leading platforms. Section~\ref{sec:discussion} provides a unified discussion and outlook, highlighting the microscopic mechanism of dissipation-enhanced superabsorption, the stabilizing role of coherent driving, and perspectives for optimized control and scalable implementations. Finally, Section~\ref{sec:conclusions} summarizes the main findings of this work.

\section{Theoretical framework}
\label{sec:theory}

A primary goal of QB research is to identify the microscopic mechanisms that enable efficient energy storage, and to understand how collective quantum phenomena enhance performance. One widely studied architecture is the cavity–mediated battery~\cite{ferraro2018, andolina2019PRL, campaioli2018chapter, quach2022superabsorption}, where an ensemble of qubits interacts with a single electromagnetic mode. This setup incorporates both collective effects, which can enhance charging performance, as well as unavoidable sources of irreversibility, such as dissipation channels and entanglement with the cavity. To study these competing effects, we compare two complementary light–matter models: the Dicke Hamiltonian (non‑RWA), commonly adopted in theoretical analyses~\cite{dicke1954original, hausinger2010analyticUSC, ashhab2010tls, forndiazRMP2019usc, kockumNRP2019usc}, and the Tavis–Cummings Hamiltonian (RWA), often employed to benchmark experimental results~\cite{tavis1968exact, agarwal2012tavis}.

As illustrated in Fig.~\ref{fig:system},  the quantum battery consists of an ensemble of $N$ qubits, which are collectively coupled to a single mode of a cavity. The qubits act as storage units, while the cavity mediates the transfer of collective excitation and serves as an intermediary charger. Charging can occur by initializing the qubits in their excited state or by applying a resonant Gaussian pulse to inject energy into the cavity. Dissipation is included through photon leakage from the cavity at rate $\kappa$, as well as through qubit  dephasing and 
relaxation at rates $\gamma^z$ and $\gamma^-$, respectively. Together, these processes define the open-system dynamics analyzed in this work. 

Each qubit has transition frequency $\omega_q$ between ground $\left|g\right>$ and excited $\left|e\right>$ states, and the cavity has frequency $\omega_c$, with qubit–cavity detuning $\Delta_c = \omega_q - \omega_c$. All qubits couple identically to the cavity mode with strength $g$. The full Hamiltonian of the driven system reads
\begin{equation}
\label{eq:full-hamiltonian}
\hat{H}(t) = \hat{H}_{\text{cav}} + \hat{H}_{\text{qub}} + \hat{H}_{\text{int}} + \hat{H}_{\text{drive}}(t),
\end{equation}
where
\begin{align}
\label{eq:cavity-hamiltonian}
\hat{H}_{\text{cav}} &= \hbar\, \omega_c\, \hat{a}^\dagger \hat{a}, \\
\label{eq:qubits-hamiltonian}
\hat{H}_{\text{qub}} &= \frac{\hbar}{2}\,\omega_q \sum_{j=1}^N \hat{\sigma}^z_j,
\end{align}
with $a^\dagger$ ($a$) the photon creation (annihilation) operators, and $\sigma^\alpha_j$ the $\alpha$-Pauli operator of the $j$th qubit, with $\alpha=x,\, y,\, z$. 

To charge the system, an external Gaussian pulse is applied to the cavity mode. The corresponding external driving Hamiltonian is given by
\begin{eqnarray}
\label{eq:H_drive}
\hat{H}_{\text{drive}}(t) = \hbar\, \eta(t)\, (\hat{a}^\dagger + \hat{a})&,& \\
\label{eq:gaussian-pulse}
\eta(t) = \eta_0 \exp\!\left(-\frac{(t-t_0)^2}{2\sigma^2}\right)&,&
\end{eqnarray}
where $\eta_0$ is the drive amplitude, $\sigma$ its width, and $t_0$ the pulse center.  
In the laboratory frame, the full expression for the drive usually includes a carrier oscillation of the type 
$\cos(\omega_d t)$, i.e. Eq.~\eqref{eq:gaussian-pulse} takes the form  $\eta(t)\cos(\omega_d t)$, where $\omega_d$ is the drive frequency. 
However, since our simulations are performed in the rotating frame of the cavity mode and the drive is resonant 
($\Delta_d \equiv \omega_d-\omega_c= 0$),
only the Gaussian envelope  $\eta(t)$ remains in Eq.~\eqref{eq:H_drive}.

\subsection{Light-matter interaction models}

Taking into account all light-matter interaction terms leads to the Dicke Hamiltonian (non-RWA), given by
\begin{equation}
\label{eq:interaction-dicke}
\hat{H}^{\text{(D)}}_{\text{int}} = \hbar\, g \sum_{j=1}^N \left(\hat{a}^\dagger\, \hat{\sigma}^-_j + \hat{a}\, \hat{\sigma}^+_j + \hat{a}\,^\dagger \hat{\sigma}^+_j + \hat{a}\, \hat{\sigma}^-_j\right),
\end{equation}
where $\hat{\sigma}^+_j = \left|e\right>_j\left<g\right|$ and $\hat{\sigma}^-_j = \left|g\right>_j\left<e\right|$.
Counter‑rotating terms $\hat{a}^\dagger\hat{\sigma}^+_j$ and $\hat{a}\,\hat{\sigma}^-_j$ simultaneously create or annihilate an excitation in both the qubit and the cavity. While these terms are negligible in weak coupling, they become relevant in the ultrastrong coupling (USC) regime, roughly characterized by $g/\min(\omega_q,\omega_c)\!\sim\!1$~\cite{forndiazRMP2019usc, kockumNRP2019usc}. This leads to effects such as virtual excitations and Bloch–Siegert shifts~\cite{hausinger2010analyticUSC, ashhab2010tls}. Retaining these terms allows us to investigate  the effect of light–matter correlations during the charging dynamics~\cite{yoshihara2017superconduct, forndiazRMP2019usc}.

In the weak‑coupling regime, 
the counter‑rotating terms oscillate rapidly and can be neglected under the RWA, yielding the Tavis–Cummings Hamiltonian,
\begin{equation}
\label{eq:interaction-tavis-cummings}
\hat{H}^{\text{(TC)}}_{\text{int}} = \hbar\, g \sum_{j=1}^N \left(\hat{a}^\dagger\, \hat{\sigma}^-_j + \hat{a}\, \hat{\sigma}^+_j\right).
\end{equation}
This Hamiltonian conserves the total number of excitations, making the dynamics integrable and supporting the emergence of coherent Rabi oscillations~\cite{tavis1968exact,agarwal2012tavis,fujii2017rwa}. 

\subsection{Thermodynamic observables and entanglement}
\label{subsec:observables}

To characterize the charging dynamics, we compute a set of observables that capture both thermodynamic behavior and quantum correlations.

The \textit{maximum energy} stored in the $N$-qubit ensemble,
\begin{equation}
\label{eq:def-energy}
    E_{\max} = \max\limits_{t_f}\!\left[\langle \hat{H}(t) \rangle - \langle \hat{H}(0) \rangle\right],
\end{equation}

measures the largest energy that is stored in the battery during the charging  procedure (
$t=t_f$), referenced to the initial ground state $\langle \hat{H}(0)\rangle$. It coincides with the maximum increase of the internal energy of the qubit subsystem, a standard definition adopted in QB analyses~\cite{ferraro2018,andolina2019PRL,campaioli2018chapter}.

The \textit{charging time} $\tau$ is the first time at which $E_{\text{qub}}(t)$ reaches $E_{\max}$, setting a characteristic speed for the process~\cite{ferraro2018, campaioli2018chapter},
\begin{equation}
\label{eq:def-charging-time}
    \tau = \inf \left\{\, t \geq 0 \;:\; E(t) = E_{\max}\,\right\}.
\end{equation}
The \textit{average maximum charging power}, or \textit{maximum power} for short, $P_{\max}$ is obtained from the ratio
\begin{equation}
\label{eq:def-power}
    \bar{P}_{\max} = \frac{E_{\max}}{\tau},
\end{equation}
balancing the energy storage capacity against the time needed to reach it. This definition is now widely used in quantum battery studies to assess quantum advantage in collective charging~\cite{ferraro2018, campaioli2018chapter}.

The \textit{entanglement entropy} provides a quantitative measure of correlations whithin the composite system through the von Neumann entropy of the reduced subsystems~\cite{nielsen2010quantum, breuer2002theory}. We denote $S_q(t)$ and $S_c(t)$ the entanglement entropies of the $N$-qubit ensemble and the cavity, respectively, defined as
\begin{eqnarray}
\label{eq:def-entropy-qubits}
    S_{q}(t) &=& -\mathrm{Tr}\!\left[\hat{\rho}_{\text{qub}}(t)\,\ln\hat{\rho}_{\text{qub}}(t)\right],\\
\label{eq:def-entropy-cavity}
    S_{c}(t) &=& -\mathrm{Tr}\!\left[\hat{\rho}_{\text{cav}}(t)\,\ln \hat{\rho}_{\text{cav}}(t)\right],
\end{eqnarray}
where the reduced density matrices are obtained from the total state $\hat{\rho}_{\text{tot}}(t)$ as
\begin{eqnarray}
\label{eq:def-density-matrix-qubits}
    \hat{\rho}_{\text{qub}}(t) &=& \mathrm{Tr}_{\text{cav}}\!\left(\hat{\rho}_{\text{tot}}(t)\right),\\
\label{eq:def-density-matrix-cavity}
    \hat{\rho}_{\text{cav}}(t) &=& \mathrm{Tr}_{\text{qub}}\!\left(\hat{\rho}_{\text{tot}}(t)\right).
\end{eqnarray}

For closed dynamics the global state remains pure; hence $S_q=S_c$, and the entropy directly measures qubit–cavity entanglement~\cite{nielsen2010quantum}. Under dissipation, however, both $S_q$ and $S_c$ capture the combined effects of coherent qubit-cavity correlations and incoherent mixing induced by the environment, thereby providing a unified measure of entanglement and noise~\cite{vinjanampathy2016quantum, anders2017quantumthermodynamics, binder2018generalquantum, reina2014correlations}.

We characterize the information content during the charging protocol by means of two complementary entropy observables. The first is the \emph{maximum entanglement entropy},
\begin{equation}
\label{eq:entropies_max}
  S_q \equiv \max_t S_q(t), \qquad
  S_c \equiv \max_t S_c(t),
\end{equation}
which quantifies the largest build-up of correlations during the charging process. The behavior of $S_q$ with respect to the system parameters allows us to identify the regimes in which  strong collective effects and superabsorption emerge, thereby serving as a witness of quantum advantage.

The second observable is the \emph{final entanglement entropy},
\begin{equation}
\label{eq:entropies_final}
  S_q^{f} \equiv S_q(t_f), \qquad
  S_c^{f} \equiv S_c(t_f),
\end{equation}
evaluated at a time $t_f \gg \tau$ well after the charging peak, once non-stationary processes have passed. These final entropies capture the residual quantum correlations and classical mixedness remaining in the battery and cavity after the protocol has been carried out. This allows us to probe the stability and resistance to decoherence of the stored energy.

\subsection{Dissipation channels and master equation}

The qubit–cavity system is typically not isolated, but rather interacts with the external environment,  inducing irreversible processes that degrade coherence and reduce charging efficiency. When modeling the dissipative quantum dynamics of the QB, we consider 
the following primary dissipation channels:

\textit{Cavity leakage}: photons escape into external modes at a rate $\kappa$, representing energy loss from the cavity. This is modeled by the collapse operator
\begin{equation}
\label{eq:cavity-leakage}
    \hat{C}_{\text{cav}} = \sqrt{\kappa}\, \hat{a}.
\end{equation}
\textit{TLS relaxation}: describes the spontaneous decay of a qubit from $\left|e\right>$ to $\left|g\right>$, releasing energy into the environment at a rate $\gamma^-$. This process 
discharges the battery by removing stored energy~\cite{carmichael1993opensystems, walls2008quantumoptics}. It is modeled by
\begin{equation}
\label{eq:tls-relaxation}
    \hat{C}_j^- = \sqrt{\gamma^-}\, \hat{\sigma}^-_j, \qquad j = 1, \ldots, N.
\end{equation}

\textit{TLS dephasing}: random environmental fluctuations randomize the phase between $\left|g\right>$ and $\left|e\right>$ without energy exchange. This suppresses coherence and qubit-cavity (bipartite) entanglement, reducing collective charging efficiency. It occurs at rate $\gamma^z$ and is described by
\begin{equation}
\label{eq:tls-dephasing}
    \hat{C}_j^z = \sqrt{\gamma^z}\, \hat{\sigma}^z_j, \qquad j = 1, \ldots, N.
\end{equation}

The dynamics of the driven qubit–cavity system under dissipation can be modeled using the Lindblad master equation in the Born–Markov approximation~\cite{masterEq,gorini1976completely,lindblad1976generators,breuer2002theory}:
\begin{equation}
\label{eq:lindblad}
    \frac{d\hat{\rho}}{dt} = -\frac{i}{\hbar}\,[\hat{H}(t), \hat{\rho}] 
    + \mathcal{L}[\hat{C}_{cav}]  + \sum_{j=1}^{N}\left( \mathcal{L}[\hat{C}^-_j] + \mathcal{L}[\hat{C}^z_j]\right),
\end{equation}
where $\hat{\rho}\equiv\hat{\rho}(t)$ is the density operator of the full system, $\hat{H}(t)$ is given in Eq.~\eqref{eq:full-hamiltonian}, and the Lindblad terms are given by $\mathcal{L}[\hat{C}]=\hat{C}\, \hat{\rho}\, \hat{C}^\dagger - \frac{1}{2}\{\hat{C}^\dagger\, \hat{C}, \hat{\rho}\}$, where the $\hat{C}$ operators   are as defined in Eqs.~\eqref{eq:cavity-leakage}–\eqref{eq:tls-dephasing}. For the rest of the paper we adopt natural units, setting $\hbar = 1$. 

The Lindblad equation provides a rigorous Markovian description of open-system dynamics, ensuring the positivity and trace preservation of the density operator $\hat{\rho}(t)$~\cite{breuer2002theory,masterEq,gorini1976completely,lindblad1976generators}. This framework enables a consistent comparison of RWA and non-RWA dynamics under Markovian dissipation.

\section{Numerical implementation}
\label{sec:numerical}

Simulating the Lindblad master equation for an open system comprising many qubits is computationally prohibitive due to the exponential growth of the state space. Evolving a density matrix of size $2^N \times 2^N$ becomes numerically intractable for $N \gtrsim 10$~\cite{breuer2002theory}. This challenge has also been recognized in QB studies, where exact diagonalization has been limited to small $N$~\cite{ferraro2018, andolina2019PRL}. More generally, as the Hilbert space dimension grows as $2^N$, the Liouvillian space has a dimension $4^N$, rendering brute-force integration infeasible. Approximate approaches have therefore been explored to address this issue. Cumulant expansions (mean-field truncations) can capture low-order correlations, but miss finer quantum details~\cite{shammah2018piqs}. Quantum trajectory methods (Monte Carlo wavefunctions) unravel the master equation into stochastic pure-state evolutions, reducing the complexity by averaging over many realizations~\cite{dalibard1992wave,dum1992montecarlo}.

In this work, we use the open-source Permutational Invariant Quantum Solver (PIQS) library~\cite{shammah2018piqs}, integrated into QuTiP~\cite{johansson2012qutip, johansson2013qutip, lambert2024qutip5quantumtoolbox}, which is designed for ensembles of indistinguishable TLSs. The key idea is that, when all qubits are identical and couple symmetrically to the environment, the dynamics are confined to the symmetric Dicke manifold.  Rather than having a dimension  $2^N$, this subspace has dimension $N+1$ (spanned by Dicke states $\left|J=N/2,m\right>$ with $m=-J,\ldots,+J$), providing a dramatic reduction in complexity~\cite{tavis1968exact}. By exploiting this symmetry, PIQS constructs the Liouvillian directly in the Dicke basis, yielding polynomial scaling in $N$ rather than exponential scaling~\cite{shammah2018piqs}. The qubit ensemble is represented via the collective spin operators $\hat{J}_{x,y,z}$ and $\hat{J}_\pm$. The qubit Hamiltonian is proportional to $\hat{J}_z$, the qubit–cavity interactions couple $\hat{J}_\pm$ to $a$ and $a^\dagger$, and dissipation channels are reformulated consistently: collective relaxation ($\gamma^-$) and dephasing ($\gamma^z$) rates are proportional to Lindblad collapse operators, $\hat{J}_-$ and $\hat{J}_z$, respectively, while cavity leakage remains a local channel proportional to $a$ ($\propto a$). This representation enables the efficient construction and integration of the Liouvillian while preserving full quantum correlations.

For the bosonic mode, we truncate the cavity Hilbert space to the Fock basis $\{|n\rangle\}_{n=0}^{n_{\max}}$ with $n_{\max}=10$. An analytical bound and an explicit convergence test are provided in Appendix~\ref{app:cutoff}.

We simulated QBs with qubit numbers ranging from $N=5$ to $N=50$ in steps of five, and coupling strengths $0 \leq g/\omega_q \leq 1$. The parallel charging protocol \cite{campaioli2017enhancing} was implemented and dissipation was included through cavity leakage ($\kappa/\omega_q \in [10^{-3},1]$), qubit relaxation ($\gamma^-/\omega_q \in [10^{-3},1]$), and qubit pure dephasing ($\gamma^z/\omega_q \in [10^{-3},1]$). 

We performed two complementary sets of simulations under the parallel charging protocol, where all the qubits all coupled to a single cavity mode within the same cavity, for both driven and undriven scenarios. The first set focused on quasi-ideal dynamics, where all dissipation channels were fixed at their minimum values. This enabled us to isolate the impact of coherent qubit–cavity interactions by examining how energy and entropy scaled with the coupling strength $g$ and system size $N$. The second set incorporates open-system dynamics with fixed $g/\omega_q=0.1$. Here we distinguish two regimes: (i) cavity effects, in which we analyze the interplay between cavity loss $\kappa$ and system size $N$, while keeping qubit relaxation and dephasing to a minimum; and (ii) decoherence effects, in which we explore the combined impact of qubit relaxation $\gamma^-$ and pure dephasing $\gamma^z$ across all system sizes,  while fixing cavity loss to its minimum value. 

In the \textit{undriven} scenario, the $N$-qubit ensemble is initialized in its ground state, while the cavity mode is prepared in an excited Fock state.
In the \textit{driven} setup, both the qubits and the cavity start in their ground states. A resonant Gaussian pulse ($\Delta_d= 0$), 
with amplitude $\eta_0=\omega_q$, width $\sigma=2.0/\omega_q$, and center $t_0=5.0/\omega_q$ then injects energy into the system through the cavity. These pulse parameters ensure a controlled, resonant charging process: the amplitude is comparable to the light–matter coupling for efficient, balanced excitation; the width provides a smooth, broadband drive faster than the dissipative timescales ($
\omega_q/\gamma^{\{z,-\}}\!\sim\!10^3$); and the early pulse timing leaves an ample evolution window. This guarantees stability and highlights intrinsic differences between Dicke and Tavis--Cummings batteries rather than artifacts of the drive.

For each parameter configuration, the time evolution of the thermodynamic observables introduced in Sec.~\ref{subsec:observables} is computed. From these data, we extract the maximum stored energy,  $E_{\max}$, the charging time, $\tau$, the maximum power, $\bar{P}_{\max}$, and the maximum (final) entanglement entropies, $S_q\,(S_q^{f})$ and $S_c \,(S_c^{f})$. Heatmaps are then generated over the dissipation parameter space  and scaling exponents $\alpha$ are obtained from log–log fits of observables versus system size.

\section{Results}
\label{sec:results}

We analyze the dynamics of the QB across different regimes. First, we characterize the dynamics in a scenario where the dissipation channels are included but remain very weak compared to the system frequencies, i.e. $\kappa,\, \gamma^-, \, \gamma^z\ll \omega_q$, such that coherence is long-lived. For this reason, we refer to this as a \textit{quasi-ideal} cavity. This regime effectively captures near-coherent evolution while allowing a realistic description of residual losses, and serves as a benchmark for optimal performance. Next, we incorporate open-system dynamics and analyze the effects of cavity photon leakage and environmental decoherence via dephasing and relaxation. This analysis demonstrates how specific dissipation channels impact the scaling and stability of energy storage. Finally, we provide a detailed examination of the scaling laws of thermodynamic observables, extracting the scaling exponents of $E_{\max}$, $\tau$, and  $\bar{P}_{\max}$. We also identify regimes of enhanced performance and entanglement entropy suppression.
\begin{figure}[!h]
    \centering
    \includegraphics[width=1.0\linewidth]{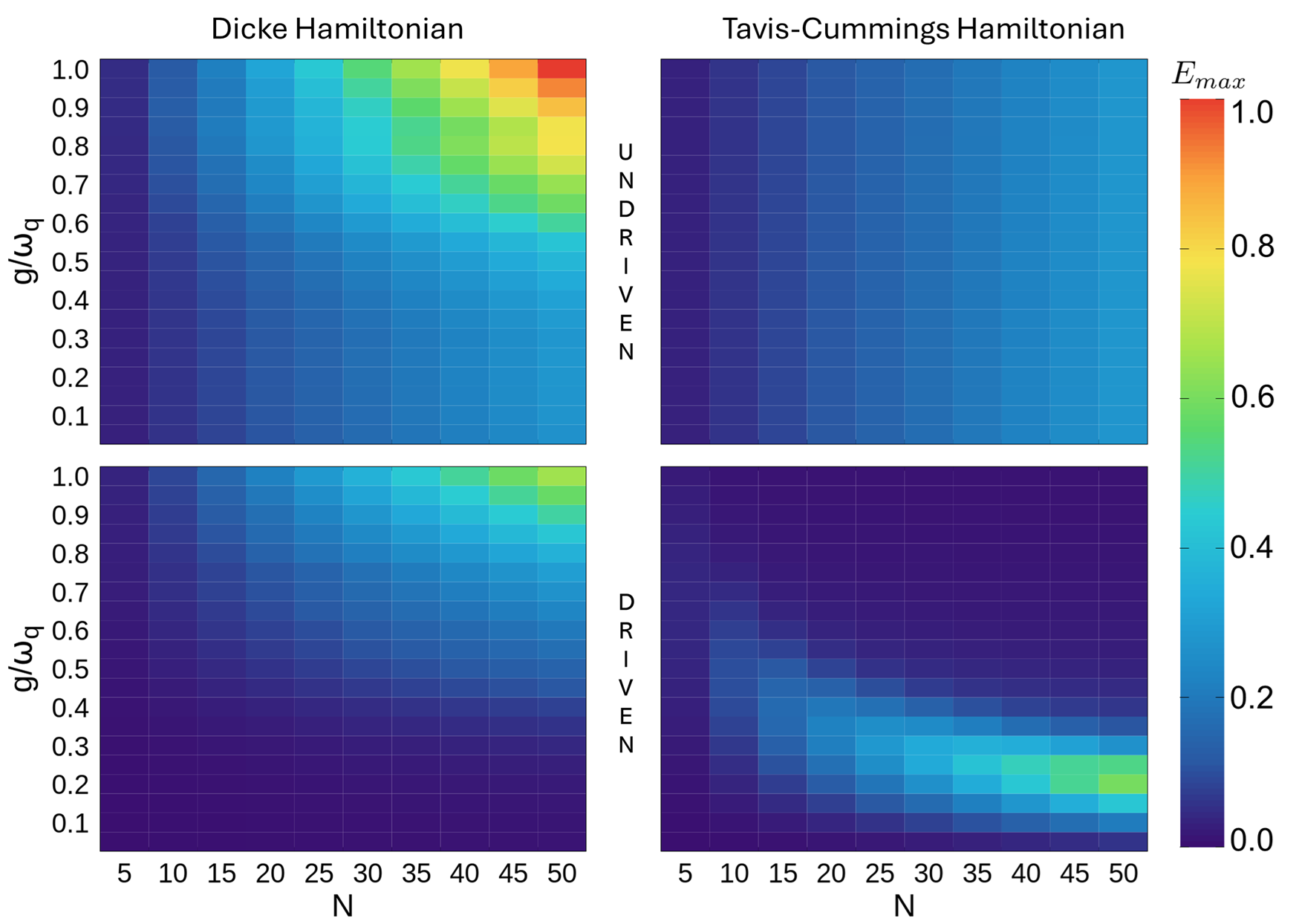}
    \caption{\textbf{Maximum stored energy $E_{\max}$} as a function of qubit number $N$ and coupling strength $g/\omega_q$; 
$\kappa/\omega_q = \gamma^z/\omega_q = \gamma^-/\omega_q=10^{-3}$.
    Results are shown for Dicke (left) and Tavis--Cummings (right) Hamiltonians in the undriven (top) and driven (bottom) setups. Color scales are row-normalized: for each setup (driven or undriven), we divide by the maximum $E_{\max}$ across both scenarios and for all $(N,g/\omega_q)$ conditions.}
    \label{fig:energy-nodiss}
\end{figure}

\subsection{Quasi-ideal cavity dynamics}

In the undriven scenario, the Tavis--Cummings model exhibits full energy revivals with Rabi oscillations at frequency $\omega \propto \sqrt{N}\,g$~\cite{tavis1968exact}, where the entanglement entropy oscillates periodically and nearly synchronously with the stored energy, reflecting coherent and reversible exchange. The Dicke model only reproduces this behavior for small $g$ and $N$. At stronger coupling ($g \sim \omega_q$), energy transfer becomes faster but less reversible, with delayed and broadened entropy peaks that remain large even as energy decays, evidencing persistent entanglement between the qubits and the cavity field. These observations are based on the real-time evolution of energy and entanglement entropy, not shown here for brevity. However, it reproduces the standard revival dynamics that are well established in the literature~\cite{eberly1980periodic,phoenix1991collapse, shore1993jaynes, garraway2011decay,zhang2014revival}.
\begin{figure}[!h]
    \centering
\includegraphics[width=1.0\linewidth]{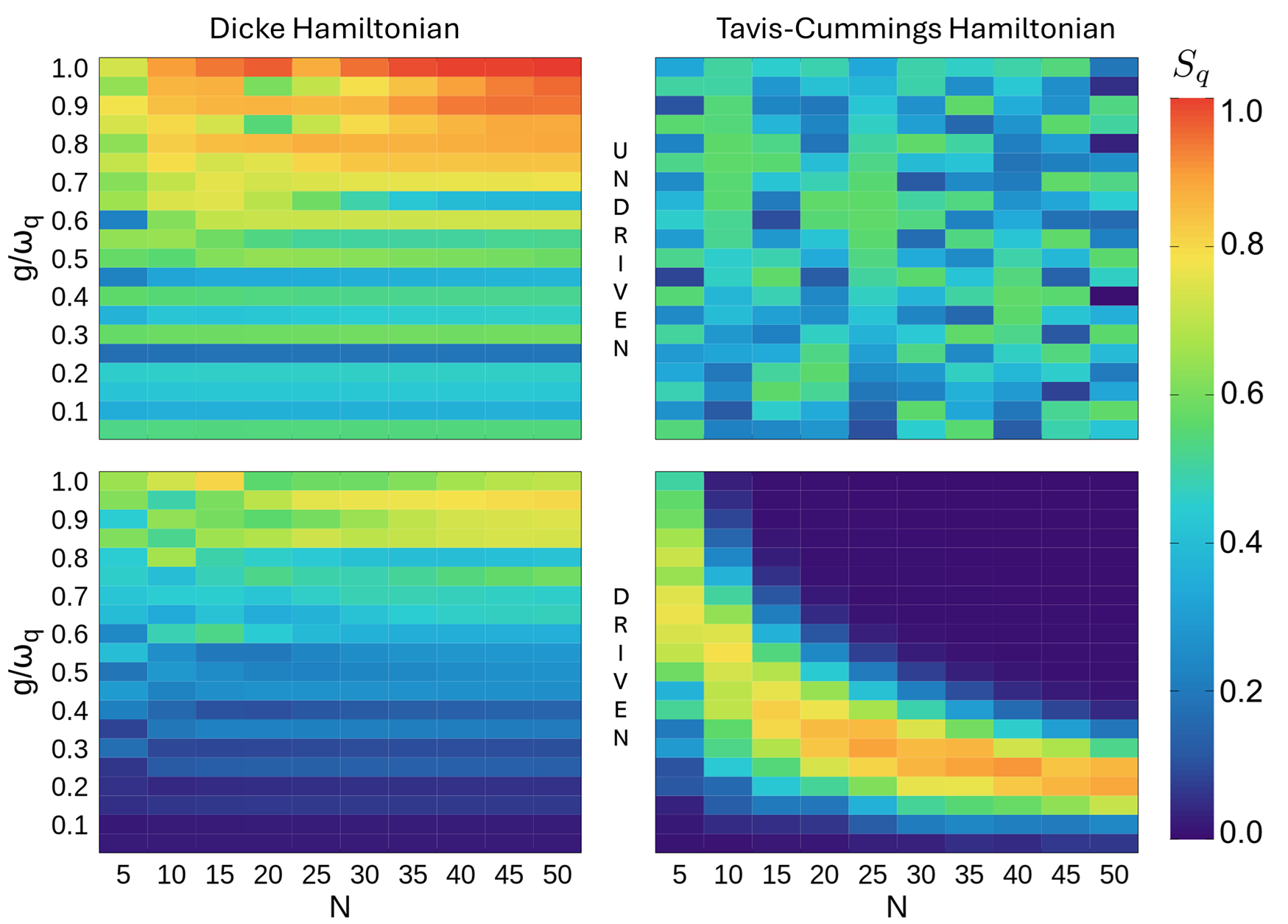}
\caption{\textbf{Entanglement entropy $S_q$}, as a function of qubit number $N$ and coupling strength $g/\omega_q$, with the same parameters as in Fig.~\ref{fig:energy-nodiss}. The panels compare the Dicke (left) and Tavis--Cummings (right) Hamiltonians for the undriven (top) and driven (bottom) setups. The color scales are row-normalized as in Fig.~\ref{fig:energy-nodiss}, but using the maximum value of $S_q$.}
\label{fig:entropy-nodiss}
\end{figure}
\begin{figure}[!h]
    \centering
\includegraphics[width=1.0\linewidth]{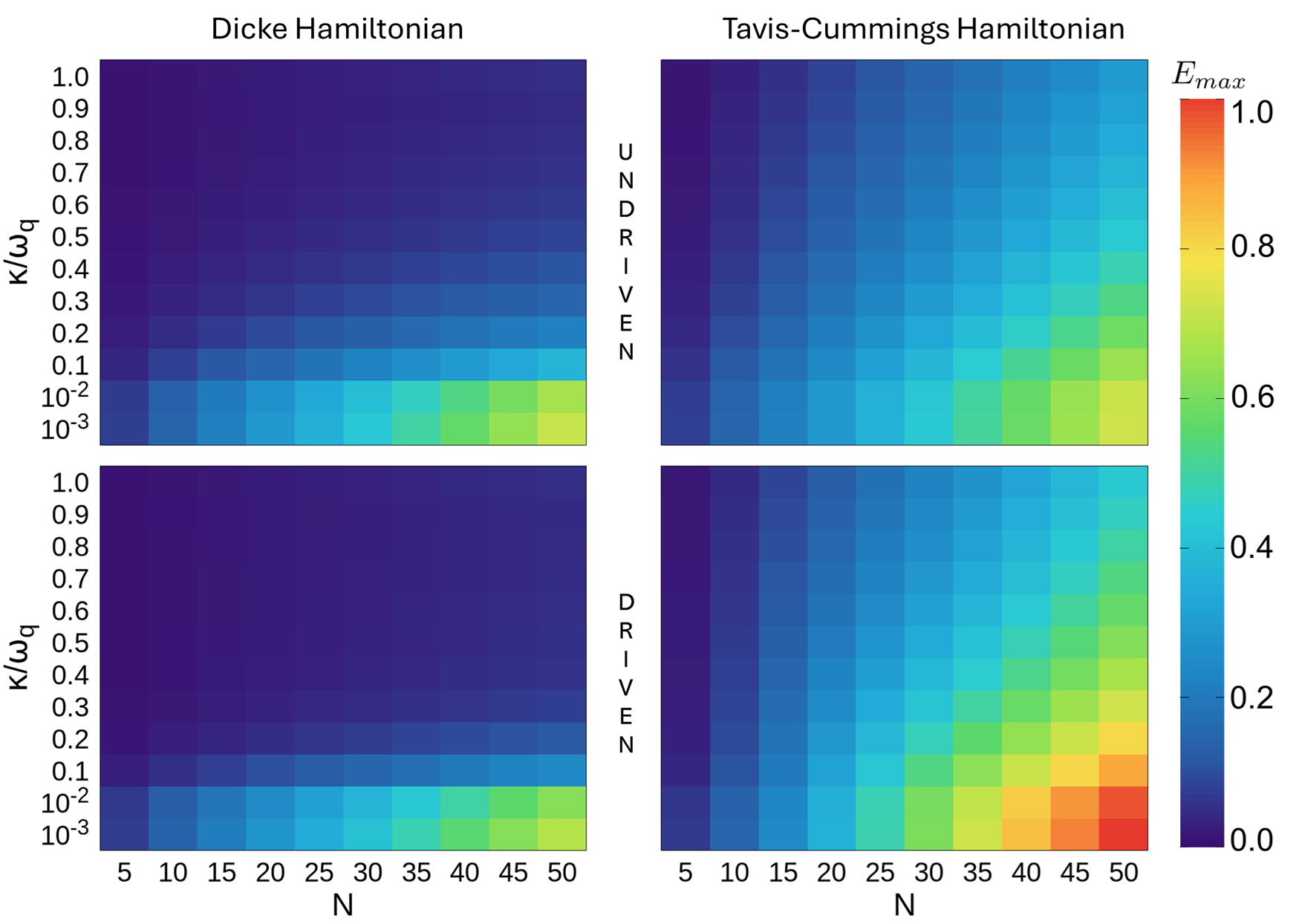}
\caption{\textbf{Maximum stored energy $E_{\max}$} as a function of $\kappa$ and qubit number $N$; $g/\omega_q = 0.1$, $\gamma^z/\omega_q = \gamma^-/\omega_q=10^{-3}$. The color scale is row-normalized, as in Fig.~\ref{fig:energy-nodiss}, using the maximum value of $E_{max}$. The $\kappa$ rates are varied over the discrete set of values 
$\{10^{-3},\,10^{-2},\,0.1,\,0.2,\,0.3,\,0.4,\,0.5,\,0.6,\,0.7,\,0.8,\,0.9,\,1.0\}$.}
\label{fig:energy_cav}
\end{figure}

In the Dicke model, the stored energy, see Fig.~\ref{fig:energy-nodiss},  exhibits growth with $N$ when $g/\omega_q \gtrsim 0.5$, highlighting the role of counter-rotating processes in enhancing capacity. In contrast, the Tavis--Cummings model displays nearly vertical contours, indicating that energy absorption remains approximately constant across different coupling strengths and depends primarily on the system size $N$. When driven, the Dicke model retains its advantage, reaching maximum  energy value at $g\sim \omega_q$. 
In contrast, the Tavis--Cummings model only 
shows localized enhancement around $g/\omega_q \sim 0.1$--0.4 for large $N$, where the coupling strength and driving frequency are optimally matched.

Figure~\ref{fig:entropy-nodiss} plots the qubit entanglement entropy as function of the coupling strength and the system size. The Dicke model exhibits strong and structured qubit--cavity correlations for $g/\omega_q \gtrsim 0.5$, which persist as $N$ increases. When driving is introduced, the dynamics of the Dicke model remain largely unchanged; in the Tavis--Cummings model, however, laser pumping dramatically enhances correlations within a narrow resonance band~\cite{bastidas2010entanglement}, where the Gaussian pulse bandwidth $\Delta\omega \approx 1/\sigma = \omega_q/2$ matches the collective Rabi frequency $g\sqrt{N}$, producing the localized entanglement enhancement observed in Fig.~\ref{fig:entropy-nodiss}.

Thus, coherent driving introduces controlled energy injection without altering the Dicke dynamics qualitatively, whereas in the Tavis--Cummings model, it selectively amplifies correlations near resonance. This establishes coherent driving as an effective mechanism for initiating collective charging and entanglement.

\subsection{Open-system dynamics}

Firstly, we examine the effect of cavity leakage, 
which accounts for photon losses from the bosonic mode, reflecting the cavity's level of non-ideality. Secondly, we examine the impact of dephasing 
and energy relaxation 
on the $N$-qubit ensemble.  
By  independently varying their strengths, we can identify regimes in which moderate dissipation improves performance, stabilizes entanglement entropy production and  enhances scalability.
\begin{figure}[!h]
    \centering
\includegraphics[width=1.0\linewidth]{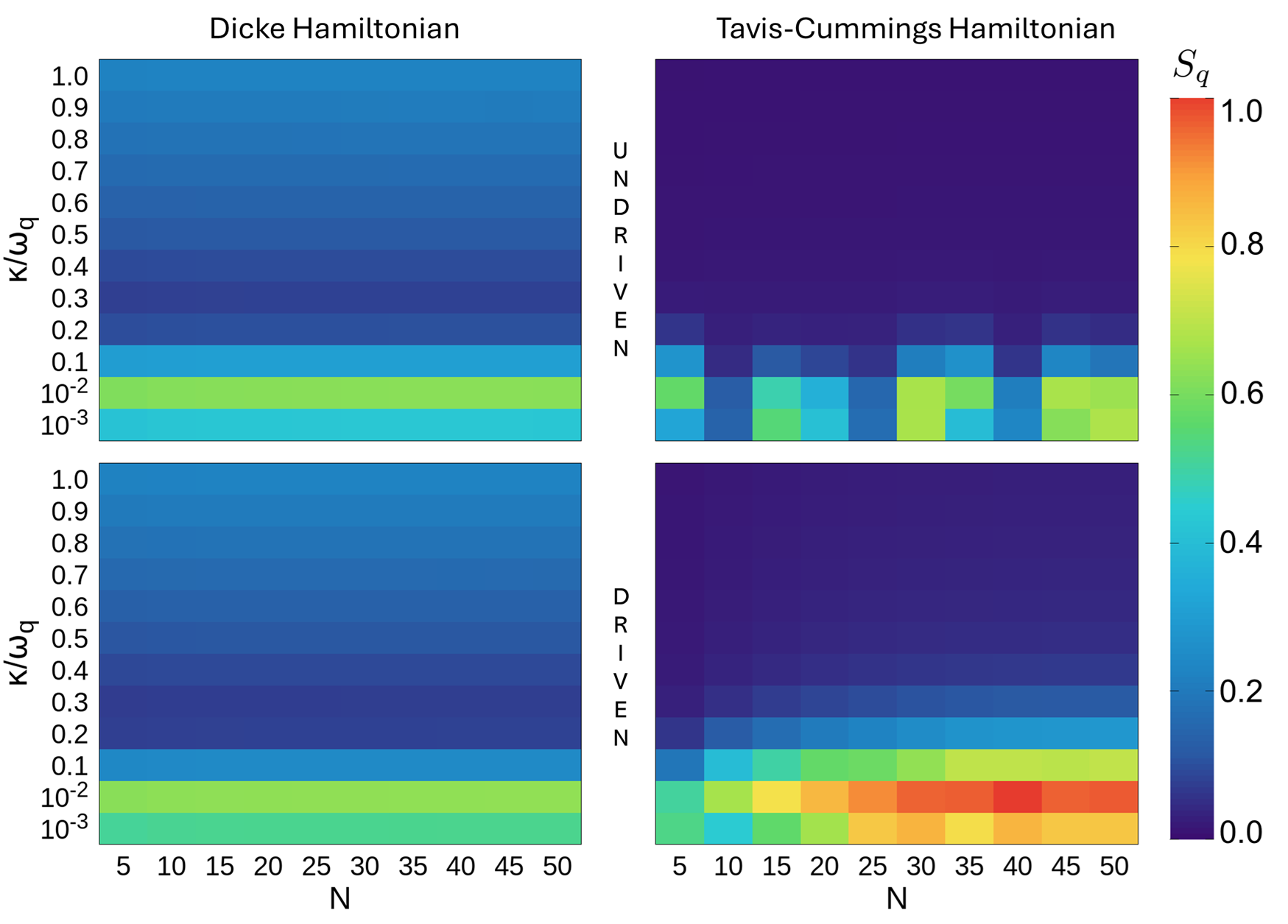}
\caption{\textbf{Maximum entanglement entropy $S_q$} as a function of $\kappa$ and qubit number $N$, with same parameters and $\kappa$ values as in Fig.~\ref{fig:energy_cav}. Results are shown for Dicke (left) and Tavis--Cummings (right) Hamiltonians under undriven (top) and driven (bottom) conditions. Color scale is row-normalized, as in Fig.~\ref{fig:energy-nodiss}, using the maximum value of $S_q$.}
\label{fig:entropy_cav}
\end{figure}

\subsubsection{Cavity leakage 
effects}

We begin the analysis by 
setting  the coupling strength to $g/\omega_q = 0.1$ and fixing the dephasing and relaxation rates to $\gamma^z/\omega_q = \gamma^-/\omega_q=10^{-3}$. 
As the dominant channel of energy loss in cavity-mediated QBs, leakage competes directly  with the charging dynamics by enabling excitations to escape into external modes. By varying $\kappa$ over a broad range, we can determine the 
effect of cavity imperfections on energy storage capacity and qubit–cavity correlations.
Figure~\ref{fig:energy_cav} shows that the Tavis--Cummings Hamiltonian maintains its characteristic linear scaling with $N$ for the maximum stored energy, across all values of cavity leakage $\kappa$, with nearly linear growth even at large dissipation rates. In contrast, the Dicke model  quickly saturates;  energy increases with $N$ only for very small $\kappa$, and decreases as leakage increases, indicating strong susceptibility to photon loss. Under driving, this asymmetry becomes more pronounced. The Tavis--Cummings model retains its linear scaling, while the Dicke model essentially flattens at low energy once appreciable cavity leakage occurs.
 These results demonstrate that counter-rotating processes, which in the quasi-ideal limit enhanced capacity, here accelerate the degradation of energy storage under cavity losses.

In the Dicke model, the maximum entanglement entropy, see Fig.~\ref{fig:entropy_cav}, displays a pronounced minimum in the region $\kappa/\omega_q \approx 0.2-0.4$, in both undriven and driven scenarios. For leakage rates that are smaller or larger, the entropy rises again. This shows that moderate photon losses briefly suppress qubit-cavity correlations before stronger dissipation re-induces mixedness through residual cavity coupling. The trend is more evident in the undriven scenario, where the entropy peaks at low $\kappa$, while driving smooths the contrast and slightly enhances correlations at larger $N$. In contrast, in the Tavis--Cummings model, the entropy is almost zero for most parameters, with small oscillatory bands near $\kappa/\omega_q \approx 0.1-0.2$ that disappear under driving. The applied drive then produces a monotonic increase in $S_q$ with system size for decreasing $\kappa$, resulting in maximal correlations at large $N$ and low leakage.

\subsubsection{Dephasing 
and Relaxation 
interplay}


We now turn to the combined influence of qubit dephasing and relaxation. For this analysis, we set the coupling strength to $g/\omega_q = 0.1$ and fix the cavity leakage rate to $\kappa/\omega_q = 10^{-3}$, allowing us to isolate the interplay between $\gamma^-$ and $\gamma^z$. Both the relaxation and dephasing rates, $\gamma^-/\omega_q$ and $\gamma^z/\omega_q$, 
are varied over the same discrete set of values 
$\{10^{-3},\,10^{-2},\,0.1,\,0.2,\,0.3,\,0.4,\,0.5,\,0.6,\,0.7,\,0.8,\,0.9,\,1.0\}$.

\begin{figure*}[!h]
    \centering
\includegraphics[width=0.7\linewidth]{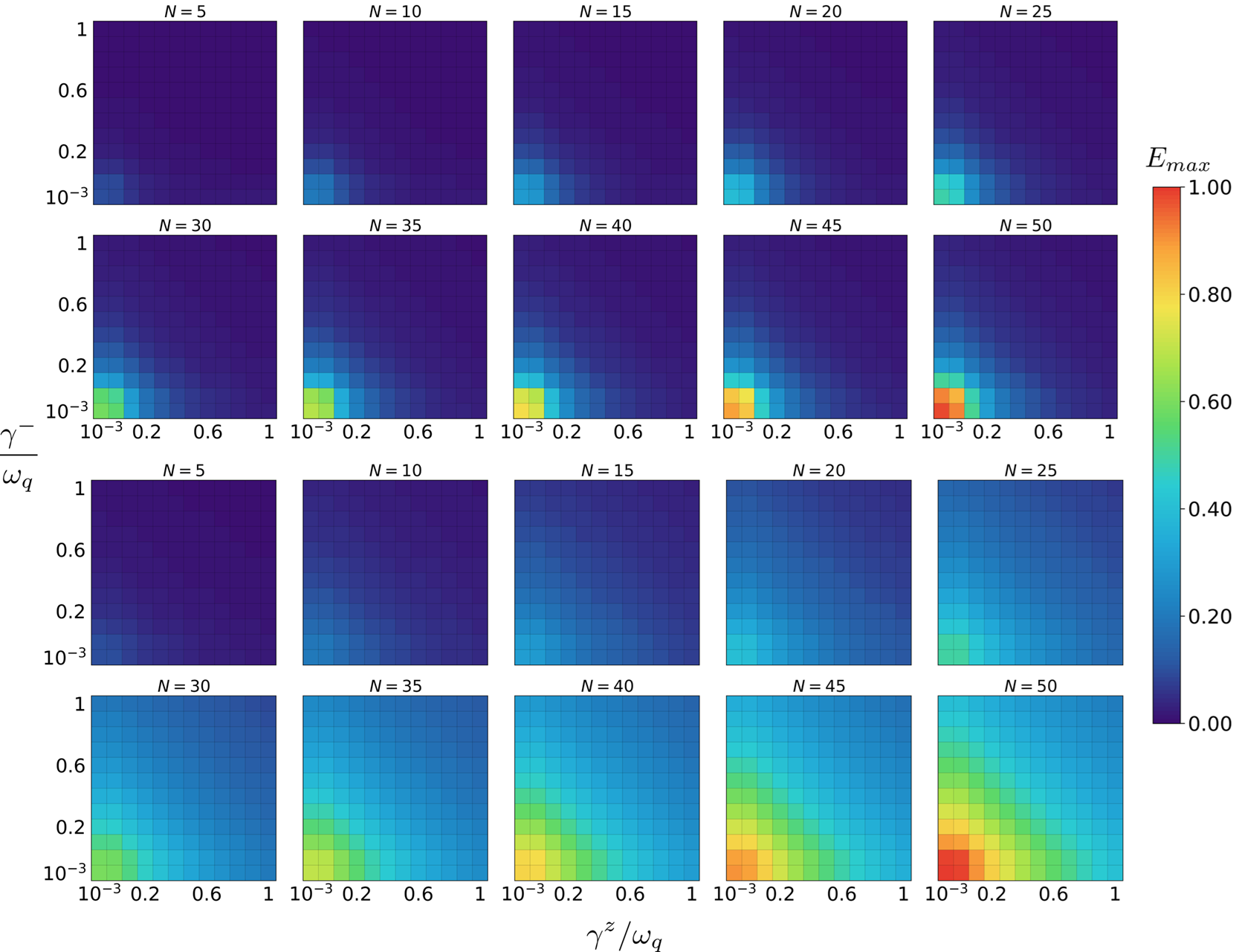}
\caption{Heatmaps of the maximum stored energy $E_{\max}$ as a function of  $\gamma^-$ and $\gamma^z$, for different qubit number $N$;  Dicke (top panels) and Tavis--Cummings (bottom panels) models. 
$g/\omega_q=0.1$ and $\kappa/\omega_q=10^{-3}$. Both the relaxation and dephasing rates, $\gamma^-/\omega_q$ and $\gamma^z/\omega_q$, are varied over the same discrete set of values 
$\{10^{-3},\,10^{-2},\,0.1,\,0.2,\,0.3,\,0.4,\,0.5,\,0.6,\,0.7,\,0.8,\,0.9,\,1.0\}$. Color scale is model-normalized, 
    using the maximum $E_{\max}$ within the corresponding panel set.}
\label{fig:decoh_energy}
\end{figure*}

For maximum stored energy, see Fig.~\ref{fig:decoh_energy} (upper panels), the Dicke model shows a clear region of enhanced energy storage extending up to approximately $\gamma^-/\omega_q,\,\gamma^z/\omega_q \approx 0.4$ as system size increases, showing that collective coupling amplifies the tolerance to moderate noise. Within this range, both dissipation channels contribute constructively without strongly degrading energy storage capacity. Beyond this threshold, energy storage rapidly decays as losses dominate.

In contrast, the Tavis–Cummings model (Fig.~\ref{fig:decoh_energy}, lower panels) shows a broader stability domain, with significant energy retention persisting up to $\gamma^-/\omega_q,\,\gamma^z/\omega_q \approx 0.8$, also depending on system size. Although the overall dependence is smoother and less sharply peaked than in the Dicke case, the RWA dynamics exhibit greater tolerance to dissipation, maintaining substantial energy storage even under relatively strong dephasing and relaxation.
\begin{figure*}[!hb]
    \centering
\includegraphics[width=0.7\linewidth]{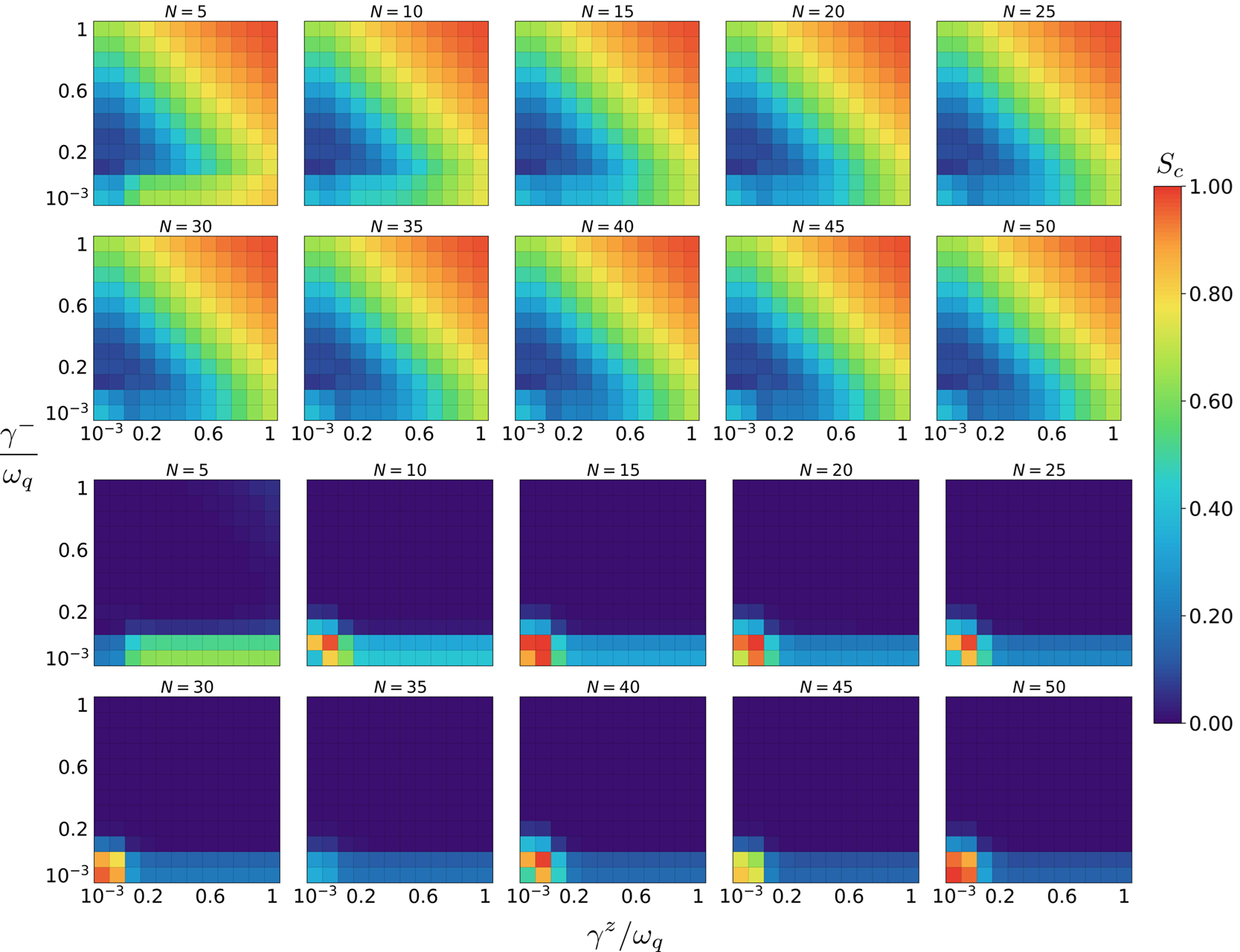}
\caption{Heatmaps of the cavity entanglement entropy $S_{\mathrm{c}}$ as a function of $\gamma^-$ and  $\gamma^z$, for different qubit number $N$.   $g/\omega_q=0.1$ and $\kappa/\omega_q=10^{-3}$. The Dicke model (top) exhibits broad regions of finite entropy, while the Tavis--Cummings model (bottom) strongly suppresses entropy generation. Color scale is model-normalized, as in Fig.~\ref{fig:decoh_energy}, using the maximum value of $S_{c}$ or $S_{c}^{RWA}$, with the relaxation and dephasing rates varied over the same range as in that figure.}
\label{fig:decoh_caventropy}
\end{figure*}
\begin{figure*}[!h]
    \centering
\includegraphics[width=0.7\linewidth]{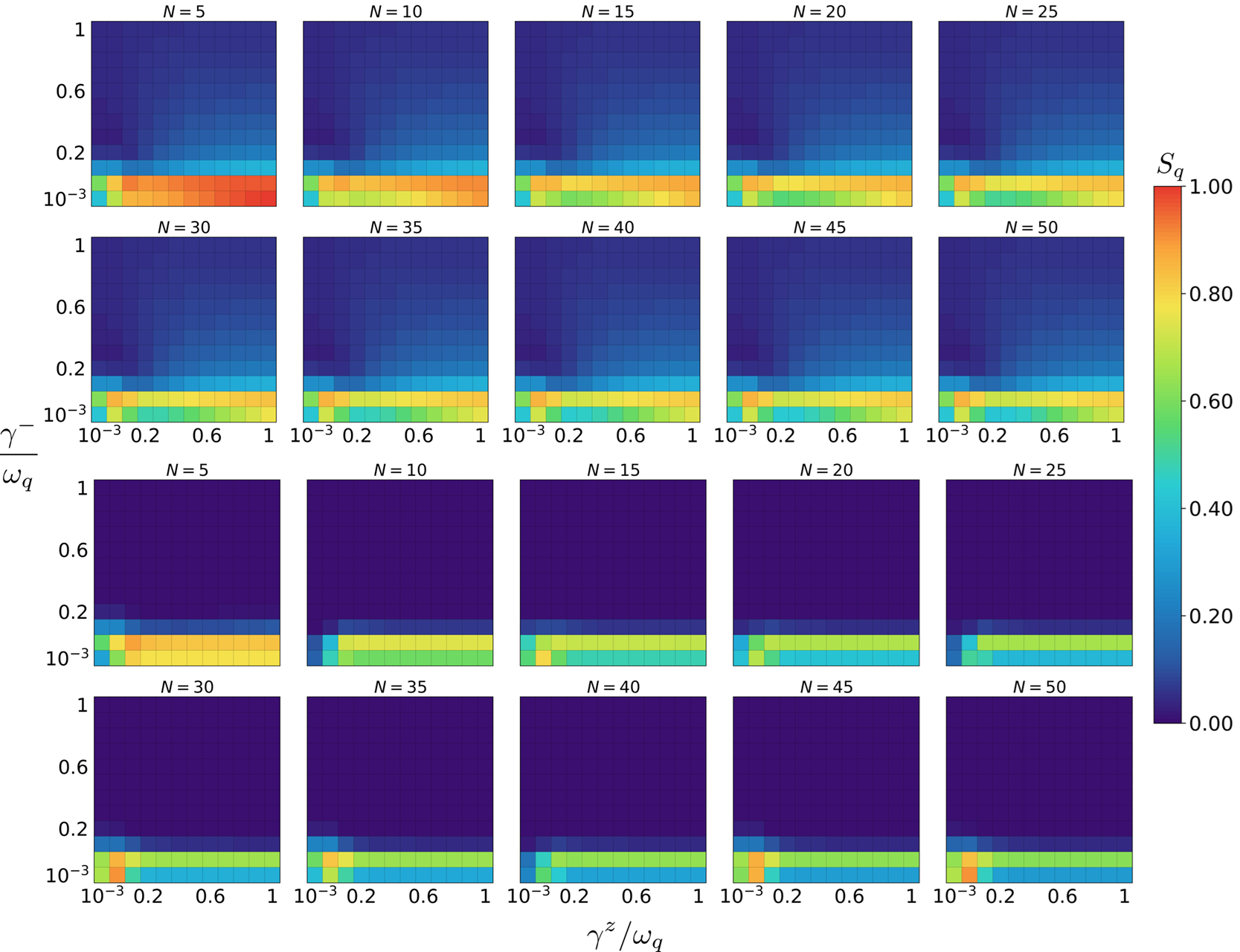}
\caption{Heatmaps of the qubit entanglement entropy $S_q$ as a function of $\gamma^-$ and $\gamma^z$, for different qubit number $N$. Parameters used are as in Fig.~\ref{fig:decoh_energy}. In the Dicke model (top), entropy is sustained under low relaxation and moderate dephasing, whereas the Tavis--Cummings model (bottom) predicts negligible entropy across almost all regimes. Color scale is model-normalized, 
    using the maximum value of $S_{q}$ or $S_{q}^{RWA}$, with the relaxation and dephasing rates varied over the same range as in that figure.}
\label{fig:decoh_tlsentropy}
\end{figure*}
\begin{figure*}[!hb]
    \centering
\includegraphics[width=0.7\linewidth]{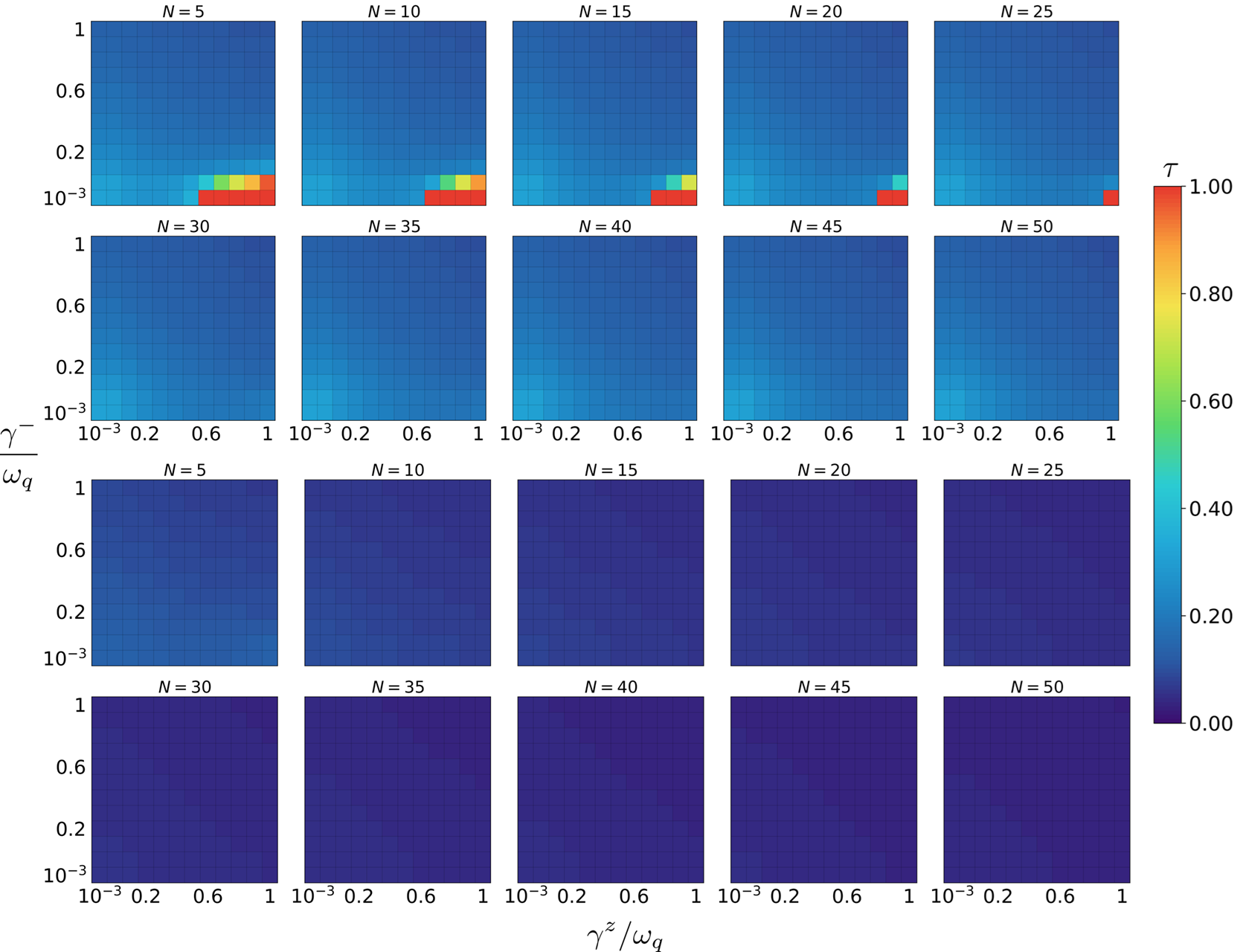}
\caption{Heatmaps of the charging time $\tau$ as a function of $\gamma^-$ and $\gamma^z$, for different qubit number $N$.
The Dicke model (top) shows nearly constant and short charging times, while the Tavis-Cummings model (bottom) overestimates dissipation-induced delays. Color scale is model-normalized, 
    using the maximum value of $\tau$ or $\tau^{RWA}$, with the relaxation and dephasing rates varied over the same range as in that figure.}
\label{fig:decoh_chargingtime}
\end{figure*}

Figure~\ref{fig:decoh_caventropy} plots the cavity entanglement entropy; this exhibits markedly different behavior for the two models. In the Dicke model (top panels), $S_c$ exhibits a broad diagonal gradient: entropy remains low when either relaxation or dephasing is small, but increases significantly once both channels act together, roughly along the diagonal $\gamma^- \approx \gamma^z$ a banded pattern emerges, with $S_c$ oscillating between a suppression band $\gamma^-/\omega_q,\,\gamma^z/\omega_q \approx 0.2-0.6$ and growth as the dissipation rates increase. This non-monotonic behavior reflects the competition between the collective coherent exchange frequency $\sqrt{N}\,g$ (amplified by counter-rotating processes) and the dissipation rates $\gamma^-,\gamma^z$.

In the Tavis–Cummings model, see Fig.~\ref{fig:decoh_caventropy} (lower panels), the cavity entanglement entropy remains strongly suppressed across most of the dissipation plane, with non-trivial values confined to a narrow horizontal band for $\gamma^-/\omega_q \lesssim 0.2$ and $\gamma^z/\omega_q \lesssim 1.0$. As the system size increases, small localized hot spots emerge within this band around $\gamma^-/\omega_q,\gamma^z/\omega_q \lesssim 0.2$, showing strong cavity mixedness. Here, dephasing acts as a tunable channel that modulates entropy generation within the cavity.


In the Dicke model, the non-trivial values of the qubit entanglement entropy are confined to a narrow band extending up to $\gamma^-/\omega_q \lesssim 0.2$ and $\gamma^z/\omega_q \lesssim 1.0$; see Fig.~\ref{fig:decoh_tlsentropy}. Within this sector, a crossover region for $\gamma^-/\omega_q,\gamma^z/\omega_q \lesssim 0.2$, indicating the coexistence of low and moderate entanglement entropy production when both decay channels are comparable. Along the dephasing-dominated axis ($0.2 \lesssim \gamma^z/\omega_q \lesssim 1.0$), the hot ridge observed at small system sizes progressively cools down as $N$ increases, showing that collective coupling suppresses excessive mixing. Moreover, for $\gamma^-/\omega_q,\gamma^z/\omega_q \in [0.2,0.6]$, a distinct colder stripe emerges as system size increases, indicating the formation of a regime of moderate and stabilized entanglement entropy production. This behavior suggests that increasing the number of qubits enhances the resistance of the ensemble to both relaxation and dephasing, leading to controlled entanglement production.

In the Tavis--Cummings model (see Fig.~\ref{fig:decoh_tlsentropy}), the overall trend is similar:  entropy values remain low across most of the dissipation plane and are confined to a narrow horizontal band for $\gamma^-/\omega_q \lesssim 0.2$. As with the Dicke model,  dependence on system size is less pronounced, confirming that, even in the presence of dephasing, excitation-number conservation within the RWA limits entanglement generation.


\begin{figure*}[!ht]
    \centering
\includegraphics[width=0.7\linewidth]{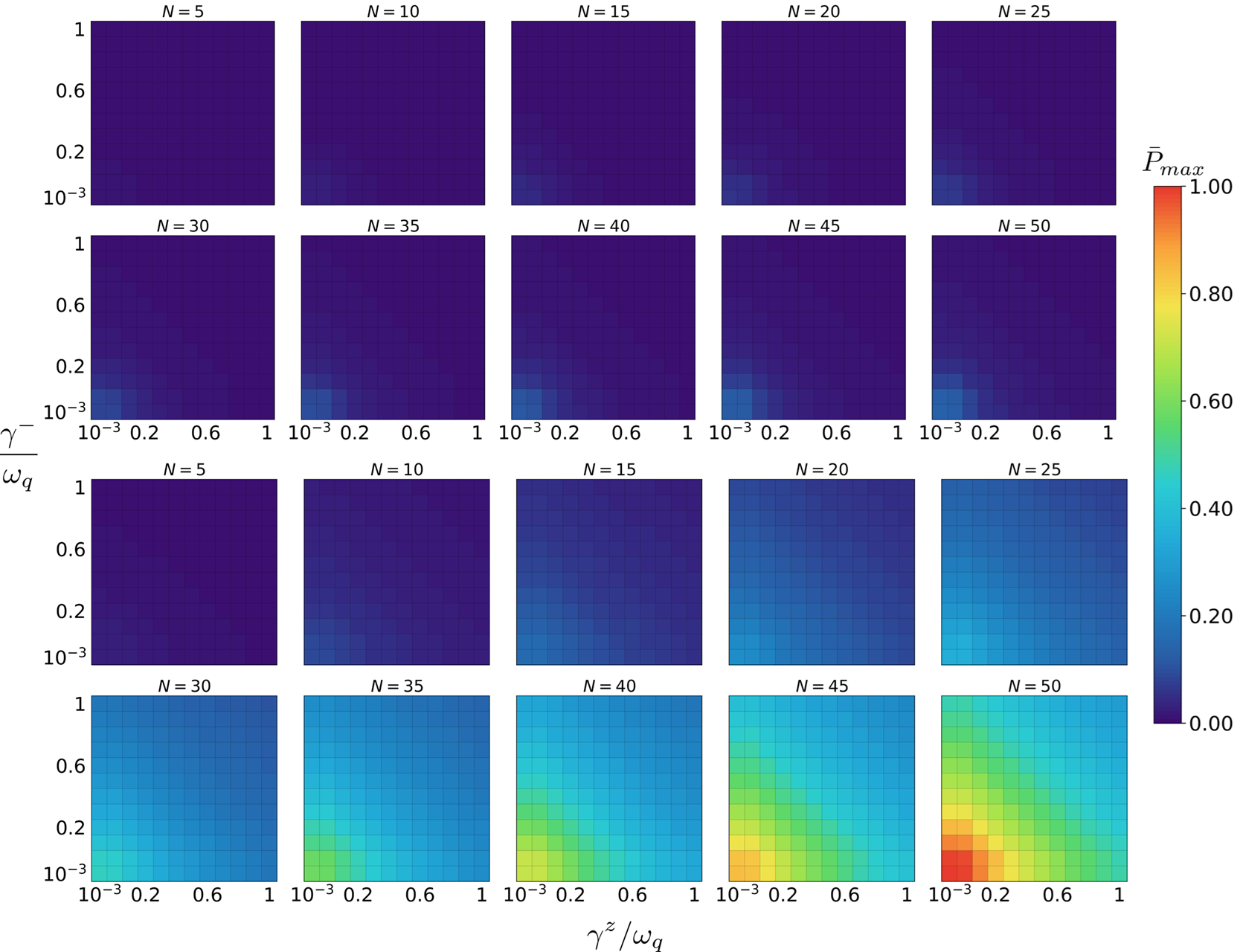}
\caption{Heatmaps of the maximum power $\bar{P}_{\max}$ as a function of  $\gamma^-$ and  $\gamma^z$, for different qubit number $N$.
    The Dicke model (top) shows enhancement only in very restricted regions, while the Tavis--Cummings model (bottom) predicts more robust and broadly distributed power scaling. Color scale is  model-normalized, 
    using the maximum value of $\bar{P}_{max}$ or $\bar{P}_{max}^{RWA}$, with the relaxation and dephasing rates varied over the same range as in that figure.}
\label{fig:decoh_power}
\end{figure*}

As shown in Fig.~\ref{fig:decoh_chargingtime}, charging time shows a homogeneous behavior in the Dicke model (upper panels), remaining small and nearly independent of the dissipation rates. This implies that the battery charges efficiently, even in the presence of decoherence channels. In contrast, the Tavis--Cummings (lower panels) model predicts significant delays: for $\gamma^-/\omega_q \lesssim 0.2$ and $0.6 \lesssim \gamma^z/\omega_q $, the charging time increases noticeably, for relatively small number of qubits ($N \lesssim 25$) suggesting that dissipation slows down the charging process. However, as system size increases ($N \gtrsim 30$), this dissipation-induced delay tends to disappear.

Figure~\ref{fig:decoh_power} shows that the maximum power closely follows the
trends in stored energy because $\tau$ is nearly homogeneous in the Dicke case (see~Fig.~\ref{fig:decoh_chargingtime}). In the Dicke model (upper panels), $\bar P_{\max}$ is only appreciable within a narrow window of relaxation and dephasing ($\gamma^-/\omega_q,\,\gamma^z/\omega_q\, \!\lesssim\! 0.2$); outside this region, power is strongly suppressed as dissipation reduces $E_{\max}$. By contrast, the Tavis--Cummings model (lower panels) exhibits good performance over a broader domain as the system size increases ($N \gtrsim 30$), extending up to $\gamma^-/\omega_q,\gamma^z/\omega_q \!\sim\! 0.8$. RWA dynamics maintain scalable performance under dephasing and relaxation. Therefore, achieving high power in the non-RWA regime requires minimizing relaxation while operating at moderate dephasing. In the RWA  regime, however, a broader sector yields  better  overall performance.

The results presented in Figures~\ref{fig:decoh_energy}--\ref{fig:decoh_power} demonstrate that better overall  performance is associated with a moderate build–up of qubit entanglement entropy, which is degraded by cavity mixedness: $E_{\max}$ and $\bar P_{\max}$ are optimized when the qubit ensemble yields an intermediate value of $S_q$ within a dephasing–dominated corridor. However,  a large value of $S_{\mathrm{cav}}$ consistently anti-correlates with energy retention, power, and charging speed.

In the Dicke case, this favorable window is narrow and sensitive—approximately $\gamma^-/\omega_q,\,\gamma^z/\omega_q\, \!\lesssim\! 0.2$, and power mirrors $E_{\max}$ because $\tau$ remains nearly constant. With increasing $N$, the hot band in $S_q$ cools, indicating stabilized, controlled qubit entanglement. In contrast, under the Tavis--Cummings dynamics, the optimal sector is broader, extending up to $\gamma^-/\omega_q,\gamma^z/\omega_q\!\sim\!0.8$, and the weak delay band in $\tau$ only mildly degrades the power maps, yielding more size–resilient performance.

From a practical point of view, optimal operation requires keeping the cavity at low $S_{\mathrm{cav}}$, minimizing relaxation, and tuning dephasing to regulate $S_q$. This strategy enables collective absorption while avoiding excessive mixing. This leaves us with an optimal operating window of $\gamma^-/\omega_q \!\lesssim\! 0.20$ and $\gamma^z/\omega_q \!\lesssim\! 0.50$. The coherent drive enlarges these favorable regions without altering this hierarchy, primarily acting  as a stabilizer of entanglement entropy growth rather than a source of additional scaling advantage.


\subsection{Dynamical scaling laws of entanglement and thermodynamic observables}
\label{sec:scalinglaws}

To systematically determine whether cavity-coupled QBs provide a collective advantage, it is also important to compare performance scales as the number of qubits increases. Finite-size scaling laws show whether collective effects, such as super-extensivity, persist under dissipation. In this context, we model the dependence of an observable, $\mathcal{O}(N)$, on the system size by a power law
\begin{equation}
    \mathcal{O}(N) \sim N^\alpha,
    \label{eq:scaling-law}
\end{equation}
and extract the exponent $\alpha$ from log--log fits. Throughout, 
$g$ is kept fixed as $N$ increases; therefore $\alpha$ characterizes dynamical scaling laws, while quantum collective effects manifest as deviations from linearity~\cite{campaioli2018chapter,ferraro2018}. Hence, $\alpha$ serves as a compact figure of merit: $\alpha>1$ signals super-extensive enhancement (quantum advantage), $\alpha\approx 1$ corresponds to classical extensivity, and $\alpha<1$ indicates degraded scaling. The following analysis focuses on the optimal regime of low relaxation $\gamma^-/\omega_q \in [10^{-3},0.10]$ and moderate dephasing $\gamma^z/\omega_q \in [10^{-3},0.50]$. The fitted values of the scaling exponents can be found in Appendix~\ref{app:scalinglaws}.

Three characteristic domains emerge: (i) tuned-decay relaxation and dephasing ($\gamma^-/\omega_q,\, \gamma^z/\omega_q \in [10^{-3},10^{-2}]$), (ii) intermediate regime ($\gamma^-/\omega_q = \{10^{-2}\}$ and $\gamma^z/\omega_q \in [0.10-0.50]$), and (iii) dephasing-dominated ($\gamma^-/\omega_q = \{10^{-3}\}$ and $\gamma^z/\omega_q \in [0.10,0.50]$). 
\begin{figure}[!ht]
    \centering
\includegraphics[width=1.0\linewidth]{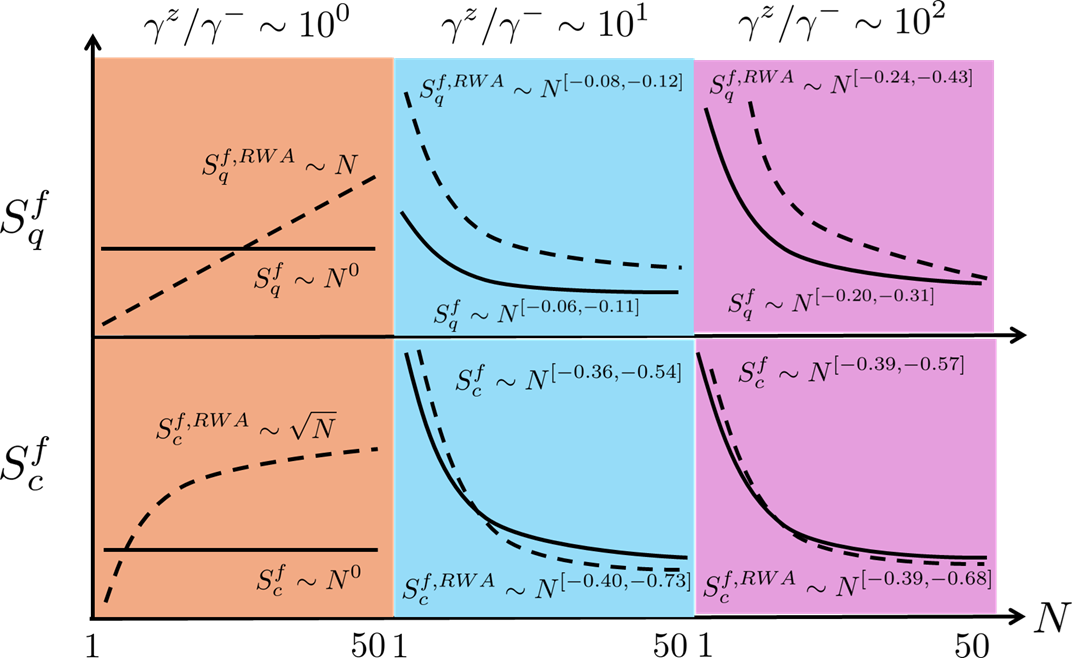}
    \caption{
        \textbf{Scaling behavior of final entropies across dissipation regimes.}
        The vertical panels indicate the different dissipation regimes (see the legend), and each graph compares the two  interaction models: the Tavis--Cummings model (dashed, RWA) and the Dicke model (solid, non-RWA). 
        The curves show the log--log scaling of the final qubit and cavity entropies, $S_q^f$ and $S_c^f$, from which the exponents $\alpha_{S_q^f}$ and $\alpha_{S_c^f}$ are extracted:
        $\alpha>0$ denotes growth of residual correlations with $N$, $\alpha\approx0$ saturation, and $\alpha<0$ suppression.
    }
\label{fig:scaling_behavior_entropies}
\end{figure}

Figure~\ref{fig:scaling_behavior_entropies} summarizes the scaling of the final subsystem entropies, $S_q^f$ and $S_c^f$, across dissipation regimes for both interaction models. 
These residual entropies quantify the correlations and mixedness that remain once the charging dynamics have settled (around $t_f \sim 50/\omega_q$). The scaling exponent $\alpha$ indicates whether larger devices end up more correlated ($\alpha>0$) or increasingly purified ($\alpha<0$) at the end of the protocol. 
The chosen dissipation parameters target low cavity mixedness while allowing for moderate qubit entanglement growth. 
For completeness, supplementary fitted plots of $\alpha_{S_q^f}$ and $\alpha_{S_c^f}$ are provided in  Figs.~\ref{fig:scaling_tuned-decay}–\ref{fig:scaling_dephasing-dominated} of  Appendix~\ref{app:scalinglaws}.

For the Dicke model, under both driven and undriven protocols, the entropy curves in Fig.~\ref{fig:scaling_behavior_entropies} display nearly flat log–log slopes ($\alpha_{S_q^f}, \alpha_{S_c^f} \!\approx\! 0$), showing that long–time correlations saturate rather than scale with $N$. As the dephasing starts dominating over relaxation both entropies exhibit negative exponents—$\alpha_{S_q^f} \!\in [-0.20, -0.31]$ for the qubit ensemble and $\alpha_{S_c^f} \!\in [-0.39, -0.57]$ for the cavity—indicating a partial purification of the steady state as the system size increases. This suppression of entanglement confirms that, in the Dicke model, dephasing stabilizes the long-time dynamics: entanglement generation remains bounded, leading to entanglement-clean operation that preserves coherence while preventing uncontrolled mixing. The drive does not qualitatively modify this behavior, reinforcing the interpretation that counter-rotating processes act as internal stabilizers of coherence in the presence of moderate noise.

In contrast, the Tavis–Cummings model shows a more intricate dependence on dissipation. At low dephasing (tuned–decay regime), the cavity entropy presents weak positive slopes ($\alpha_{S_c^f} \!\approx\! 0.2$–$0.4$) while the qubit entanglement entropy remains almost size-independent ($\alpha_{S_q^f} \!\approx\! 0$). As dephasing increases, these trends invert: both entropies develop bounded negative exponents—$\alpha_{S_c^f} \!\in [-0.39, -0.68]$ and $\alpha_{S_q^f} \!\in [-0.24, -0.43]$ showing that decoherence ultimately limits entanglement growth. This transition from positive to negative scaling marks the point where dephasing ceases to act as a destructive mechanism and instead regulates the entanglement buildup. Hence, moderate noise proves beneficial: it allows correlations to emerge sufficiently to enhance collective response, but then constrains them, yielding a steady operating regime in which coherence and stability coexist.

\begin{figure}[!ht]
    \centering
    \includegraphics[width=1.0\linewidth]{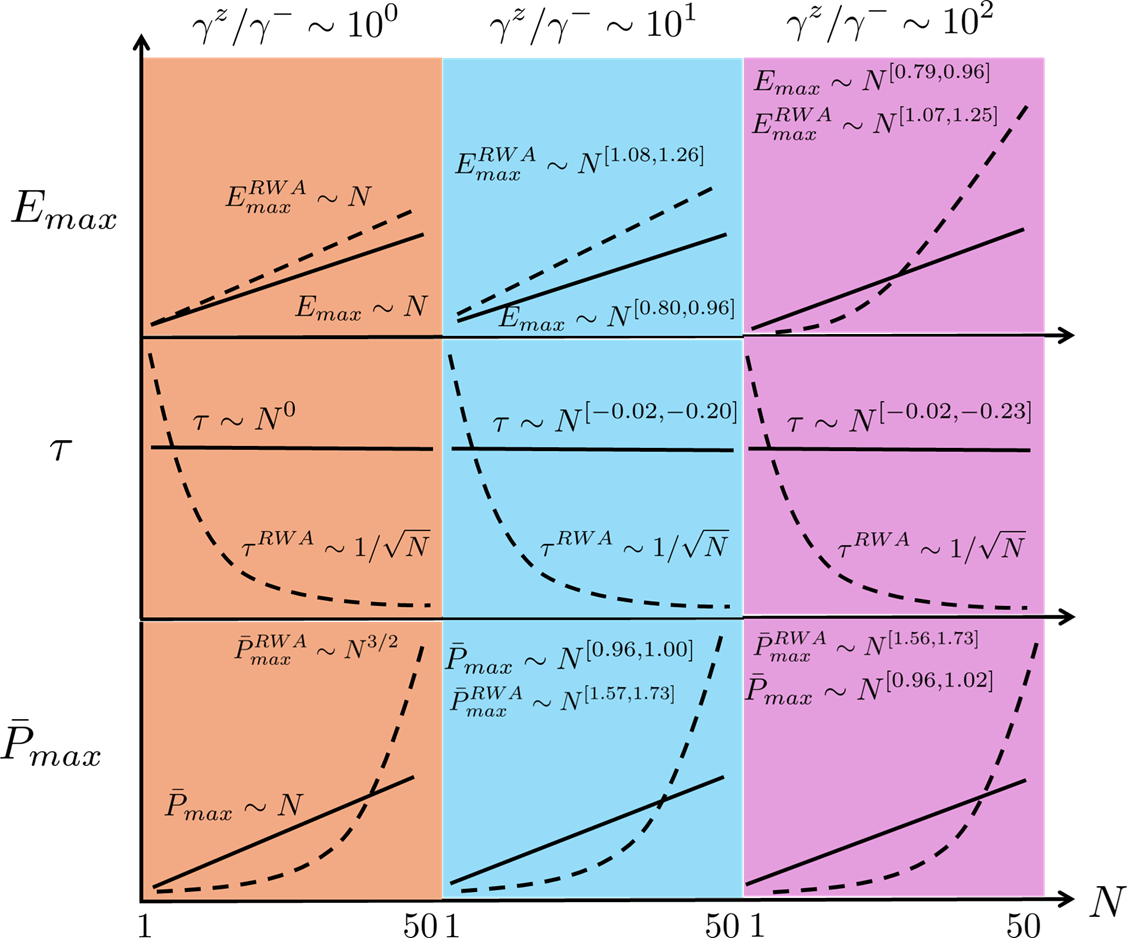}
    \caption{
        \textbf{Dynamical scaling behavior of thermodynamic observables across dissipation regimes.}
        The vertical panels indicate the different dissipation regimes and interaction models, as in Fig.~\ref{fig:scaling_behavior_entropies}. 
        The curves show the log--log scaling of the maximum stored energy $E_{\max}$, charging time $\tau$, and maximum power $\bar{P}_{\max}$ with qubit number $N$, from which the exponents $\alpha_{E_{\max}}$, $\alpha_{\tau}$ and $\alpha_{\bar{P}_{\max}}$ are obtained. 
        Values $\alpha>1$ signal superextensive (quantum-enhanced) scaling, $\alpha\approx1$ classical extensivity, and $\alpha<1$ degraded performance.
    }
    \label{fig:scaling_behavior_thermo}
\end{figure}

Building on these entropy results, Figure~\ref{fig:scaling_behavior_thermo} presents the scaling of thermodynamic observables — maximum stored energy $E_{\max}$, charging time $\tau$, and maximum power $\bar{P}_{\max}$ — as functions of $N$ and noise regimes. 

In the tuned–decay regime ($\gamma_z/\gamma^- \!\sim\! 10^0$), the Dicke model exhibits purely extensive thermodynamic behavior, with linear energy scaling ($\alpha_{E_{\max}} = 1.00$), negligible dependence of the charging time on system size ($\alpha_{\tau} = 0$), and correspondingly linear power growth ($\alpha_{\bar{P}_{\max}} = 1.00$). In contrast, the Tavis–Cummings model already shows a collective enhancement: the power scales superextensively ($\alpha_{\bar{P}_{\max}} \!\approx\! 1.50$) as a consequence of faster charging times ($\alpha_{\tau} \!\approx\! -0.50$), while energy storage remains approximately linear ($\alpha_{E_{\max}} \!\approx\! 1$). These results highlight the emergence of a weak quantum advantage in the RWA dynamics, which correlates with the moderate growth of entanglement entropy observed in the same regime.

As dephasing begins to dominate ($\gamma_z/\gamma^- \!\sim\! 10^1$), the Dicke model remains close to extensive scaling, with $\alpha_{E_{\max}} \!\in [0.80, 0.96]$, $\alpha_{\tau} \!\in [-0.02, -0.20]$, and $\alpha_{\bar{P}_{\max}} \!\in [0.96, 1.00]$. The slight improvement in $\tau$ partially compensates for the small reduction in energy scaling, resulting in stable yet only mildly accelerated charging. In the same regime, the Tavis–Cummings model consolidates its advantage: the energy scaling becomes superextensive ($\alpha_{E_{\max}} \!\in [1.08, 1.26]$), the power exponent increases up to $\alpha_{\bar{P}_{\max}} \!\in [1.57, 1.73]$, and the charging time keeps its favorable contribution to the thermodynamic performance enhancement ($\alpha_{\tau} \!\approx\! -0.48$). These trends demonstrate that the emergence of entanglement, as quantified by the fitted scaling exponents of the final entropies (see Fig.~\ref{fig:scaling_behavior_entropies}), directly supports the observed enhancement in collective performance.

In the dephasing-dominated regime ($\gamma_z/\gamma^- \!\sim\! 10^2$), the Dicke model sustains nearly extensive dynamical scaling, with $\alpha_{E_{\max}} \!\in [0.79, 0.96]$, $\alpha_{\bar{P}_{\max}} \!\in [0.96, 1.02]$, and $\alpha_{\tau} \!\in [-0.02, -0.23]$. Conversely, the Tavis–Cummings model reaches its strongest quantum advantage: both energy and power scale superextensively ($\alpha_{E_{\max}} \!\in [1.07, 1.25]$, $\alpha_{\bar{P}_{\max}} \!\in [1.56, 1.73]$), with accelerated charging times ($\alpha_{\tau} \!\approx\! -0.49$). The concurrent bounded suppression of entanglement entropies ($\alpha_{S_q^f}, \alpha_{S_c^f} < 0$) confirms that moderate dephasing enhances performance by stabilizing coherence while maintaining cooperative energy transfer. 
Overall, dephasing plays a dual role: it mitigates uncontrolled entanglement entropy growth and simultaneously reinforces superextensive dynamical scaling, enabling a steady and scalable quantum advantage in the open Tavis–Cummings battery.

Together, Figs.~\ref{fig:scaling_behavior_entropies} and~\ref{fig:scaling_behavior_thermo} demonstrate the intrinsic connection between entanglement scaling and thermodynamic enhancement in cavity-driven QBs. The Tavis--Cummings model achieves genuine quantum advantage in energy and power scaling, but only in regimes where qubit and cavity entanglement entropies increase with $N$, indicating the buildup of entanglement. Conversely, the Dicke model achieves cleaner, entropy-suppressed charging at the expense of reduced collective scaling. 
The coherent drive stabilizes entanglement entropy production within the optimal window, maintaining a balance in which controlled entanglement supports super-extensive performance without excessive noise. This interplay links the observed scaling laws directly to the phenomenon of superabsorption, in which cooperative quantum correlations accelerate energy transfer. 
Bipartite entanglement therefore emerges as the key resource underlying the collective enhancement of quantum battery performance.

\section{Quantum hardware implementation}
\label{sec:hardware}

In solid-state and circuit-QED platforms, the coherence and relaxation properties are typically characterized by the longitudinal relaxation time $T_1$ and the transverse coherence time $T_2$. The relaxation rate $\gamma^{-}/\omega_q
= 2\pi/T_1$ quantifies energy loss through spontaneous decay, whereas the dephasing rate arises from fluctuations in the qubit transition frequency that randomize phase coherence without exchanging energy. These quantities are related through the standard expression
\begin{equation}
    \label{eq:T1T2relation}
    \frac{1}{T_2} = \frac{1}{2T_1} + \frac{1}{T_{\phi}},
\end{equation}
where \(T_{\phi}\) is the pure-dephasing time associated exclusively with phase noise. Therefore, \(\gamma^{z}= 2\pi/T_{\phi}\) and the ratio \(\gamma^{z}/\gamma^-\) provides a convenient metric for identifying the dominant decoherence mechanism across different experimental platforms.
We identify three main regimes:

\begin{itemize}
\item  Dephasing-dominated: $\gamma^{z}/\gamma^- \sim 10^{2}$. 
\item Intermediate: $\gamma^{z}/\gamma^- \sim 10^{1}$.
\item Tuned decay channels: $\gamma^{z}/\gamma^- \sim 10^{0}$.
\end{itemize}

In the optimal operating window that yields the best performance, \(\gamma^-/\omega_q \in [10^{-3},\,2\times 10^{-1}]\) and \(\gamma^{z}/\omega_q \in [0.1,\,0.4]\), i.e., \(\gamma^{z}/\gamma^- \sim 10^{0}\!-\!10^{2}\). There are several experimentally relevant platforms that naturally realize the required balance of relaxation and dephasing, and that are compatible with Dicke/Tavis--Cummings QBs. 

We have made the following hardware-matching identification, although this is not exhaustive: 

{\it Dephasing-dominated} regime.--- 
Solid-state defect ensembles, such as NV centers in diamond~\cite{jarmola2012, stanwix2011}, SiC divacancies~\cite{seo2016}, and rare-earth ions in crystals \cite{alexander2022} typically exhibit very long \(T_{1}\) (ms--s) but shorter coherence due to dephasing (µs--ms). Trapped-ion qubits~\cite{zhang2025} likewise feature negligible relaxation ($T_1\sim 10^3$~s) with coherence limited by technical dephasing ($T_2\sim 10$~s). Organic microcavities with molecular dyes~\cite{quach2022superabsorption} have demonstrated QB-like superabsorption at room temperature, with typical $T_1\sim 1$~ns and $T_2\sim 100$~fs. 

{\it Intermediate} regime.---Self-assembled In(Ga)As quantum-dot excitons in microcavities~\cite{bardot2005, accanto2012} commonly exhibit ps--to-hundreds-ps pure dephasing with ns radiative lifetimes, providing a standard Tavis--Cummings platform with tunable dissipation. Electrons trapped on superfluid helium surfaces and coupled to high-impedance superconducting resonators (\(\gamma^{z}/\gamma^- \sim 10^{1}\)) have recently reached the strong-coupling regime, achieving vacuum Rabi splittings  
by combining long electron coherence times with large dipole moments~\cite{koolstra2025strongcouplingmicrowavephoton}, offering a promising platform for implementing collectively coupled quantum batteries within the optimal relaxation–dephasing window identified in this work.

{\it Tuned decay channels} regime.--- Superconducting transmons in circuit QED typically have comparable $T_{1}\sim T_{2}\sim$ tens--hundreds of µs and already host multi-qubit QB protocols and demonstrations \cite{Chang-Kang2022, Fu-Quan2023}.

Collectively, these systems span the entire ratio range that we probe, and each one meets the key requirement identified by our maps—low \(S_{\mathrm{c}}\) with moderate \(S_{\mathrm{q}}\)—either natively (NV/SiC/REI/ions and organic materials) or with standard engineering (QDs and transmons) to tune towards the desired \((\gamma^-/\omega_q,\gamma^{z}/\omega_q)\) sector.

\section{Discussion and Outlook}
\label{sec:discussion}

Our study shows a fundamental mechanism for understanding 
quantum superabsorption in open, cavity-coupled QBs. While previous studies have identified superabsorption~\cite{higgins2014superabsorption, quach2020dark} and suggested a link to qubit–cavity entanglement~\cite{ferraro2018, campaioli2017enhancing, quach2020dark, gherardini2020stabilizing}, they have not  provided a quantitative description of how correlations scale with system size or persist under noise. In this work, we  explicitly quantify these correlations   through the scaling laws of subsystem entropies,  demonstrating that bipartite entanglement partly underlies the observed superextensive energy and power scaling. We demonstrate a distinct noise-enhanced regime in which moderate dephasing and low relaxation stabilize cooperative faster charging, leading to   dissipation-enhanced superabsorption. Within this window, we also show that external coherent driving can act as an entanglement entropy growth stabilizer, maintaining favorable scaling and enabling stable quantum advantage. 

\begin{figure}
    \centering
    \includegraphics[width=1.0\linewidth]{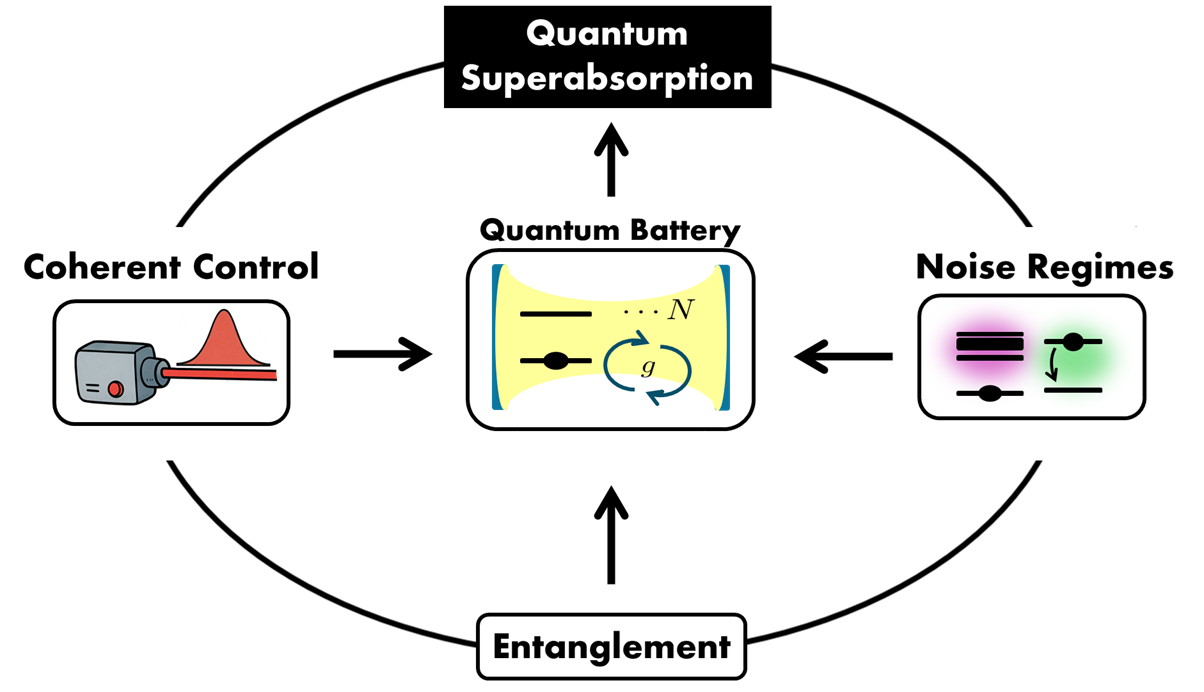}
    \caption{
    \textbf{Mechanism of dissipation-enhanced quantum superabsorption.}
    Qubit–cavity entanglement generates the cooperative correlations necessary for collective energy superabsorption, resulting in superextensive thermodynamic performance. A noise-enhanced regime forms the basis of dissipation-assisted superabsorption, where scaling remains favorable.  Coherent driving acts as a stabilizer of entanglement entropy growth. The trade-off between quantum advantage and controlled noise enables stable superabsorption.
    }
    \label{fig:quantumsuperabsorption_cyle}
\end{figure}

The interplay between these processes is summarized schematically in Fig.~\ref{fig:quantumsuperabsorption_cyle}, which highlights the self-consistent cycle connecting qubit–cavity entanglement, thermodynamic enhancement, and noise-assisted scaling. These processes are balanced by a fundamental trade-off between performance and entanglement entropy generation.

To our knowledge, this work provides the first quantitative framework that: (i) extracts finite-size scaling exponents for qubit and cavity entropies under Markovian dynamics; (ii) correlates those exponents with superextensive finite-size scaling in order to diagnose superabsorption; (iii) identifies a low-relaxation, moderate-dephasing window in which coherent driving stabilizes entanglement entropy growth while sustaining collective enhancement.

Even though systems with tens of qubits already demonstrate remarkable collective effects and performance enhancements, increasing the number of qubits to larger ensembles is expected to  improve charging efficiency further and yield more favorable scaling laws. Recent experimental detection of  superabsorption in organic microcavities~\cite{quach2022superabsorption} has shown that collective coupling can enhance energy uptake even for modest ensemble sizes. This study, however, did not quantify the entanglement responsible for the cooperative behavior. Our results indicate that collective effects, including super-extensive power and enhanced energy scaling, become increasingly pronounced with system size. We show that these advantages originate from bipartite entanglement between the qubit ensemble and the cavity mode. By deriving and quantifying the corresponding scaling laws, our work clarifies the microscopic mechanism behind the emergent collective advantage observed in both theoretical and experimental QBs.
%
%
Improving QBs' performance lies in optimizing the coherent drive used for charging. In our model, we considered Gaussian pulses mimicking ultrafast laser control as a minimal protocol; however, more refined control of the driving field could enhance both efficiency and stability. Tailored pulse shaping and optimal control techniques could be employed to maximize energy transfer while minimizing entanglement entropy growth. Such coherent control strategies are particularly relevant for laser-driven or circuit-based implementations~\cite{bastidas2010entanglement, dou2022staqb, hu2021staqb, salazar2023circuitdesign}, where the drive determines how excitations are distributed. Recent theoretical studies have shown that properly designed driving fields can accelerate charging, suppress decoherence, and maintain quantum correlations during the process~\cite{mazzoncini2023ocqb,hu2021staqb,qi2025staqb}. 
Thus, engineered dissipation provides a means for stabilizing many-body states and dynamics in platforms ranging from atoms and ions to superconducting circuits~\cite{diehl2008engineered, barreiro2011open, harrington2022engineered}.
Developing laser control protocols that exploit these principles offers a promising pathway toward stable, high-efficiency collective charging of QBs.

Although our analysis focused on Markovian noise, several physical environments exhibit memory effects that could be exploited to preserve coherence. Structured reservoirs, such as photonic crystals, can lead the local density of states to slow, enhance or even inhibit spontaneous emission, thereby prolonging coherence~\cite{lodahl2004control}.  Correlated spin baths underlying central-spin decoherence generate non-Markovian dynamics~\cite{cywinski2009prl}. This is just one of many scenarios in which non-Markovian effects dictate quantum dynamics~\cite{JH_CoherentControl_TLS_nonMarkovian,RMP_BLP_2016,Review_nonMark_Dynamics}. 
In this context, data-driven approaches are emerging, with machine-learning models able to detect, classify, and quantify non-Markovian processes, as well as inferring bath parameters directly from qubit dynamics~\cite{luchnikov2020mlNM, scarpetta2025machinelearning,altherr2021nonMarkovMetrology, martina2021mlNM}.  

Future work will focus on the collective quantum enhancement effects associated with non-Markovian qubit dynamics in the QB system, as well as the corresponding superabsorption and scaling-law behavior of entanglement and thermodynamic observables. 

\section{Conclusions}
\label{sec:conclusions}

A comparative analysis of cavity-coupled quantum batteries under Markovian dissipation shows that there is a fundamental balance between thermodynamic advantage and entanglement entropy generation. In the Tavis--Cummings model, collective coupling sustains superextensive scaling of stored energy and power despite decoherence, enabled by bipartite entanglement  which grows with system size. In contrast, the Dicke model exhibits cleaner, entropy-suppressed charging, albeit with extensive or sublinear scaling. 

By quantifying the scaling of qubit and cavity entropies, we demonstrate that long-time mixedness follows a power law with $N$, delineating two regimes. The first is a  balanced relaxation–dephasing regime where $\alpha_S>0$ and collective superabsorption emerges. The second is a dephasing-dominated regime where $\alpha_S<0$ and the dynamics purify. These results show that dephasing can generate and control quantum correlations, highlighting entanglement as the essential underlying  resource of dissipation-assisted superextensive charging.

Remarkably, the strongest quantum advantage emerges within the dephasing-dominated window. Here, moderate dephasing suppresses uncontrolled oscillations and stabilizes collective energy transfer. This sustains superextensive scaling while limiting entanglement entropy growth, demonstrating that controlled decoherence can act as a functional resource rather than a hindrance. Furthermore, the dissipation ratios associated with this optimal regime align with state-of-the-art experimental cavity and circuit QED platforms, including transmon qubits and NV-center ensembles. This highlights the practical relevance of our findings.

In this context, coherent driving acts as  a regulatory mechanism rather than a direct source of additional energy. Introducing a resonant drive to the cavity does not significantly alter the energy or power scaling exponents  beyond what collective interactions already provide. Its primary function is to stabilize the charging dynamics and control  entanglement growth. Coherent driving prevents excessive entanglement entropy growth while maintaining multipartite correlations across the ensemble. In this sense, the drive acts as a coherence stabilizer, extending the operational lifetime of the quantum advantage without compromising scalability.

In summary, this work establishes an understanding  of the performance and stability of open QBs. We identify the regime of low relaxation and moderate dephasing as the optimal operational window in which collective effects, controlled entanglement entropy growth and quantum advantage can coexist. Deriving entropy scaling laws provides a new framework for quantifying the evolution of quantum correlations with system size  in the presence of noise.  This  bridges the gap between thermodynamic performance and information-theoretic stability. Our findings demonstrate that dissipation, when balanced with coherent control, can enable scalable quantum advantage rather than being purely destructive.

\section{Acknowledgments}

We thank M. Hjorth-Jensen, G. F. Lange and J. Bergli for a critical reading of the manuscript and valuable discussions. 
The authors thank the Norwegian Partnership Programme for Global Academic Cooperation (NORPART), Grant NORPART-2021/10436---QTECNOS, the Department of Physics, the CCSE and the MSN at the University of Oslo. We also thank the Laboratory for Scientific Computing at CIBioFi. J.H.R. thanks Universidad del Valle for support (CI 71405).

\newpage
\appendix

\section{Cavity photon-number cutoff and numerical convergence}
\label{app:cutoff}

 We derive
analytical bounds on the cavity photon number to justify the
Fock-space truncation used in the numerical simulations. We also performed explicit convergence checks as $n_{\max}$  increased.

\subsection{Tavis--Cummings (TC) model}
\label{app:TC_bound}

\textit{Undriven case.} The total excitation operator $\hat{\mathcal N}=\hat a^\dagger \hat a+\sum_{j=1}^N \hat\sigma_j^+\hat\sigma_j^-$ satisfies $[\hat H_{\rm TC},\hat{\mathcal N}]=0$, i.e., $\hat{\mathcal N}$ is conserved. Therefore, if the system is initialized in the one-excitation sector (e.g. $|1\rangle_{\rm cav}\otimes|G\rangle$), the dynamics remains in $\mathcal N=1$ and Fock states $|n\ge 2\rangle$ are never populated. Hence a cutoff with $n_{\max}=1$ is exact for the undriven TC dynamics under this initialization.

\textit{Driven case.} When a coherent drive $\hat H_{\rm drive}(t)=\eta(t)(\hat a^\dagger+\hat a)$ is present, $\hat{\mathcal N}$ is no longer conserved, but its change is fully controlled by the drive, as follows
\begin{equation}
\frac{d\langle \hat{\mathcal N}\rangle}{dt}
=-i\langle[\hat{\mathcal N},\hat H_{\rm drive}(t)]\rangle
=-i\,\eta(t)\,\langle \hat a^\dagger-\hat a\rangle .
\end{equation}
Taking absolute values and integrating over the pulse yields the bound
\begin{equation}
\Delta \mathcal N \lesssim \int_{-\infty}^{\infty} |\eta(t)|\,dt,
\label{eq:Nmax_general}
\end{equation}
up to an order-unity prefactor set by $|\langle \hat a^\dagger-\hat a\rangle|$.
For the Gaussian envelope $\eta(t)=\eta_0\exp[-(t-t_0)^2/(2\sigma^2)]$,
\begin{equation}
\ \mathcal N_{\max}\sim \eta_0\,\sigma\sqrt{2\pi} ,
\label{eq:Nmax_final}
\end{equation}
so for our parameters ($\eta_0=\omega_q$, $\sigma=2/\omega_q$) we obtain $\mathcal N_{\max}\approx 2\sqrt{2\pi}\approx 5$ injected excitations. 
%
\subsection{Dicke (D) model}
\label{app:Dicke_virtual}

In the Dicke model, the counter-rotating (CR) interaction
$\hat V_{\rm CR}=g\sum_{j=1}^N(\hat a^\dagger \hat\sigma_j^+ + \hat a\,\hat\sigma_j^-)$ does not conserve $\hat{\mathcal N}$ and can generate virtual photon--qubit pairs. In the regime $g/\omega_c\ll 1$ (here $g/\omega_c=0.1$), the photon population induced solely by counter-rotating processes can be estimated perturbatively.
We split the Hamiltonian into $\hat H=\hat H_0+\hat V_{\rm CR}$, with
$\hat H_0=\omega_c \hat a^\dagger \hat a + (\omega_q/2)\sum_j \hat\sigma_j^z$, where $V_{\rm CR}$ is treated as a perturbation. The relevant counter-rotating process starting from the vacuum sector creates one photon and excites one qubit, as follows
\begin{equation}
\hat a^\dagger \hat\sigma_j^+\,|0\rangle_{\rm cav}\otimes|G\rangle
= |1\rangle_{\rm cav}\otimes|e_j\rangle,
\end{equation}
with an energy cost $\Delta E \simeq \omega_c+\omega_q$. Thus the amplitude of this virtual admixture is of order $g/(\omega_c+\omega_q)$ for each qubit. Since the $N$ states $|1\rangle_{\rm cav}\otimes|e_j\rangle$ are orthogonal, their contributions add in probability, yielding a mean virtual photon number
\begin{equation}
\Delta n_{\rm virtual}
\equiv \langle \hat a^\dagger \hat a\rangle_{\rm CR}
\sim N\left(\frac{g}{\omega_c+\omega_q}\right)^2
\lesssim N\left(\frac{g}{\omega_c}\right)^2,
\label{eq:Dn_virtual}
\end{equation}
where the last inequality corresponds to the resonance condition $\omega_c=\omega_q$. For our parameters $g/\omega_c=0.1$ and $N=50$, Eq.~\eqref{eq:Dn_virtual}
gives $\Delta n_{\rm virtual}\lesssim 0.5$, i.e. only a fraction of one photon on top of the excitation-number bound given by Eq.~\eqref{eq:Nmax_final}. 

\subsection{Numerical convergence test}
\label{app:conv}

In addition to the analytical estimates above, we performed an explicit convergence test with respect to the cavity photon-number cutoff $n_{\max}$. We selected a set of parameters, $g/\omega_c=0.1$ and $\gamma^-/\omega_q,\,\gamma^z/\omega_q,\,\kappa/\omega_q=10^{-3}$, such that we maximize photon retention and therefore provide the strictest requirement on the Fock-space truncation.
Table~\ref{tab:cutoff_conv} shows representative observables for both the Dicke and Tavis--Cummings models as $n_{\max}$ is increased.

\begin{table}[!ht]
\centering
\caption{Convergence with respect to the cavity photon-number cutoff $n_{\max}$. We report the maximum stored energy $E_{\max}$ and the cavity entropy $S_c$ for the Dicke (D) and Tavis--Cummings (TC) models for $g/\omega_c=0.1$ and the lowest
dissipation rates ($\gamma^-/\omega_q,\,\gamma^z/\omega_q,\,\kappa/\omega_q=10^{-3}$) used in the manuscript. The results quickly saturate  and are effectively converged by $n_{\max}=10$.}
\begin{tabular}{|c|c|c|c|c|}
\hline
$n_{\max}$ &
$E_{\max}^{(\mathrm{D})}$ &
$E_{\max}^{(\mathrm{TC})}$ &
$S_{c}^{(\mathrm{D})}$ &
$S_{c}^{(\mathrm{TC})}$ \\
\hline\hline
2  & 0.03459 & 0.15814 & 0.06171 & 0.06331 \\
5  & 0.23127 & 0.80604 & 0.10332 & 0.25676 \\\hline
10 & 0.49374 & 0.99073 & 0.08294 & 0.62638 \\\hline
15 & 0.49452 & 0.99294 & 0.08393 & 0.63075 \\
20 & 0.49452 & 0.99294 & 0.08393 & 0.63075 \\
25 & 0.49452 & 0.99294 & 0.08393 & 0.63075 \\
\hline
\end{tabular}
\label{tab:cutoff_conv}
\end{table}

All quantities rapidly saturate; already at $n_{\max}=10$, the values agree with the $n_{\max}=25$ reference within $\lesssim 1\%$ (worst case, $\approx 1.2\%$ for $S_c$ in the Dicke model), which is acceptable. Accordingly, we use $n_{\max}=10$ as the cavity cutoff throughout the manuscript.

\newpage
\section{Scaling Law Exponents}
\label{app:scalinglaws}

\begin{figure*}
    \centering
    \includegraphics[width=0.85\linewidth]{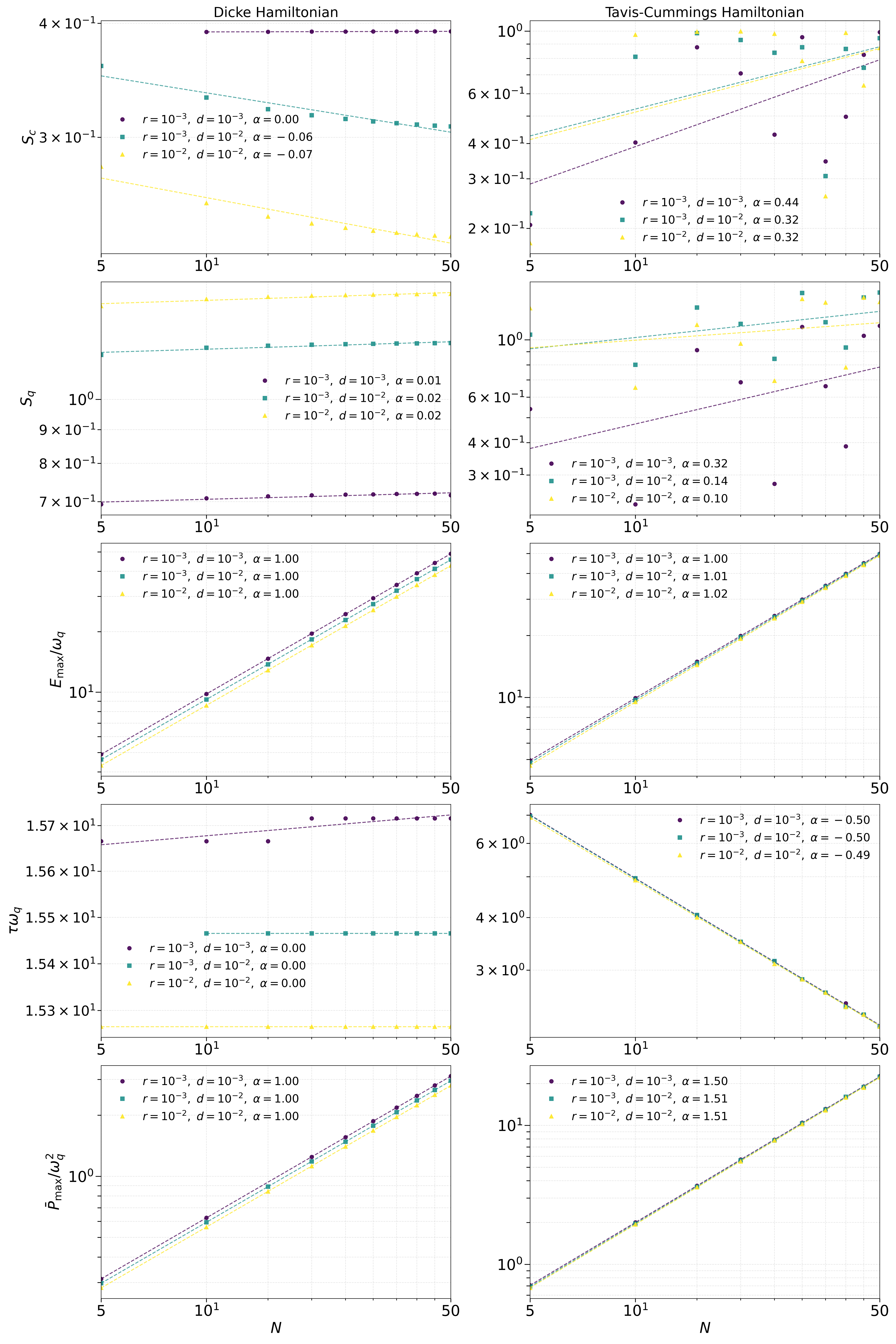}
    \caption{\textbf{Tuned-decay regime $\gamma^z/\gamma^-\,\sim 10^0$}. Entanglement entropies and thermodynamic observables for the Dicke (left) and Tavis-Cummings (right) models are shown as a function of system size $N$, for $\gamma^-/\omega_q,\,\gamma^z/\omega_q, = \{10^{-3},10^{-2}\}$. In the legend of the figures, we have set   $\gamma^-/\omega_q \equiv r$ and $\gamma^z/\omega_q \equiv d$; $r\leq d$. The system parameters were fixed at $g/\omega_q = 0.1$ and $\kappa/\omega_q=10^{-3}$.}
    \label{fig:scaling_tuned-decay}
\end{figure*}

\begin{figure*}
    \centering
    \includegraphics[width=0.85\linewidth]{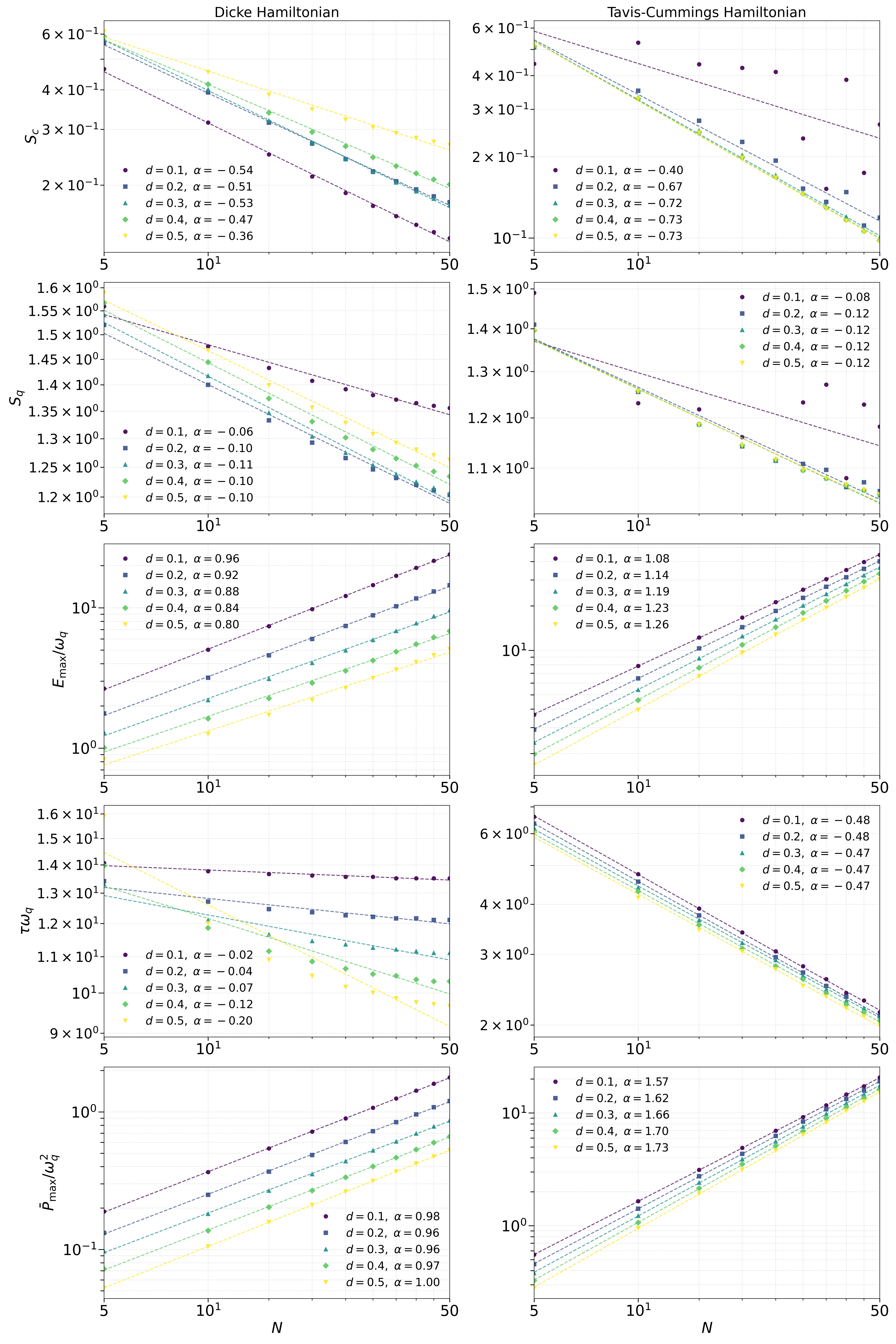}
    \caption{\textbf{Intermediate regime $\gamma^z/\gamma^-\,\sim 10^1$}. Entanglement entropies and thermodynamic observables for the Dicke (left) and Tavis-Cummings (right) models are shown as function of system size $N$,  for $\gamma^-/\omega_q = 10^{-2}$ and $\gamma^z/\omega_q = \{0.1, 0.2, 0.3, 0.4, 0.5\}$. The system parameters were fixed as in Fig.~\ref{fig:scaling_tuned-decay}.}
    \label{fig:scaling_intermediate}
\end{figure*}

\begin{figure*}
    \centering
    \includegraphics[width=0.85\linewidth]{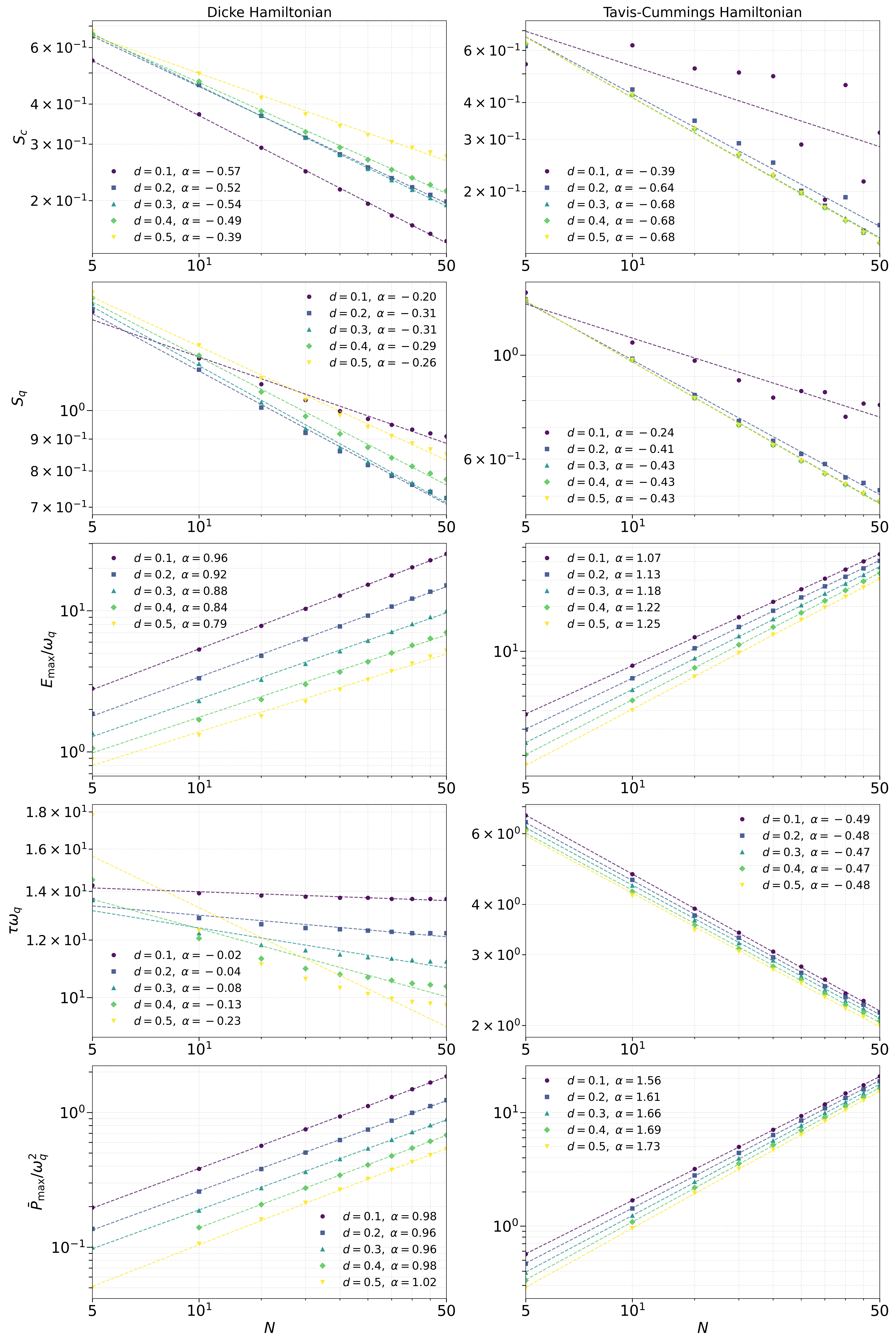}
    \caption{\textbf{Dephasing-dominated regime $\gamma^z/\gamma^-\,\sim 10^2$}. Entanglement entropies and thermodynamic observables for the Dicke (left) and Tavis-Cummings (right) models are shown as a function of system size $N$, for $\gamma^-/\omega_q =10^{-3}$ and $\gamma^z/\omega_q = \{0.1, 0.2, 0.3, 0.4, 0.5\}$.
The system parameters were fixed as in Fig.~\ref{fig:scaling_tuned-decay}.}
    \label{fig:scaling_dephasing-dominated}
\end{figure*}

This appendix supplements  the main text by presenting the detailed log–log fits that were used to obtain the scaling exponents supporting the results of Sec.~\ref{sec:results}. 
Figures~\ref{fig:scaling_tuned-decay}–\ref{fig:scaling_dephasing-dominated} display the system–size dependence of the cavity $S_c^f$ and qubit $S_q^f$ entanglement entropies, as well as the thermodynamic quantities $E_{\max}$, $\tau$, and $\bar{P}_{\max}$ for the Dicke (non-RWA) and Tavis–Cummings (RWA) models. Each curve represents a power–law fit of Eq.~\eqref{eq:scaling-law} given by $\log\left(O(N)\right) \sim \alpha \log\left(N\right)$, from which the corresponding $\alpha$ exponents  are extracted. These fits demonstrate how collective effects and dissipation jointly determine the scaling behavior of entanglement and thermodynamic observables across different regimes.

Outlier removal was applied to all figures below by computing the residuals with respect to a preliminary fit and  estimating their dispersion through the median absolute deviation.
Points whose normalized residual $|z| > 2$ were then discarded~\cite{ronchetti2009robust_statistics, rousseeuw2003robust_regression}. 

Figure~\ref{fig:scaling_tuned-decay} shows the log–log fits for the tuned–decay regime, where relaxation and dephasing act on comparable timescales, $\gamma_z/\gamma^- \sim 10^0$. Under this dissipation condition, the Dicke model exhibits extensive thermodynamic behavior, i.e. linear scaling in both energy storage capacity and average maximum power ($\alpha_{E}, \alpha_{\bar{P}} = 1$), due to the absence of favorable scaling in charging time ($\alpha_{\tau} = 0$). In contrast, the Tavis–Cummings model exhibis linear scaling in energy capacity ($\alpha_{E} \approx 1$) and a quantum advantage in maximum power ($\alpha_{\bar{P}} \approx 1.5$), due to  faster charging times ($\alpha_{\tau} \approx -0.5$). 

For entanglement entropies, the Dicke Hamiltonian exhibits nearly flat log–log trends ($\alpha_{S_q^f}, \alpha_{S_c^f} \approx 0$), confirming that 
extensive thermodynamic behavior is accompanied by the absence of entanglement.
 In contrast, the Tavis–Cummings model displays non-null entropy slopes, albeit with less regular fits. While the quality of these fits suggests that scaling is still developing in this balanced dissipation regime, moderate entanglement entropy growth appears to correlate with the emerging quantum advantage in power. As the noise regime shifts towards stronger dephasing, correlations between entropy and performance become clearer and more consistent, as shown in the following regimes.

Figure~\ref{fig:scaling_intermediate} shows the log–log fits for the intermediate dissipation regime, where dephasing begins to dominate relaxation, $\gamma_z/\gamma^- \sim 10^1$. In this case, the Dicke model shows significant entanglement decay in the cavity ($\alpha_{S_c^f} \in [-0.36, -0.54]$), accompanied by slight suppression of correlations in the $N$-qubit ensemble ($\alpha_{S_q^f} \in [-0.06, -0.11]$). The thermodynamic performance remains almost extensive, with a slight reduction in the energy-scaling exponent ($\alpha_{E} \in [0.80, 0.96]$) being offset by a modest improvement in charging times ($\alpha_{\tau} \in [-0.02, -0.20]$), resulting in almost extensive power scaling ($\alpha_{\bar{P}} \in [0.96, 1.00]$). In contrast, the Tavis–Cummings model enters a clearer quantum-advantage regime: the energy scaling becomes slightly superextensive ($\alpha_{E} \approx 1.08$–$1.26$), while the power exponent rises to $\alpha_{\bar{P}} \in [1.57, 1.73]$, supported by sublinear charging times ($\alpha_{\tau} \approx -0.48$). The entanglement entropy fits in this regime exhibit bounded negative slopes: $\alpha_{S_c^f} \in [-0.40,-0.73]$ and $\alpha_{S_q^f} \in [-0.08,-0.12]$,  suggesting that controlled entanglement decay accompanies the enhancement of collective performance. 

Figure~\ref{fig:scaling_dephasing-dominated} shows the log–log fits for the dephasing-dominated regime ($\gamma_z/\gamma^- \sim 10^2$). This defines the optimal operational window discussed in Sec.~\ref{sec:results}. In this limit, dephasing suppresses residual oscillations while preserving collective energy exchange, resulting in stable and reproducible scaling behavior. For the Dicke model, both the cavity and qubit entanglement entropies exhibit negative exponents ($\alpha_{S_c^f} \in [-0.39, -0.57]$, $\alpha_{S_q^f} \in [-0.20, -0.31]$), indicating partial purification of the steady state as the system size increases. This entropy suppression correlates with nearly extensive behavior, where the energy and power remain close to linear scaling ($\alpha_{E} \in [0.79, 0.96]$, $\alpha_{\bar{P}} \in [0.96, 1.02]$), and the charging time shows only weak and inconsistent size dependence ($\alpha_{\tau} \in [-0.02, -0.23]$). In contrast, the Tavis–Cummings model presents energy and power scaling $\alpha_{E} \in [1.07, 1.25]$, $\alpha_{\bar{P}} \in [1.56, 1.73]$ alongside faster charging times ($\alpha_{\tau} \approx -0.49$). This is accompanied by bounded yet strong entropy suppression in both the cavity ($\alpha_{S_c} \in [-0.39, -0.68]$) and the $N$-qubit ensemble ($\alpha_{S_q} \in [-0.24, -0.43]$). These trends show that moderate dephasing not only stabilizes correlations but also enhances collective performance, enabling superextensive finite-size scaling with controlled entanglement growth. This regime therefore realizes the coexistence of moderate noise, stabilized coherence, and scalable quantum advantage, demonstrating that dissipation can act as a constructive resource in open quantum batteries.

\begin{thebibliography}{89}%
\makeatletter
\providecommand \@ifxundefined [1]{%
 \@ifx{#1\undefined}
}%
\providecommand \@ifnum [1]{%
 \ifnum #1\expandafter \@firstoftwo
 \else \expandafter \@secondoftwo
 \fi
}%
\providecommand \@ifx [1]{%
 \ifx #1\expandafter \@firstoftwo
 \else \expandafter \@secondoftwo
 \fi
}%
\providecommand \natexlab [1]{#1}%
\providecommand \enquote  [1]{``#1''}%
\providecommand \bibnamefont  [1]{#1}%
\providecommand \bibfnamefont [1]{#1}%
\providecommand \citenamefont [1]{#1}%
\providecommand \href@noop [0]{\@secondoftwo}%
\providecommand \href [0]{\begingroup \@sanitize@url \@href}%
\providecommand \@href[1]{\@@startlink{#1}\@@href}%
\providecommand \@@href[1]{\endgroup#1\@@endlink}%
\providecommand \@sanitize@url [0]{\catcode `\\12\catcode `\$12\catcode `\&12\catcode `\#12\catcode `\^12\catcode `\_12\catcode `\%12\relax}%
\providecommand \@@startlink[1]{}%
\providecommand \@@endlink[0]{}%
\providecommand \url  [0]{\begingroup\@sanitize@url \@url }%
\providecommand \@url [1]{\endgroup\@href {#1}{\urlprefix }}%
\providecommand \urlprefix  [0]{URL }%
\providecommand \Eprint [0]{\href }%
\providecommand \doibase [0]{https://doi.org/}%
\providecommand \selectlanguage [0]{\@gobble}%
\providecommand \bibinfo  [0]{\@secondoftwo}%
\providecommand \bibfield  [0]{\@secondoftwo}%
\providecommand \translation [1]{[#1]}%
\providecommand \BibitemOpen [0]{}%
\providecommand \bibitemStop [0]{}%
\providecommand \bibitemNoStop [0]{.\EOS\space}%
\providecommand \EOS [0]{\spacefactor3000\relax}%
\providecommand \BibitemShut  [1]{\csname bibitem#1\endcsname}%
\let\auto@bib@innerbib\@empty
\bibitem [{\citenamefont {Higgins}\ \emph {et~al.}(2014)\citenamefont {Higgins}, \citenamefont {Lovett}, \citenamefont {Gauger}, \citenamefont {Kohl},\ and\ \citenamefont {Benjamin}}]{higgins2014superabsorption}%
  \BibitemOpen
  \bibfield  {author} {\bibinfo {author} {\bibfnamefont {K.~D.~B.}\ \bibnamefont {Higgins}}, \bibinfo {author} {\bibfnamefont {B.~W.}\ \bibnamefont {Lovett}}, \bibinfo {author} {\bibfnamefont {E.~M.}\ \bibnamefont {Gauger}}, \bibinfo {author} {\bibfnamefont {M.~M.}\ \bibnamefont {Kohl}},\ and\ \bibinfo {author} {\bibfnamefont {S.~C.}\ \bibnamefont {Benjamin}},\ }\href {https://doi.org/10.1038/ncomms5705} {\bibfield  {journal} {\bibinfo  {journal} {Nature Communications}\ }\textbf {\bibinfo {volume} {5}},\ \bibinfo {pages} {4705} (\bibinfo {year} {2014})}\BibitemShut {NoStop}%
\bibitem [{\citenamefont {Dicke}(1954)}]{dicke1954original}%
  \BibitemOpen
  \bibfield  {author} {\bibinfo {author} {\bibfnamefont {R.~H.}\ \bibnamefont {Dicke}},\ }\href {https://doi.org/10.1103/PhysRev.93.99} {\bibfield  {journal} {\bibinfo  {journal} {Phys. Rev.}\ }\textbf {\bibinfo {volume} {93}},\ \bibinfo {pages} {99} (\bibinfo {year} {1954})}\BibitemShut {NoStop}%
\bibitem [{\citenamefont {Horodecki}\ \emph {et~al.}(2009)\citenamefont {Horodecki}, \citenamefont {Horodecki}, \citenamefont {Horodecki},\ and\ \citenamefont {Horodecki}}]{horodecki2009quantum}%
  \BibitemOpen
  \bibfield  {author} {\bibinfo {author} {\bibfnamefont {R.}~\bibnamefont {Horodecki}}, \bibinfo {author} {\bibfnamefont {P.}~\bibnamefont {Horodecki}}, \bibinfo {author} {\bibfnamefont {M.}~\bibnamefont {Horodecki}},\ and\ \bibinfo {author} {\bibfnamefont {K.}~\bibnamefont {Horodecki}},\ }\href@noop {} {\bibfield  {journal} {\bibinfo  {journal} {Reviews of modern physics}\ }\textbf {\bibinfo {volume} {81}},\ \bibinfo {pages} {865} (\bibinfo {year} {2009})}\BibitemShut {NoStop}%
\bibitem [{\citenamefont {Carnot}(1824)}]{carnot1824reflections}%
  \BibitemOpen
  \bibfield  {author} {\bibinfo {author} {\bibfnamefont {S.}~\bibnamefont {Carnot}},\ }\href@noop {} {\bibfield  {journal} {\bibinfo  {journal} {Paris: Bachelier}\ }\textbf {\bibinfo {volume} {108}},\ \bibinfo {pages} {1824} (\bibinfo {year} {1824})}\BibitemShut {NoStop}%
\bibitem [{\citenamefont {Fermi}(2012)}]{fermi2012thermodynamics}%
  \BibitemOpen
  \bibfield  {author} {\bibinfo {author} {\bibfnamefont {E.}~\bibnamefont {Fermi}},\ }\href {https://books.google.no/books?id=xCjDAgAAQBAJ} {\emph {\bibinfo {title} {Thermodynamics}}},\ Dover Books on Physics\ (\bibinfo  {publisher} {Dover Publications},\ \bibinfo {year} {2012})\BibitemShut {NoStop}%
\bibitem [{\citenamefont {Kardar}(2007)}]{kardar2007statistical}%
  \BibitemOpen
  \bibfield  {author} {\bibinfo {author} {\bibfnamefont {M.}~\bibnamefont {Kardar}},\ }\href@noop {} {\emph {\bibinfo {title} {Statistical physics of particles}}}\ (\bibinfo  {publisher} {Cambridge University Press},\ \bibinfo {year} {2007})\BibitemShut {NoStop}%
\bibitem [{\citenamefont {Vinjanampathy}\ and\ \citenamefont {Anders}(2016)}]{vinjanampathy2016quantum}%
  \BibitemOpen
  \bibfield  {author} {\bibinfo {author} {\bibfnamefont {S.}~\bibnamefont {Vinjanampathy}}\ and\ \bibinfo {author} {\bibfnamefont {J.}~\bibnamefont {Anders}},\ }\href@noop {} {\bibfield  {journal} {\bibinfo  {journal} {Contemporary Physics}\ }\textbf {\bibinfo {volume} {57}},\ \bibinfo {pages} {545} (\bibinfo {year} {2016})}\BibitemShut {NoStop}%
\bibitem [{\citenamefont {Binder}\ \emph {et~al.}(2018{\natexlab{a}})\citenamefont {Binder}, \citenamefont {Correa}, \citenamefont {Gogolin}, \citenamefont {Anders},\ and\ \citenamefont {Adesso}}]{binder2018thermodynamics}%
  \BibitemOpen
  \bibfield  {author} {\bibinfo {author} {\bibfnamefont {F.}~\bibnamefont {Binder}}, \bibinfo {author} {\bibfnamefont {L.~A.}\ \bibnamefont {Correa}}, \bibinfo {author} {\bibfnamefont {C.}~\bibnamefont {Gogolin}}, \bibinfo {author} {\bibfnamefont {J.}~\bibnamefont {Anders}},\ and\ \bibinfo {author} {\bibfnamefont {G.}~\bibnamefont {Adesso}},\ }\href@noop {} {\bibfield  {journal} {\bibinfo  {journal} {Fundamental Theories of Physics}\ }\textbf {\bibinfo {volume} {195}} (\bibinfo {year} {2018}{\natexlab{a}})}\BibitemShut {NoStop}%
\bibitem [{\citenamefont {Herrera}\ \emph {et~al.}(2023)\citenamefont {Herrera}, \citenamefont {Reina}, \citenamefont {D'Amico},\ and\ \citenamefont {Serra}}]{herrera2023correlation}%
  \BibitemOpen
  \bibfield  {author} {\bibinfo {author} {\bibfnamefont {M.}~\bibnamefont {Herrera}}, \bibinfo {author} {\bibfnamefont {J.~H.}\ \bibnamefont {Reina}}, \bibinfo {author} {\bibfnamefont {I.}~\bibnamefont {D'Amico}},\ and\ \bibinfo {author} {\bibfnamefont {R.~M.}\ \bibnamefont {Serra}},\ }\href@noop {} {\bibfield  {journal} {\bibinfo  {journal} {Physical Review Research}\ }\textbf {\bibinfo {volume} {5}},\ \bibinfo {pages} {043104} (\bibinfo {year} {2023})}\BibitemShut {NoStop}%
\bibitem [{\citenamefont {Hovhannisyan}\ \emph {et~al.}(2013)\citenamefont {Hovhannisyan}, \citenamefont {Perarnau-Llobet}, \citenamefont {Huber},\ and\ \citenamefont {Ac{\'\i}n}}]{hovhannisyan2013entanglement}%
  \BibitemOpen
  \bibfield  {author} {\bibinfo {author} {\bibfnamefont {K.~V.}\ \bibnamefont {Hovhannisyan}}, \bibinfo {author} {\bibfnamefont {M.}~\bibnamefont {Perarnau-Llobet}}, \bibinfo {author} {\bibfnamefont {M.}~\bibnamefont {Huber}},\ and\ \bibinfo {author} {\bibfnamefont {A.}~\bibnamefont {Ac{\'\i}n}},\ }\href@noop {} {\bibfield  {journal} {\bibinfo  {journal} {Physical review letters}\ }\textbf {\bibinfo {volume} {111}},\ \bibinfo {pages} {240401} (\bibinfo {year} {2013})}\BibitemShut {NoStop}%
\bibitem [{\citenamefont {Campaioli}\ \emph {et~al.}(2017)\citenamefont {Campaioli}, \citenamefont {Pollock}, \citenamefont {Binder}, \citenamefont {C{\'e}leri}, \citenamefont {Goold}, \citenamefont {Vinjanampathy},\ and\ \citenamefont {Modi}}]{campaioli2017enhancing}%
  \BibitemOpen
  \bibfield  {author} {\bibinfo {author} {\bibfnamefont {F.}~\bibnamefont {Campaioli}}, \bibinfo {author} {\bibfnamefont {F.~A.}\ \bibnamefont {Pollock}}, \bibinfo {author} {\bibfnamefont {F.~C.}\ \bibnamefont {Binder}}, \bibinfo {author} {\bibfnamefont {L.}~\bibnamefont {C{\'e}leri}}, \bibinfo {author} {\bibfnamefont {J.}~\bibnamefont {Goold}}, \bibinfo {author} {\bibfnamefont {S.}~\bibnamefont {Vinjanampathy}},\ and\ \bibinfo {author} {\bibfnamefont {K.}~\bibnamefont {Modi}},\ }\href@noop {} {\bibfield  {journal} {\bibinfo  {journal} {Physical review letters}\ }\textbf {\bibinfo {volume} {118}},\ \bibinfo {pages} {150601} (\bibinfo {year} {2017})}\BibitemShut {NoStop}%
\bibitem [{\citenamefont {Campaioli}\ \emph {et~al.}(2018{\natexlab{a}})\citenamefont {Campaioli}, \citenamefont {Pollock},\ and\ \citenamefont {Modi}}]{campaioli2018quantum}%
  \BibitemOpen
  \bibfield  {author} {\bibinfo {author} {\bibfnamefont {F.}~\bibnamefont {Campaioli}}, \bibinfo {author} {\bibfnamefont {F.~A.}\ \bibnamefont {Pollock}},\ and\ \bibinfo {author} {\bibfnamefont {K.}~\bibnamefont {Modi}},\ }\href@noop {} {\bibfield  {journal} {\bibinfo  {journal} {Fundamental Theories of Physics}\ }\textbf {\bibinfo {volume} {195}},\ \bibinfo {pages} {207} (\bibinfo {year} {2018}{\natexlab{a}})}\BibitemShut {NoStop}%
\bibitem [{\citenamefont {Campaioli}\ \emph {et~al.}(2024{\natexlab{a}})\citenamefont {Campaioli}, \citenamefont {Gherardini}, \citenamefont {Quach}, \citenamefont {Polini},\ and\ \citenamefont {Andolina}}]{Campaioli_2024_Review}%
  \BibitemOpen
  \bibfield  {author} {\bibinfo {author} {\bibfnamefont {F.}~\bibnamefont {Campaioli}}, \bibinfo {author} {\bibfnamefont {S.}~\bibnamefont {Gherardini}}, \bibinfo {author} {\bibfnamefont {J.~Q.}\ \bibnamefont {Quach}}, \bibinfo {author} {\bibfnamefont {M.}~\bibnamefont {Polini}},\ and\ \bibinfo {author} {\bibfnamefont {G.~M.}\ \bibnamefont {Andolina}},\ }\href {https://doi.org/10.1103/RevModPhys.96.031001} {\bibfield  {journal} {\bibinfo  {journal} {Rev. Mod. Phys.}\ }\textbf {\bibinfo {volume} {96}},\ \bibinfo {pages} {031001} (\bibinfo {year} {2024}{\natexlab{a}})}\BibitemShut {NoStop}%
\bibitem [{\citenamefont {Juli{\`a}-Farr{\'e}}\ \emph {et~al.}(2020)\citenamefont {Juli{\`a}-Farr{\'e}}, \citenamefont {Salamon}, \citenamefont {Riera}, \citenamefont {Bera},\ and\ \citenamefont {Lewenstein}}]{julia2020bounds}%
  \BibitemOpen
  \bibfield  {author} {\bibinfo {author} {\bibfnamefont {S.}~\bibnamefont {Juli{\`a}-Farr{\'e}}}, \bibinfo {author} {\bibfnamefont {T.}~\bibnamefont {Salamon}}, \bibinfo {author} {\bibfnamefont {A.}~\bibnamefont {Riera}}, \bibinfo {author} {\bibfnamefont {M.~N.}\ \bibnamefont {Bera}},\ and\ \bibinfo {author} {\bibfnamefont {M.}~\bibnamefont {Lewenstein}},\ }\href@noop {} {\bibfield  {journal} {\bibinfo  {journal} {Physical Review Research}\ }\textbf {\bibinfo {volume} {2}},\ \bibinfo {pages} {023113} (\bibinfo {year} {2020})}\BibitemShut {NoStop}%
\bibitem [{\citenamefont {Yang}\ \emph {et~al.}(2023)\citenamefont {Yang}, \citenamefont {Yang}, \citenamefont {Alimuddin}, \citenamefont {Salvia}, \citenamefont {Fei}, \citenamefont {Zhao}, \citenamefont {Nimmrichter},\ and\ \citenamefont {Luo}}]{yang2023battery}%
  \BibitemOpen
  \bibfield  {author} {\bibinfo {author} {\bibfnamefont {X.}~\bibnamefont {Yang}}, \bibinfo {author} {\bibfnamefont {Y.-H.}\ \bibnamefont {Yang}}, \bibinfo {author} {\bibfnamefont {M.}~\bibnamefont {Alimuddin}}, \bibinfo {author} {\bibfnamefont {R.}~\bibnamefont {Salvia}}, \bibinfo {author} {\bibfnamefont {S.-M.}\ \bibnamefont {Fei}}, \bibinfo {author} {\bibfnamefont {L.-M.}\ \bibnamefont {Zhao}}, \bibinfo {author} {\bibfnamefont {S.}~\bibnamefont {Nimmrichter}},\ and\ \bibinfo {author} {\bibfnamefont {M.-X.}\ \bibnamefont {Luo}},\ }\href@noop {} {\bibfield  {journal} {\bibinfo  {journal} {Physical review letters}\ }\textbf {\bibinfo {volume} {131}},\ \bibinfo {pages} {030402} (\bibinfo {year} {2023})}\BibitemShut {NoStop}%
\bibitem [{\citenamefont {Ito}\ and\ \citenamefont {Watanabe}(2020)}]{ito2020collectively}%
  \BibitemOpen
  \bibfield  {author} {\bibinfo {author} {\bibfnamefont {K.}~\bibnamefont {Ito}}\ and\ \bibinfo {author} {\bibfnamefont {G.}~\bibnamefont {Watanabe}},\ }\href@noop {} {\bibfield  {journal} {\bibinfo  {journal} {arXiv preprint arXiv:2008.07089}\ } (\bibinfo {year} {2020})}\BibitemShut {NoStop}%
\bibitem [{\citenamefont {Liu}\ \emph {et~al.}(2019)\citenamefont {Liu}, \citenamefont {Segal},\ and\ \citenamefont {Hanna}}]{liu2019loss}%
  \BibitemOpen
  \bibfield  {author} {\bibinfo {author} {\bibfnamefont {J.}~\bibnamefont {Liu}}, \bibinfo {author} {\bibfnamefont {D.}~\bibnamefont {Segal}},\ and\ \bibinfo {author} {\bibfnamefont {G.}~\bibnamefont {Hanna}},\ }\href@noop {} {\bibfield  {journal} {\bibinfo  {journal} {The Journal of Physical Chemistry C}\ }\textbf {\bibinfo {volume} {123}},\ \bibinfo {pages} {18303} (\bibinfo {year} {2019})}\BibitemShut {NoStop}%
\bibitem [{\citenamefont {Gherardini}\ \emph {et~al.}(2020)\citenamefont {Gherardini}, \citenamefont {Campaioli}, \citenamefont {Caruso},\ and\ \citenamefont {Binder}}]{gherardini2020stabilizing}%
  \BibitemOpen
  \bibfield  {author} {\bibinfo {author} {\bibfnamefont {S.}~\bibnamefont {Gherardini}}, \bibinfo {author} {\bibfnamefont {F.}~\bibnamefont {Campaioli}}, \bibinfo {author} {\bibfnamefont {F.}~\bibnamefont {Caruso}},\ and\ \bibinfo {author} {\bibfnamefont {F.~C.}\ \bibnamefont {Binder}},\ }\href@noop {} {\bibfield  {journal} {\bibinfo  {journal} {Physical Review Research}\ }\textbf {\bibinfo {volume} {2}},\ \bibinfo {pages} {013095} (\bibinfo {year} {2020})}\BibitemShut {NoStop}%
\bibitem [{\citenamefont {Xu}\ \emph {et~al.}(2021)\citenamefont {Xu}, \citenamefont {Zhang}, \citenamefont {Liu}, \citenamefont {Liu},\ and\ \citenamefont {Fan}}]{xu2021control}%
  \BibitemOpen
  \bibfield  {author} {\bibinfo {author} {\bibfnamefont {K.}~\bibnamefont {Xu}}, \bibinfo {author} {\bibfnamefont {Y.}~\bibnamefont {Zhang}}, \bibinfo {author} {\bibfnamefont {T.}~\bibnamefont {Liu}}, \bibinfo {author} {\bibfnamefont {W.}~\bibnamefont {Liu}},\ and\ \bibinfo {author} {\bibfnamefont {H.}~\bibnamefont {Fan}},\ }\href@noop {} {\bibfield  {journal} {\bibinfo  {journal} {Physical Review E}\ }\textbf {\bibinfo {volume} {104}},\ \bibinfo {pages} {054118} (\bibinfo {year} {2021})}\BibitemShut {NoStop}%
\bibitem [{\citenamefont {Reina}\ \emph {et~al.}(2002)\citenamefont {Reina}, \citenamefont {Quiroga},\ and\ \citenamefont {Johnson}}]{reina2002decoherence}%
  \BibitemOpen
  \bibfield  {author} {\bibinfo {author} {\bibfnamefont {J.~H.}\ \bibnamefont {Reina}}, \bibinfo {author} {\bibfnamefont {L.}~\bibnamefont {Quiroga}},\ and\ \bibinfo {author} {\bibfnamefont {N.~F.}\ \bibnamefont {Johnson}},\ }\href {https://doi.org/10.1103/PhysRevA.65.032326} {\bibfield  {journal} {\bibinfo  {journal} {Phys. Rev. A}\ }\textbf {\bibinfo {volume} {65}},\ \bibinfo {pages} {032326} (\bibinfo {year} {2002})}\BibitemShut {NoStop}%
\bibitem [{\citenamefont {Quach}\ and\ \citenamefont {Munro}(2020)}]{quach2020dark}%
  \BibitemOpen
  \bibfield  {author} {\bibinfo {author} {\bibfnamefont {J.~Q.}\ \bibnamefont {Quach}}\ and\ \bibinfo {author} {\bibfnamefont {W.~J.}\ \bibnamefont {Munro}},\ }\href@noop {} {\bibfield  {journal} {\bibinfo  {journal} {Physical Review Applied}\ }\textbf {\bibinfo {volume} {14}},\ \bibinfo {pages} {024092} (\bibinfo {year} {2020})}\BibitemShut {NoStop}%
\bibitem [{\citenamefont {Reina}(2002)}]{JHR_DPhil}%
  \BibitemOpen
  \bibfield  {author} {\bibinfo {author} {\bibfnamefont {J.~H.}\ \bibnamefont {Reina}},\ }\emph {\bibinfo {title} {Quantum information processing in nanostructures}},\ \href {https://ora.ox.ac.uk/objects/uuid:6375c7c4-ecf6-4e88-a0f5-ff7493393d37} {Ph.D. thesis},\ \bibinfo  {school} {University of Oxford} (\bibinfo {year} {2002})\BibitemShut {NoStop}%
\bibitem [{\citenamefont {Alicki}\ and\ \citenamefont {Fannes}(2013)}]{alicki2013entanglement}%
  \BibitemOpen
  \bibfield  {author} {\bibinfo {author} {\bibfnamefont {R.}~\bibnamefont {Alicki}}\ and\ \bibinfo {author} {\bibfnamefont {M.}~\bibnamefont {Fannes}},\ }\href@noop {} {\bibfield  {journal} {\bibinfo  {journal} {Physical Review E}\ }\textbf {\bibinfo {volume} {87}},\ \bibinfo {pages} {042123} (\bibinfo {year} {2013})}\BibitemShut {NoStop}%
\bibitem [{\citenamefont {Ferraro}\ \emph {et~al.}(2018)\citenamefont {Ferraro}, \citenamefont {Campisi}, \citenamefont {Andolina}, \citenamefont {Pellegrini},\ and\ \citenamefont {Polini}}]{ferraro2018}%
  \BibitemOpen
  \bibfield  {author} {\bibinfo {author} {\bibfnamefont {D.}~\bibnamefont {Ferraro}}, \bibinfo {author} {\bibfnamefont {M.}~\bibnamefont {Campisi}}, \bibinfo {author} {\bibfnamefont {G.~M.}\ \bibnamefont {Andolina}}, \bibinfo {author} {\bibfnamefont {V.}~\bibnamefont {Pellegrini}},\ and\ \bibinfo {author} {\bibfnamefont {M.}~\bibnamefont {Polini}},\ }\href@noop {} {\bibfield  {journal} {\bibinfo  {journal} {Physical Review Letters}\ }\textbf {\bibinfo {volume} {120}},\ \bibinfo {pages} {117702} (\bibinfo {year} {2018})}\BibitemShut {NoStop}%
\bibitem [{\citenamefont {Quach}\ \emph {et~al.}(2022)\citenamefont {Quach}, \citenamefont {Han}, \citenamefont {Ganzer}, \citenamefont {Bosch~Aguilera}, \citenamefont {Mentink}, \citenamefont {Tong}, \citenamefont {Tollerud}, \citenamefont {Davis}, \citenamefont {Munro}, \citenamefont {Romero},\ and\ \citenamefont {Davis}}]{quach2022superabsorption}%
  \BibitemOpen
  \bibfield  {author} {\bibinfo {author} {\bibfnamefont {J.~Q.}\ \bibnamefont {Quach}}, \bibinfo {author} {\bibfnamefont {X.~L.}\ \bibnamefont {Han}}, \bibinfo {author} {\bibfnamefont {L.}~\bibnamefont {Ganzer}}, \bibinfo {author} {\bibfnamefont {A.}~\bibnamefont {Bosch~Aguilera}}, \bibinfo {author} {\bibfnamefont {J.~H.}\ \bibnamefont {Mentink}}, \bibinfo {author} {\bibfnamefont {X.}~\bibnamefont {Tong}}, \bibinfo {author} {\bibfnamefont {J.~O.}\ \bibnamefont {Tollerud}}, \bibinfo {author} {\bibfnamefont {J.~A.}\ \bibnamefont {Davis}}, \bibinfo {author} {\bibfnamefont {W.~J.}\ \bibnamefont {Munro}}, \bibinfo {author} {\bibfnamefont {J.}~\bibnamefont {Romero}},\ and\ \bibinfo {author} {\bibfnamefont {T.~J.}\ \bibnamefont {Davis}},\ }\href@noop {} {\bibfield  {journal} {\bibinfo  {journal} {Science Advances}\ }\textbf {\bibinfo {volume} {8}},\ \bibinfo {pages} {eabk3160} (\bibinfo {year} {2022})}\BibitemShut {NoStop}%
\bibitem [{\citenamefont {Kamin}\ \emph {et~al.}(2020)\citenamefont {Kamin}, \citenamefont {Campaioli}, \citenamefont {Binder},\ and\ \citenamefont {Kurtz}}]{kamin2020nonmarkovian}%
  \BibitemOpen
  \bibfield  {author} {\bibinfo {author} {\bibfnamefont {F.}~\bibnamefont {Kamin}}, \bibinfo {author} {\bibfnamefont {F.}~\bibnamefont {Campaioli}}, \bibinfo {author} {\bibfnamefont {F.~C.}\ \bibnamefont {Binder}},\ and\ \bibinfo {author} {\bibfnamefont {S.}~\bibnamefont {Kurtz}},\ }\href@noop {} {\bibfield  {journal} {\bibinfo  {journal} {Physical Review E}\ }\textbf {\bibinfo {volume} {102}},\ \bibinfo {pages} {052109} (\bibinfo {year} {2020})}\BibitemShut {NoStop}%
\bibitem [{\citenamefont {Ghosh}\ \emph {et~al.}(2021)\citenamefont {Ghosh}, \citenamefont {Chanda},\ and\ \citenamefont {Sen~(De)}}]{ghosh2021enhancedperformance}%
  \BibitemOpen
  \bibfield  {author} {\bibinfo {author} {\bibfnamefont {A.}~\bibnamefont {Ghosh}}, \bibinfo {author} {\bibfnamefont {T.}~\bibnamefont {Chanda}},\ and\ \bibinfo {author} {\bibfnamefont {A.}~\bibnamefont {Sen~(De)}},\ }\href@noop {} {\bibfield  {journal} {\bibinfo  {journal} {Physical Review A}\ }\textbf {\bibinfo {volume} {104}},\ \bibinfo {pages} {032207} (\bibinfo {year} {2021})}\BibitemShut {NoStop}%
\bibitem [{\citenamefont {Tavis}\ and\ \citenamefont {Cummings}(1968)}]{tavis1968exact}%
  \BibitemOpen
  \bibfield  {author} {\bibinfo {author} {\bibfnamefont {M.}~\bibnamefont {Tavis}}\ and\ \bibinfo {author} {\bibfnamefont {F.~W.}\ \bibnamefont {Cummings}},\ }\href@noop {} {\bibfield  {journal} {\bibinfo  {journal} {Physical Review}\ }\textbf {\bibinfo {volume} {170}},\ \bibinfo {pages} {379} (\bibinfo {year} {1968})}\BibitemShut {NoStop}%
\bibitem [{\citenamefont {Agarwal}\ \emph {et~al.}(2012)\citenamefont {Agarwal}, \citenamefont {Rafsanjani},\ and\ \citenamefont {Eberly}}]{agarwal2012tavis}%
  \BibitemOpen
  \bibfield  {author} {\bibinfo {author} {\bibfnamefont {S.}~\bibnamefont {Agarwal}}, \bibinfo {author} {\bibfnamefont {S.~M.~H.}\ \bibnamefont {Rafsanjani}},\ and\ \bibinfo {author} {\bibfnamefont {J.~H.}\ \bibnamefont {Eberly}},\ }\href {https://doi.org/10.1103/PhysRevA.85.043815} {\bibfield  {journal} {\bibinfo  {journal} {Phys. Rev. A}\ }\textbf {\bibinfo {volume} {85}},\ \bibinfo {pages} {043815} (\bibinfo {year} {2012})}\BibitemShut {NoStop}%
\bibitem [{\citenamefont {Fujii}(2017)}]{fujii2017rwa}%
  \BibitemOpen
  \bibfield  {author} {\bibinfo {author} {\bibfnamefont {K.}~\bibnamefont {Fujii}},\ }\href {https://doi.org/10.4236/jmp.2017.812124} {\bibfield  {journal} {\bibinfo  {journal} {Journal of Modern Physics}\ }\textbf {\bibinfo {volume} {8}},\ \bibinfo {pages} {2042} (\bibinfo {year} {2017})}\BibitemShut {NoStop}%
\bibitem [{\citenamefont {Jaako}\ \emph {et~al.}(2016)\citenamefont {Jaako}, \citenamefont {Xiang}, \citenamefont {Garcia-Ripoll},\ and\ \citenamefont {Rabl}}]{jaako2016ultrastrong}%
  \BibitemOpen
  \bibfield  {author} {\bibinfo {author} {\bibfnamefont {T.}~\bibnamefont {Jaako}}, \bibinfo {author} {\bibfnamefont {Z.-L.}\ \bibnamefont {Xiang}}, \bibinfo {author} {\bibfnamefont {J.~J.}\ \bibnamefont {Garcia-Ripoll}},\ and\ \bibinfo {author} {\bibfnamefont {P.}~\bibnamefont {Rabl}},\ }\href@noop {} {\bibfield  {journal} {\bibinfo  {journal} {Physical Review A}\ }\textbf {\bibinfo {volume} {94}},\ \bibinfo {pages} {033850} (\bibinfo {year} {2016})}\BibitemShut {NoStop}%
\bibitem [{\citenamefont {Andolina}\ \emph {et~al.}(2019)\citenamefont {Andolina}, \citenamefont {Keck}, \citenamefont {Mari}, \citenamefont {Giovannetti},\ and\ \citenamefont {Polini}}]{andolina2019PRL}%
  \BibitemOpen
  \bibfield  {author} {\bibinfo {author} {\bibfnamefont {G.~M.}\ \bibnamefont {Andolina}}, \bibinfo {author} {\bibfnamefont {M.}~\bibnamefont {Keck}}, \bibinfo {author} {\bibfnamefont {A.}~\bibnamefont {Mari}}, \bibinfo {author} {\bibfnamefont {V.}~\bibnamefont {Giovannetti}},\ and\ \bibinfo {author} {\bibfnamefont {M.}~\bibnamefont {Polini}},\ }\href {https://doi.org/10.1103/PhysRevLett.122.047702} {\bibfield  {journal} {\bibinfo  {journal} {Phys. Rev. Lett.}\ }\textbf {\bibinfo {volume} {122}},\ \bibinfo {pages} {047702} (\bibinfo {year} {2019})}\BibitemShut {NoStop}%
\bibitem [{\citenamefont {Campaioli}\ \emph {et~al.}(2018{\natexlab{b}})\citenamefont {Campaioli}, \citenamefont {Pollock},\ and\ \citenamefont {Vinjanampathy}}]{campaioli2018chapter}%
  \BibitemOpen
  \bibfield  {author} {\bibinfo {author} {\bibfnamefont {F.}~\bibnamefont {Campaioli}}, \bibinfo {author} {\bibfnamefont {F.~A.}\ \bibnamefont {Pollock}},\ and\ \bibinfo {author} {\bibfnamefont {S.}~\bibnamefont {Vinjanampathy}},\ }in\ \href {https://doi.org/10.1007/978-3-319-99046-0_8} {\emph {\bibinfo {booktitle} {Fundamental Theories of Physics}}},\ Vol.\ \bibinfo {volume} {195},\ \bibinfo {editor} {edited by\ \bibinfo {editor} {\bibfnamefont {F.}~\bibnamefont {Binder}}, \bibinfo {editor} {\bibfnamefont {L.~A.}\ \bibnamefont {Correa}}, \bibinfo {editor} {\bibfnamefont {C.}~\bibnamefont {Gogolin}}, \bibinfo {editor} {\bibfnamefont {J.}~\bibnamefont {Anders}},\ and\ \bibinfo {editor} {\bibfnamefont {G.}~\bibnamefont {Adesso}}}\ (\bibinfo  {publisher} {Springer},\ \bibinfo {year} {2018})\ pp.\ \bibinfo {pages} {207--225}\BibitemShut {NoStop}%
\bibitem [{\citenamefont {Hausinger}\ and\ \citenamefont {Grifoni}(2010)}]{hausinger2010analyticUSC}%
  \BibitemOpen
  \bibfield  {author} {\bibinfo {author} {\bibfnamefont {J.}~\bibnamefont {Hausinger}}\ and\ \bibinfo {author} {\bibfnamefont {M.}~\bibnamefont {Grifoni}},\ }\href {https://doi.org/10.1103/PhysRevA.82.062320} {\bibfield  {journal} {\bibinfo  {journal} {Phys. Rev. A}\ }\textbf {\bibinfo {volume} {82}},\ \bibinfo {pages} {062320} (\bibinfo {year} {2010})}\BibitemShut {NoStop}%
\bibitem [{\citenamefont {Ashhab}\ and\ \citenamefont {Nori}(2010)}]{ashhab2010tls}%
  \BibitemOpen
  \bibfield  {author} {\bibinfo {author} {\bibfnamefont {S.}~\bibnamefont {Ashhab}}\ and\ \bibinfo {author} {\bibfnamefont {F.}~\bibnamefont {Nori}},\ }\href {https://doi.org/10.1103/PhysRevA.81.042311} {\bibfield  {journal} {\bibinfo  {journal} {Phys. Rev. A}\ }\textbf {\bibinfo {volume} {81}},\ \bibinfo {pages} {042311} (\bibinfo {year} {2010})}\BibitemShut {NoStop}%
\bibitem [{\citenamefont {Forn-Díaz}\ \emph {et~al.}(2019)\citenamefont {Forn-Díaz}, \citenamefont {Lamata}, \citenamefont {Rico}, \citenamefont {Kono},\ and\ \citenamefont {Solano}}]{forndiazRMP2019usc}%
  \BibitemOpen
  \bibfield  {author} {\bibinfo {author} {\bibfnamefont {P.}~\bibnamefont {Forn-Díaz}}, \bibinfo {author} {\bibfnamefont {L.}~\bibnamefont {Lamata}}, \bibinfo {author} {\bibfnamefont {E.}~\bibnamefont {Rico}}, \bibinfo {author} {\bibfnamefont {J.}~\bibnamefont {Kono}},\ and\ \bibinfo {author} {\bibfnamefont {E.}~\bibnamefont {Solano}},\ }\href {https://doi.org/10.1103/RevModPhys.91.025005} {\bibfield  {journal} {\bibinfo  {journal} {Rev. Mod. Phys.}\ }\textbf {\bibinfo {volume} {91}},\ \bibinfo {pages} {025005} (\bibinfo {year} {2019})}\BibitemShut {NoStop}%
\bibitem [{\citenamefont {Kockum}\ \emph {et~al.}(2019)\citenamefont {Kockum}, \citenamefont {Miranowicz}, \citenamefont {De~Liberato}, \citenamefont {Savasta},\ and\ \citenamefont {Nori}}]{kockumNRP2019usc}%
  \BibitemOpen
  \bibfield  {author} {\bibinfo {author} {\bibfnamefont {A.~F.}\ \bibnamefont {Kockum}}, \bibinfo {author} {\bibfnamefont {A.}~\bibnamefont {Miranowicz}}, \bibinfo {author} {\bibfnamefont {S.}~\bibnamefont {De~Liberato}}, \bibinfo {author} {\bibfnamefont {S.}~\bibnamefont {Savasta}},\ and\ \bibinfo {author} {\bibfnamefont {F.}~\bibnamefont {Nori}},\ }\href {https://doi.org/10.1038/s42254-018-0006-2} {\bibfield  {journal} {\bibinfo  {journal} {Nat. Rev. Phys.}\ }\textbf {\bibinfo {volume} {1}},\ \bibinfo {pages} {19} (\bibinfo {year} {2019})}\BibitemShut {NoStop}%
\bibitem [{\citenamefont {Yoshihara}\ \emph {et~al.}(2017)\citenamefont {Yoshihara}, \citenamefont {Fuse}, \citenamefont {Ashhab}, \citenamefont {Kakuyanagi}, \citenamefont {Saito},\ and\ \citenamefont {Semba}}]{yoshihara2017superconduct}%
  \BibitemOpen
  \bibfield  {author} {\bibinfo {author} {\bibfnamefont {F.}~\bibnamefont {Yoshihara}}, \bibinfo {author} {\bibfnamefont {T.}~\bibnamefont {Fuse}}, \bibinfo {author} {\bibfnamefont {S.}~\bibnamefont {Ashhab}}, \bibinfo {author} {\bibfnamefont {K.}~\bibnamefont {Kakuyanagi}}, \bibinfo {author} {\bibfnamefont {S.}~\bibnamefont {Saito}},\ and\ \bibinfo {author} {\bibfnamefont {K.}~\bibnamefont {Semba}},\ }\href {https://doi.org/10.1038/nphys3906} {\bibfield  {journal} {\bibinfo  {journal} {Nat. Phys.}\ }\textbf {\bibinfo {volume} {13}},\ \bibinfo {pages} {44} (\bibinfo {year} {2017})}\BibitemShut {NoStop}%
\bibitem [{\citenamefont {Nielsen}\ and\ \citenamefont {Chuang}(2010)}]{nielsen2010quantum}%
  \BibitemOpen
  \bibfield  {author} {\bibinfo {author} {\bibfnamefont {M.~A.}\ \bibnamefont {Nielsen}}\ and\ \bibinfo {author} {\bibfnamefont {I.~L.}\ \bibnamefont {Chuang}},\ }\href@noop {} {\emph {\bibinfo {title} {Quantum computation and quantum information}}}\ (\bibinfo  {publisher} {Cambridge university press},\ \bibinfo {year} {2010})\BibitemShut {NoStop}%
\bibitem [{\citenamefont {Breuer}\ and\ \citenamefont {Petruccione}(2002)}]{breuer2002theory}%
  \BibitemOpen
  \bibfield  {author} {\bibinfo {author} {\bibfnamefont {H.-P.}\ \bibnamefont {Breuer}}\ and\ \bibinfo {author} {\bibfnamefont {F.}~\bibnamefont {Petruccione}},\ }\href@noop {} {\emph {\bibinfo {title} {The theory of open quantum systems}}}\ (\bibinfo  {publisher} {OUP Oxford},\ \bibinfo {year} {2002})\BibitemShut {NoStop}%
\bibitem [{\citenamefont {Anders}\ and\ \citenamefont {Esposito}(2017)}]{anders2017quantumthermodynamics}%
  \BibitemOpen
  \bibfield  {author} {\bibinfo {author} {\bibfnamefont {J.}~\bibnamefont {Anders}}\ and\ \bibinfo {author} {\bibfnamefont {M.}~\bibnamefont {Esposito}},\ }\href {https://doi.org/10.1088/1367-2630/aa7bb1} {\bibfield  {journal} {\bibinfo  {journal} {New Journal of Physics}\ }\textbf {\bibinfo {volume} {19}},\ \bibinfo {pages} {010201} (\bibinfo {year} {2017})}\BibitemShut {NoStop}%
\bibitem [{\citenamefont {Binder}\ \emph {et~al.}(2018{\natexlab{b}})\citenamefont {Binder}, \citenamefont {Correa}, \citenamefont {Gogolin}, \citenamefont {Anders},\ and\ \citenamefont {Adesso}}]{binder2018generalquantum}%
  \BibitemOpen
  \bibfield  {author} {\bibinfo {author} {\bibfnamefont {F.}~\bibnamefont {Binder}}, \bibinfo {author} {\bibfnamefont {L.~A.}\ \bibnamefont {Correa}}, \bibinfo {author} {\bibfnamefont {C.}~\bibnamefont {Gogolin}}, \bibinfo {author} {\bibfnamefont {J.}~\bibnamefont {Anders}},\ and\ \bibinfo {author} {\bibfnamefont {G.}~\bibnamefont {Adesso}},\ }in\ \href {https://doi.org/10.1007/978-3-319-99046-0_2} {\emph {\bibinfo {booktitle} {Thermodynamics in the Quantum Regime}}},\ \bibinfo {series} {Fundamental Theories of Physics}, Vol.\ \bibinfo {volume} {195}\ (\bibinfo  {publisher} {Springer},\ \bibinfo {year} {2018})\ pp.\ \bibinfo {pages} {43--73}\BibitemShut {NoStop}%
\bibitem [{\citenamefont {Reina}\ \emph {et~al.}(2014)\citenamefont {Reina}, \citenamefont {Susa},\ and\ \citenamefont {Fanchini}}]{reina2014correlations}%
  \BibitemOpen
  \bibfield  {author} {\bibinfo {author} {\bibfnamefont {J.~H.}\ \bibnamefont {Reina}}, \bibinfo {author} {\bibfnamefont {C.~E.}\ \bibnamefont {Susa}},\ and\ \bibinfo {author} {\bibfnamefont {F.~F.}\ \bibnamefont {Fanchini}},\ }\href {https://doi.org/10.1038/srep07443} {\bibfield  {journal} {\bibinfo  {journal} {Scientific Reports}\ }\textbf {\bibinfo {volume} {4}},\ \bibinfo {pages} {7443} (\bibinfo {year} {2014})}\BibitemShut {NoStop}%
\bibitem [{\citenamefont {Carmichael}(1993)}]{carmichael1993opensystems}%
  \BibitemOpen
  \bibfield  {author} {\bibinfo {author} {\bibfnamefont {H.~J.}\ \bibnamefont {Carmichael}},\ }\href {https://doi.org/10.1007/978-3-540-47620-7} {\emph {\bibinfo {title} {An Open Systems Approach to Quantum Optics}}}\ (\bibinfo  {publisher} {Springer},\ \bibinfo {year} {1993})\BibitemShut {NoStop}%
\bibitem [{\citenamefont {Walls}\ and\ \citenamefont {Milburn}(2008)}]{walls2008quantumoptics}%
  \BibitemOpen
  \bibfield  {author} {\bibinfo {author} {\bibfnamefont {D.~F.}\ \bibnamefont {Walls}}\ and\ \bibinfo {author} {\bibfnamefont {G.~J.}\ \bibnamefont {Milburn}},\ }\href {https://doi.org/10.1007/978-3-540-28574-8} {\emph {\bibinfo {title} {Quantum Optics}}},\ \bibinfo {edition} {2nd}\ ed.\ (\bibinfo  {publisher} {Springer},\ \bibinfo {year} {2008})\BibitemShut {NoStop}%
\bibitem [{\citenamefont {Campaioli}\ \emph {et~al.}(2024{\natexlab{b}})\citenamefont {Campaioli}, \citenamefont {Cole},\ and\ \citenamefont {Hapuarachchi}}]{masterEq}%
  \BibitemOpen
  \bibfield  {author} {\bibinfo {author} {\bibfnamefont {F.}~\bibnamefont {Campaioli}}, \bibinfo {author} {\bibfnamefont {J.~H.}\ \bibnamefont {Cole}},\ and\ \bibinfo {author} {\bibfnamefont {H.}~\bibnamefont {Hapuarachchi}},\ }\href {https://doi.org/10.1103/PRXQuantum.5.020202} {\bibfield  {journal} {\bibinfo  {journal} {PRX Quantum}\ }\textbf {\bibinfo {volume} {5}},\ \bibinfo {pages} {020202} (\bibinfo {year} {2024}{\natexlab{b}})}\BibitemShut {NoStop}%
\bibitem [{\citenamefont {Gorini}\ \emph {et~al.}(1976)\citenamefont {Gorini}, \citenamefont {Kossakowski},\ and\ \citenamefont {Sudarshan}}]{gorini1976completely}%
  \BibitemOpen
  \bibfield  {author} {\bibinfo {author} {\bibfnamefont {V.}~\bibnamefont {Gorini}}, \bibinfo {author} {\bibfnamefont {A.}~\bibnamefont {Kossakowski}},\ and\ \bibinfo {author} {\bibfnamefont {E.~C.~G.}\ \bibnamefont {Sudarshan}},\ }\href@noop {} {\bibfield  {journal} {\bibinfo  {journal} {Journal of Mathematical Physics}\ }\textbf {\bibinfo {volume} {17}},\ \bibinfo {pages} {821} (\bibinfo {year} {1976})}\BibitemShut {NoStop}%
\bibitem [{\citenamefont {Lindblad}(1976)}]{lindblad1976generators}%
  \BibitemOpen
  \bibfield  {author} {\bibinfo {author} {\bibfnamefont {G.}~\bibnamefont {Lindblad}},\ }\href@noop {} {\bibfield  {journal} {\bibinfo  {journal} {Communications in mathematical physics}\ }\textbf {\bibinfo {volume} {48}},\ \bibinfo {pages} {119} (\bibinfo {year} {1976})}\BibitemShut {NoStop}%
\bibitem [{\citenamefont {Shammah}\ \emph {et~al.}(2018)\citenamefont {Shammah}, \citenamefont {Ahmed}, \citenamefont {Lambert}, \citenamefont {Liberato},\ and\ \citenamefont {Nori}}]{shammah2018piqs}%
  \BibitemOpen
  \bibfield  {author} {\bibinfo {author} {\bibfnamefont {N.}~\bibnamefont {Shammah}}, \bibinfo {author} {\bibfnamefont {S.}~\bibnamefont {Ahmed}}, \bibinfo {author} {\bibfnamefont {N.}~\bibnamefont {Lambert}}, \bibinfo {author} {\bibfnamefont {S.~D.}\ \bibnamefont {Liberato}},\ and\ \bibinfo {author} {\bibfnamefont {F.}~\bibnamefont {Nori}},\ }\href {https://doi.org/10.1103/PhysRevA.98.063815} {\bibfield  {journal} {\bibinfo  {journal} {Phys. Rev. A}\ }\textbf {\bibinfo {volume} {98}},\ \bibinfo {pages} {063815} (\bibinfo {year} {2018})}\BibitemShut {NoStop}%
\bibitem [{\citenamefont {Dalibard}\ \emph {et~al.}(1992)\citenamefont {Dalibard}, \citenamefont {Castin},\ and\ \citenamefont {Mølmer}}]{dalibard1992wave}%
  \BibitemOpen
  \bibfield  {author} {\bibinfo {author} {\bibfnamefont {J.}~\bibnamefont {Dalibard}}, \bibinfo {author} {\bibfnamefont {Y.}~\bibnamefont {Castin}},\ and\ \bibinfo {author} {\bibfnamefont {K.}~\bibnamefont {Mølmer}},\ }\href {https://doi.org/10.1103/PhysRevLett.68.580} {\bibfield  {journal} {\bibinfo  {journal} {Phys. Rev. Lett.}\ }\textbf {\bibinfo {volume} {68}},\ \bibinfo {pages} {580} (\bibinfo {year} {1992})}\BibitemShut {NoStop}%
\bibitem [{\citenamefont {Dum}\ \emph {et~al.}(1992)\citenamefont {Dum}, \citenamefont {Zoller},\ and\ \citenamefont {Ritsch}}]{dum1992montecarlo}%
  \BibitemOpen
  \bibfield  {author} {\bibinfo {author} {\bibfnamefont {R.}~\bibnamefont {Dum}}, \bibinfo {author} {\bibfnamefont {P.}~\bibnamefont {Zoller}},\ and\ \bibinfo {author} {\bibfnamefont {H.}~\bibnamefont {Ritsch}},\ }\href {https://doi.org/10.1103/PhysRevA.45.4879} {\bibfield  {journal} {\bibinfo  {journal} {Phys. Rev. A}\ }\textbf {\bibinfo {volume} {45}},\ \bibinfo {pages} {4879} (\bibinfo {year} {1992})}\BibitemShut {NoStop}%
\bibitem [{\citenamefont {Johansson}\ \emph {et~al.}(2012)\citenamefont {Johansson}, \citenamefont {Nation},\ and\ \citenamefont {Nori}}]{johansson2012qutip}%
  \BibitemOpen
  \bibfield  {author} {\bibinfo {author} {\bibfnamefont {J.}~\bibnamefont {Johansson}}, \bibinfo {author} {\bibfnamefont {P.}~\bibnamefont {Nation}},\ and\ \bibinfo {author} {\bibfnamefont {F.}~\bibnamefont {Nori}},\ }\href {https://doi.org/https://doi.org/10.1016/j.cpc.2012.02.021} {\bibfield  {journal} {\bibinfo  {journal} {Computer Physics Communications}\ }\textbf {\bibinfo {volume} {183}},\ \bibinfo {pages} {1760} (\bibinfo {year} {2012})}\BibitemShut {NoStop}%
\bibitem [{\citenamefont {Johansson}\ \emph {et~al.}(2013)\citenamefont {Johansson}, \citenamefont {Nation},\ and\ \citenamefont {Nori}}]{johansson2013qutip}%
  \BibitemOpen
  \bibfield  {author} {\bibinfo {author} {\bibfnamefont {J.}~\bibnamefont {Johansson}}, \bibinfo {author} {\bibfnamefont {P.}~\bibnamefont {Nation}},\ and\ \bibinfo {author} {\bibfnamefont {F.}~\bibnamefont {Nori}},\ }\href {https://doi.org/https://doi.org/10.1016/j.cpc.2012.11.019} {\bibfield  {journal} {\bibinfo  {journal} {Computer Physics Communications}\ }\textbf {\bibinfo {volume} {184}},\ \bibinfo {pages} {1234} (\bibinfo {year} {2013})}\BibitemShut {NoStop}%
\bibitem [{\citenamefont {Lambert}\ \emph {et~al.}(2024)\citenamefont {Lambert}, \citenamefont {Giguère}, \citenamefont {Menczel}, \citenamefont {Li}, \citenamefont {Hopf}, \citenamefont {Suárez}, \citenamefont {Gali}, \citenamefont {Lishman}, \citenamefont {Gadhvi}, \citenamefont {Agarwal}, \citenamefont {Galicia}, \citenamefont {Shammah}, \citenamefont {Nation}, \citenamefont {Johansson}, \citenamefont {Ahmed}, \citenamefont {Cross}, \citenamefont {Pitchford},\ and\ \citenamefont {Nori}}]{lambert2024qutip5quantumtoolbox}%
  \BibitemOpen
  \bibfield  {author} {\bibinfo {author} {\bibfnamefont {N.}~\bibnamefont {Lambert}}, \bibinfo {author} {\bibfnamefont {E.}~\bibnamefont {Giguère}}, \bibinfo {author} {\bibfnamefont {P.}~\bibnamefont {Menczel}}, \bibinfo {author} {\bibfnamefont {B.}~\bibnamefont {Li}}, \bibinfo {author} {\bibfnamefont {P.}~\bibnamefont {Hopf}}, \bibinfo {author} {\bibfnamefont {G.}~\bibnamefont {Suárez}}, \bibinfo {author} {\bibfnamefont {M.}~\bibnamefont {Gali}}, \bibinfo {author} {\bibfnamefont {J.}~\bibnamefont {Lishman}}, \bibinfo {author} {\bibfnamefont {R.}~\bibnamefont {Gadhvi}}, \bibinfo {author} {\bibfnamefont {R.}~\bibnamefont {Agarwal}}, \bibinfo {author} {\bibfnamefont {A.}~\bibnamefont {Galicia}}, \bibinfo {author} {\bibfnamefont {N.}~\bibnamefont {Shammah}}, \bibinfo {author} {\bibfnamefont {P.}~\bibnamefont {Nation}}, \bibinfo {author} {\bibfnamefont {J.~R.}\ \bibnamefont {Johansson}}, \bibinfo {author} {\bibfnamefont {S.}~\bibnamefont {Ahmed}}, \bibinfo {author} {\bibfnamefont {S.}~\bibnamefont {Cross}},
  \bibinfo {author} {\bibfnamefont {A.}~\bibnamefont {Pitchford}},\ and\ \bibinfo {author} {\bibfnamefont {F.}~\bibnamefont {Nori}},\ }\href {https://arxiv.org/abs/2412.04705} {\bibinfo {title} {Qutip 5: The quantum toolbox in python}} (\bibinfo {year} {2024}),\ \Eprint {https://arxiv.org/abs/2412.04705} {arXiv:2412.04705 [quant-ph]} \BibitemShut {NoStop}%
\bibitem [{\citenamefont {Eberly}\ \emph {et~al.}(1980)\citenamefont {Eberly}, \citenamefont {Narozhny},\ and\ \citenamefont {Sanchez-Mondragon}}]{eberly1980periodic}%
  \BibitemOpen
  \bibfield  {author} {\bibinfo {author} {\bibfnamefont {J.~H.}\ \bibnamefont {Eberly}}, \bibinfo {author} {\bibfnamefont {N.~B.}\ \bibnamefont {Narozhny}},\ and\ \bibinfo {author} {\bibfnamefont {J.~J.}\ \bibnamefont {Sanchez-Mondragon}},\ }\href@noop {} {\bibfield  {journal} {\bibinfo  {journal} {Physical Review Letters}\ }\textbf {\bibinfo {volume} {44}},\ \bibinfo {pages} {1323} (\bibinfo {year} {1980})}\BibitemShut {NoStop}%
\bibitem [{\citenamefont {Phoenix}\ and\ \citenamefont {Knight}(1991)}]{phoenix1991collapse}%
  \BibitemOpen
  \bibfield  {author} {\bibinfo {author} {\bibfnamefont {S.~J.~D.}\ \bibnamefont {Phoenix}}\ and\ \bibinfo {author} {\bibfnamefont {P.~L.}\ \bibnamefont {Knight}},\ }\href@noop {} {\bibfield  {journal} {\bibinfo  {journal} {Physical Review A}\ }\textbf {\bibinfo {volume} {44}},\ \bibinfo {pages} {6023} (\bibinfo {year} {1991})}\BibitemShut {NoStop}%
\bibitem [{\citenamefont {Shore}\ and\ \citenamefont {Knight}(1993)}]{shore1993jaynes}%
  \BibitemOpen
  \bibfield  {author} {\bibinfo {author} {\bibfnamefont {B.~W.}\ \bibnamefont {Shore}}\ and\ \bibinfo {author} {\bibfnamefont {P.~L.}\ \bibnamefont {Knight}},\ }\href@noop {} {\bibfield  {journal} {\bibinfo  {journal} {Journal of Modern Optics}\ }\textbf {\bibinfo {volume} {40}},\ \bibinfo {pages} {1195} (\bibinfo {year} {1993})}\BibitemShut {NoStop}%
\bibitem [{\citenamefont {Garraway}\ and\ \citenamefont {Stenholm}(2011)}]{garraway2011decay}%
  \BibitemOpen
  \bibfield  {author} {\bibinfo {author} {\bibfnamefont {B.~M.}\ \bibnamefont {Garraway}}\ and\ \bibinfo {author} {\bibfnamefont {S.}~\bibnamefont {Stenholm}},\ }\href@noop {} {\bibfield  {journal} {\bibinfo  {journal} {Physical Review A}\ }\textbf {\bibinfo {volume} {84}},\ \bibinfo {pages} {043839} (\bibinfo {year} {2011})}\BibitemShut {NoStop}%
\bibitem [{\citenamefont {Zhang}\ \emph {et~al.}(2014)\citenamefont {Zhang}, \citenamefont {Ye},\ and\ \citenamefont {Zhang}}]{zhang2014revival}%
  \BibitemOpen
  \bibfield  {author} {\bibinfo {author} {\bibfnamefont {G.-F.}\ \bibnamefont {Zhang}}, \bibinfo {author} {\bibfnamefont {L.}~\bibnamefont {Ye}},\ and\ \bibinfo {author} {\bibfnamefont {J.-Q.}\ \bibnamefont {Zhang}},\ }\href@noop {} {\bibfield  {journal} {\bibinfo  {journal} {Optics Communications}\ }\textbf {\bibinfo {volume} {313}},\ \bibinfo {pages} {186} (\bibinfo {year} {2014})}\BibitemShut {NoStop}%
\bibitem [{\citenamefont {Bastidas}\ \emph {et~al.}(2010)\citenamefont {Bastidas}, \citenamefont {Reina}, \citenamefont {Emary},\ and\ \citenamefont {Brandes}}]{bastidas2010entanglement}%
  \BibitemOpen
  \bibfield  {author} {\bibinfo {author} {\bibfnamefont {V.~M.}\ \bibnamefont {Bastidas}}, \bibinfo {author} {\bibfnamefont {J.~H.}\ \bibnamefont {Reina}}, \bibinfo {author} {\bibfnamefont {C.}~\bibnamefont {Emary}},\ and\ \bibinfo {author} {\bibfnamefont {T.}~\bibnamefont {Brandes}},\ }\href {https://doi.org/10.1103/PhysRevA.81.012316} {\bibfield  {journal} {\bibinfo  {journal} {Phys. Rev. A}\ }\textbf {\bibinfo {volume} {81}},\ \bibinfo {pages} {012316} (\bibinfo {year} {2010})}\BibitemShut {NoStop}%
\bibitem [{\citenamefont {Jarmola}\ \emph {et~al.}(2012)\citenamefont {Jarmola}, \citenamefont {Acosta}, \citenamefont {Jensen}, \citenamefont {Chemerisov},\ and\ \citenamefont {Budker}}]{jarmola2012}%
  \BibitemOpen
  \bibfield  {author} {\bibinfo {author} {\bibfnamefont {A.}~\bibnamefont {Jarmola}}, \bibinfo {author} {\bibfnamefont {V.~M.}\ \bibnamefont {Acosta}}, \bibinfo {author} {\bibfnamefont {K.}~\bibnamefont {Jensen}}, \bibinfo {author} {\bibfnamefont {S.}~\bibnamefont {Chemerisov}},\ and\ \bibinfo {author} {\bibfnamefont {D.}~\bibnamefont {Budker}},\ }\href {https://doi.org/10.1103/PhysRevLett.108.197601} {\bibfield  {journal} {\bibinfo  {journal} {Phys. Rev. Lett.}\ }\textbf {\bibinfo {volume} {108}},\ \bibinfo {pages} {197601} (\bibinfo {year} {2012})}\BibitemShut {NoStop}%
\bibitem [{\citenamefont {Stanwix}\ \emph {et~al.}(2010)\citenamefont {Stanwix}, \citenamefont {Pham}, \citenamefont {Maze}, \citenamefont {Le~Sage}, \citenamefont {Yeung}, \citenamefont {Cappellaro}, \citenamefont {Hemmer}, \citenamefont {Yacoby}, \citenamefont {Lukin},\ and\ \citenamefont {Walsworth}}]{stanwix2011}%
  \BibitemOpen
  \bibfield  {author} {\bibinfo {author} {\bibfnamefont {P.~L.}\ \bibnamefont {Stanwix}}, \bibinfo {author} {\bibfnamefont {L.~M.}\ \bibnamefont {Pham}}, \bibinfo {author} {\bibfnamefont {J.~R.}\ \bibnamefont {Maze}}, \bibinfo {author} {\bibfnamefont {D.}~\bibnamefont {Le~Sage}}, \bibinfo {author} {\bibfnamefont {T.~K.}\ \bibnamefont {Yeung}}, \bibinfo {author} {\bibfnamefont {P.}~\bibnamefont {Cappellaro}}, \bibinfo {author} {\bibfnamefont {P.~R.}\ \bibnamefont {Hemmer}}, \bibinfo {author} {\bibfnamefont {A.}~\bibnamefont {Yacoby}}, \bibinfo {author} {\bibfnamefont {M.~D.}\ \bibnamefont {Lukin}},\ and\ \bibinfo {author} {\bibfnamefont {R.~L.}\ \bibnamefont {Walsworth}},\ }\href {https://doi.org/10.1103/PhysRevB.82.201201} {\bibfield  {journal} {\bibinfo  {journal} {Phys. Rev. B}\ }\textbf {\bibinfo {volume} {82}},\ \bibinfo {pages} {201201} (\bibinfo {year} {2010})}\BibitemShut {NoStop}%
\bibitem [{\citenamefont {Seo}\ \emph {et~al.}(2016)\citenamefont {Seo}, \citenamefont {Falk}, \citenamefont {Klimov}, \citenamefont {Miao}, \citenamefont {Galli},\ and\ \citenamefont {Awschalom}}]{seo2016}%
  \BibitemOpen
  \bibfield  {author} {\bibinfo {author} {\bibfnamefont {H.}~\bibnamefont {Seo}}, \bibinfo {author} {\bibfnamefont {A.~L.}\ \bibnamefont {Falk}}, \bibinfo {author} {\bibfnamefont {P.~V.}\ \bibnamefont {Klimov}}, \bibinfo {author} {\bibfnamefont {K.~C.}\ \bibnamefont {Miao}}, \bibinfo {author} {\bibfnamefont {G.}~\bibnamefont {Galli}},\ and\ \bibinfo {author} {\bibfnamefont {D.~D.}\ \bibnamefont {Awschalom}},\ }\href {https://doi.org/10.1038/ncomms12935} {\bibfield  {journal} {\bibinfo  {journal} {Nature Communications}\ }\textbf {\bibinfo {volume} {7}},\ \bibinfo {pages} {12935} (\bibinfo {year} {2016})}\BibitemShut {NoStop}%
\bibitem [{\citenamefont {Alexander}\ \emph {et~al.}(2022)\citenamefont {Alexander} \emph {et~al.}}]{alexander2022}%
  \BibitemOpen
  \bibfield  {author} {\bibinfo {author} {\bibfnamefont {J.}~\bibnamefont {Alexander}} \emph {et~al.},\ }\href {https://doi.org/10.1103/PhysRevB.106.245416} {\bibfield  {journal} {\bibinfo  {journal} {Phys. Rev. B}\ }\textbf {\bibinfo {volume} {106}},\ \bibinfo {pages} {245416} (\bibinfo {year} {2022})},\ \Eprint {https://arxiv.org/abs/2206.04027} {arXiv:2206.04027 [quant-ph]} \BibitemShut {NoStop}%
\bibitem [{\citenamefont {Zhang}\ \emph {et~al.}(2025)\citenamefont {Zhang}, \citenamefont {Wang}, \citenamefont {Chen},\ and\ \citenamefont {{et al.}}}]{zhang2025}%
  \BibitemOpen
  \bibfield  {author} {\bibinfo {author} {\bibfnamefont {J.}~\bibnamefont {Zhang}}, \bibinfo {author} {\bibfnamefont {P.}~\bibnamefont {Wang}}, \bibinfo {author} {\bibfnamefont {W.}~\bibnamefont {Chen}},\ and\ \bibinfo {author} {\bibnamefont {{et al.}}},\ }\bibfield  {journal} {\bibinfo  {journal} {Phys. Rev. Lett.}\ }\href {https://doi.org/10.1103/g45c-ssfx} {10.1103/g45c-ssfx} (\bibinfo {year} {2025}),\ \bibinfo {note} {in press}\BibitemShut {NoStop}%
\bibitem [{\citenamefont {Bardot}\ \emph {et~al.}(2005)\citenamefont {Bardot}, \citenamefont {Schwab}, \citenamefont {Bayer}, \citenamefont {Fafard}, \citenamefont {Wasilewski},\ and\ \citenamefont {Hawrylak}}]{bardot2005}%
  \BibitemOpen
  \bibfield  {author} {\bibinfo {author} {\bibfnamefont {C.}~\bibnamefont {Bardot}}, \bibinfo {author} {\bibfnamefont {M.}~\bibnamefont {Schwab}}, \bibinfo {author} {\bibfnamefont {M.}~\bibnamefont {Bayer}}, \bibinfo {author} {\bibfnamefont {S.}~\bibnamefont {Fafard}}, \bibinfo {author} {\bibfnamefont {Z.}~\bibnamefont {Wasilewski}},\ and\ \bibinfo {author} {\bibfnamefont {P.}~\bibnamefont {Hawrylak}},\ }\href {https://doi.org/10.1103/PhysRevB.72.035314} {\bibfield  {journal} {\bibinfo  {journal} {Phys. Rev. B}\ }\textbf {\bibinfo {volume} {72}},\ \bibinfo {pages} {035314} (\bibinfo {year} {2005})}\BibitemShut {NoStop}%
\bibitem [{\citenamefont {Accanto}\ \emph {et~al.}(2012)\citenamefont {Accanto}, \citenamefont {Masia}, \citenamefont {Moreels}, \citenamefont {Hens}, \citenamefont {Langbein},\ and\ \citenamefont {Borri}}]{accanto2012}%
  \BibitemOpen
  \bibfield  {author} {\bibinfo {author} {\bibfnamefont {N.}~\bibnamefont {Accanto}}, \bibinfo {author} {\bibfnamefont {F.}~\bibnamefont {Masia}}, \bibinfo {author} {\bibfnamefont {I.}~\bibnamefont {Moreels}}, \bibinfo {author} {\bibfnamefont {Z.}~\bibnamefont {Hens}}, \bibinfo {author} {\bibfnamefont {W.}~\bibnamefont {Langbein}},\ and\ \bibinfo {author} {\bibfnamefont {P.}~\bibnamefont {Borri}},\ }\href {https://doi.org/10.1021/nn300992a} {\bibfield  {journal} {\bibinfo  {journal} {ACS Nano}\ }\textbf {\bibinfo {volume} {6}},\ \bibinfo {pages} {5227} (\bibinfo {year} {2012})}\BibitemShut {NoStop}%
\bibitem [{\citenamefont {Koolstra}\ \emph {et~al.}(2025)\citenamefont {Koolstra}, \citenamefont {Glen}, \citenamefont {Beysengulov}, \citenamefont {Byeon}, \citenamefont {Castoria}, \citenamefont {Sammon}, \citenamefont {Lyon}, \citenamefont {Rees},\ and\ \citenamefont {Pollanen}}]{koolstra2025strongcouplingmicrowavephoton}%
  \BibitemOpen
  \bibfield  {author} {\bibinfo {author} {\bibfnamefont {G.}~\bibnamefont {Koolstra}}, \bibinfo {author} {\bibfnamefont {E.~O.}\ \bibnamefont {Glen}}, \bibinfo {author} {\bibfnamefont {N.~R.}\ \bibnamefont {Beysengulov}}, \bibinfo {author} {\bibfnamefont {H.}~\bibnamefont {Byeon}}, \bibinfo {author} {\bibfnamefont {K.~E.}\ \bibnamefont {Castoria}}, \bibinfo {author} {\bibfnamefont {M.}~\bibnamefont {Sammon}}, \bibinfo {author} {\bibfnamefont {S.~A.}\ \bibnamefont {Lyon}}, \bibinfo {author} {\bibfnamefont {D.~G.}\ \bibnamefont {Rees}},\ and\ \bibinfo {author} {\bibfnamefont {J.}~\bibnamefont {Pollanen}},\ }\href {https://arxiv.org/abs/2509.14506} {\bibinfo {title} {Strong coupling of a microwave photon to an electron on helium}} (\bibinfo {year} {2025}),\ \Eprint {https://arxiv.org/abs/2509.14506} {arXiv:2509.14506 [quant-ph]} \BibitemShut {NoStop}%
\bibitem [{\citenamefont {Hu}\ \emph {et~al.}(2022)\citenamefont {Hu}, \citenamefont {Qiu}, \citenamefont {Souza}, \citenamefont {Yuan}, \citenamefont {Zhou}, \citenamefont {Zhang}, \citenamefont {Chu}, \citenamefont {Pan}, \citenamefont {Hu}, \citenamefont {Li}, \citenamefont {Xu}, \citenamefont {Zhong}, \citenamefont {Liu}, \citenamefont {Yan}, \citenamefont {Tan}, \citenamefont {Bachelard}, \citenamefont {Villas-Boas}, \citenamefont {Santos},\ and\ \citenamefont {Yu}}]{Chang-Kang2022}%
  \BibitemOpen
  \bibfield  {author} {\bibinfo {author} {\bibfnamefont {C.-K.}\ \bibnamefont {Hu}}, \bibinfo {author} {\bibfnamefont {J.}~\bibnamefont {Qiu}}, \bibinfo {author} {\bibfnamefont {P.~J.~P.}\ \bibnamefont {Souza}}, \bibinfo {author} {\bibfnamefont {J.}~\bibnamefont {Yuan}}, \bibinfo {author} {\bibfnamefont {Y.}~\bibnamefont {Zhou}}, \bibinfo {author} {\bibfnamefont {L.}~\bibnamefont {Zhang}}, \bibinfo {author} {\bibfnamefont {J.}~\bibnamefont {Chu}}, \bibinfo {author} {\bibfnamefont {X.}~\bibnamefont {Pan}}, \bibinfo {author} {\bibfnamefont {L.}~\bibnamefont {Hu}}, \bibinfo {author} {\bibfnamefont {J.}~\bibnamefont {Li}}, \bibinfo {author} {\bibfnamefont {Y.}~\bibnamefont {Xu}}, \bibinfo {author} {\bibfnamefont {Y.}~\bibnamefont {Zhong}}, \bibinfo {author} {\bibfnamefont {S.}~\bibnamefont {Liu}}, \bibinfo {author} {\bibfnamefont {F.}~\bibnamefont {Yan}}, \bibinfo {author} {\bibfnamefont {D.}~\bibnamefont {Tan}}, \bibinfo {author} {\bibfnamefont {R.}~\bibnamefont {Bachelard}}, \bibinfo {author} {\bibfnamefont
  {C.~J.}\ \bibnamefont {Villas-Boas}}, \bibinfo {author} {\bibfnamefont {A.~C.}\ \bibnamefont {Santos}},\ and\ \bibinfo {author} {\bibfnamefont {D.}~\bibnamefont {Yu}},\ }\href {https://doi.org/10.1088/2058-9565/ac8444} {\bibfield  {journal} {\bibinfo  {journal} {Quantum Science and Technology}\ }\textbf {\bibinfo {volume} {7}},\ \bibinfo {pages} {045018} (\bibinfo {year} {2022})}\BibitemShut {NoStop}%
\bibitem [{\citenamefont {Dou}\ and\ \citenamefont {Yang}(2023)}]{Fu-Quan2023}%
  \BibitemOpen
  \bibfield  {author} {\bibinfo {author} {\bibfnamefont {F.-Q.}\ \bibnamefont {Dou}}\ and\ \bibinfo {author} {\bibfnamefont {F.-M.}\ \bibnamefont {Yang}},\ }\href {https://doi.org/10.1103/PhysRevA.107.023725} {\bibfield  {journal} {\bibinfo  {journal} {Phys. Rev. A}\ }\textbf {\bibinfo {volume} {107}},\ \bibinfo {pages} {023725} (\bibinfo {year} {2023})}\BibitemShut {NoStop}%
\bibitem [{\citenamefont {Dou}\ \emph {et~al.}(2022)\citenamefont {Dou}, \citenamefont {Wang},\ and\ \citenamefont {Sun}}]{dou2022staqb}%
  \BibitemOpen
  \bibfield  {author} {\bibinfo {author} {\bibfnamefont {F.-Q.}\ \bibnamefont {Dou}}, \bibinfo {author} {\bibfnamefont {Y.-J.}\ \bibnamefont {Wang}},\ and\ \bibinfo {author} {\bibfnamefont {J.-A.}\ \bibnamefont {Sun}},\ }\href {https://doi.org/10.1007/s11467-021-1130-5} {\bibfield  {journal} {\bibinfo  {journal} {Frontiers of Physics}\ }\textbf {\bibinfo {volume} {17}},\ \bibinfo {pages} {13201} (\bibinfo {year} {2022})}\BibitemShut {NoStop}%
\bibitem [{\citenamefont {Hu}\ \emph {et~al.}(2021)\citenamefont {Hu}, \citenamefont {Qi},\ and\ \citenamefont {Jing}}]{hu2021staqb}%
  \BibitemOpen
  \bibfield  {author} {\bibinfo {author} {\bibfnamefont {H.}~\bibnamefont {Hu}}, \bibinfo {author} {\bibfnamefont {S.}~\bibnamefont {Qi}},\ and\ \bibinfo {author} {\bibfnamefont {J.}~\bibnamefont {Jing}},\ }\bibfield  {journal} {\bibinfo  {journal} {arXiv preprint arXiv:2104.12143}\ }\href {https://doi.org/10.48550/arXiv.2104.12143} {10.48550/arXiv.2104.12143} (\bibinfo {year} {2021})\BibitemShut {NoStop}%
\bibitem [{\citenamefont {Salazar}\ \emph {et~al.}(2023)\citenamefont {Salazar}, \citenamefont {Calderón-Losada},\ and\ \citenamefont {Reina}}]{salazar2023circuitdesign}%
  \BibitemOpen
  \bibfield  {author} {\bibinfo {author} {\bibfnamefont {W.~E.}\ \bibnamefont {Salazar}}, \bibinfo {author} {\bibfnamefont {O.}~\bibnamefont {Calderón-Losada}},\ and\ \bibinfo {author} {\bibfnamefont {J.~H.}\ \bibnamefont {Reina}},\ }\href {https://arxiv.org/abs/2310.20416} {\bibinfo {title} {Linear-nonlinear duality for circuit design on quantum computing platforms}} (\bibinfo {year} {2023}),\ \Eprint {https://arxiv.org/abs/2310.20416} {arXiv:2310.20416 [quant-ph]} \BibitemShut {NoStop}%
\bibitem [{\citenamefont {Mazzoncini}\ \emph {et~al.}(2023)\citenamefont {Mazzoncini}, \citenamefont {Cavina}, \citenamefont {Andolina}, \citenamefont {Erdman},\ and\ \citenamefont {Giovannetti}}]{mazzoncini2023ocqb}%
  \BibitemOpen
  \bibfield  {author} {\bibinfo {author} {\bibfnamefont {F.}~\bibnamefont {Mazzoncini}}, \bibinfo {author} {\bibfnamefont {V.}~\bibnamefont {Cavina}}, \bibinfo {author} {\bibfnamefont {G.~M.}\ \bibnamefont {Andolina}}, \bibinfo {author} {\bibfnamefont {P.~A.}\ \bibnamefont {Erdman}},\ and\ \bibinfo {author} {\bibfnamefont {V.}~\bibnamefont {Giovannetti}},\ }\href {https://doi.org/10.1103/PhysRevA.107.032218} {\bibfield  {journal} {\bibinfo  {journal} {Phys. Rev. A}\ }\textbf {\bibinfo {volume} {107}},\ \bibinfo {pages} {032218} (\bibinfo {year} {2023})}\BibitemShut {NoStop}%
\bibitem [{\citenamefont {Qi}\ and\ \citenamefont {Jing}(2025)}]{qi2025staqb}%
  \BibitemOpen
  \bibfield  {author} {\bibinfo {author} {\bibfnamefont {S.-f.}\ \bibnamefont {Qi}}\ and\ \bibinfo {author} {\bibfnamefont {J.}~\bibnamefont {Jing}},\ }\href {https://doi.org/10.1016/j.physleta.2024.130124} {\bibfield  {journal} {\bibinfo  {journal} {Phys. Lett. A}\ }\textbf {\bibinfo {volume} {530}},\ \bibinfo {pages} {130124} (\bibinfo {year} {2025})}\BibitemShut {NoStop}%
\bibitem [{\citenamefont {Diehl}\ \emph {et~al.}(2008)\citenamefont {Diehl}, \citenamefont {Micheli}, \citenamefont {Kantian}, \citenamefont {Kraus}, \citenamefont {B\"uchler},\ and\ \citenamefont {Zoller}}]{diehl2008engineered}%
  \BibitemOpen
  \bibfield  {author} {\bibinfo {author} {\bibfnamefont {S.}~\bibnamefont {Diehl}}, \bibinfo {author} {\bibfnamefont {A.}~\bibnamefont {Micheli}}, \bibinfo {author} {\bibfnamefont {A.}~\bibnamefont {Kantian}}, \bibinfo {author} {\bibfnamefont {B.}~\bibnamefont {Kraus}}, \bibinfo {author} {\bibfnamefont {H.~P.}\ \bibnamefont {B\"uchler}},\ and\ \bibinfo {author} {\bibfnamefont {P.}~\bibnamefont {Zoller}},\ }\href {https://doi.org/10.1038/nphys1073} {\bibfield  {journal} {\bibinfo  {journal} {Nat. Phys.}\ }\textbf {\bibinfo {volume} {4}},\ \bibinfo {pages} {878} (\bibinfo {year} {2008})}\BibitemShut {NoStop}%
\bibitem [{\citenamefont {Barreiro}\ \emph {et~al.}(2011)\citenamefont {Barreiro}, \citenamefont {M\"uller}, \citenamefont {Schindler}, \citenamefont {Nigg}, \citenamefont {Monz}, \citenamefont {Chwalla}, \citenamefont {Hennrich}, \citenamefont {Roos}, \citenamefont {Zoller},\ and\ \citenamefont {Blatt}}]{barreiro2011open}%
  \BibitemOpen
  \bibfield  {author} {\bibinfo {author} {\bibfnamefont {J.~T.}\ \bibnamefont {Barreiro}}, \bibinfo {author} {\bibfnamefont {M.}~\bibnamefont {M\"uller}}, \bibinfo {author} {\bibfnamefont {P.}~\bibnamefont {Schindler}}, \bibinfo {author} {\bibfnamefont {D.}~\bibnamefont {Nigg}}, \bibinfo {author} {\bibfnamefont {T.}~\bibnamefont {Monz}}, \bibinfo {author} {\bibfnamefont {M.}~\bibnamefont {Chwalla}}, \bibinfo {author} {\bibfnamefont {M.}~\bibnamefont {Hennrich}}, \bibinfo {author} {\bibfnamefont {C.~F.}\ \bibnamefont {Roos}}, \bibinfo {author} {\bibfnamefont {P.}~\bibnamefont {Zoller}},\ and\ \bibinfo {author} {\bibfnamefont {R.}~\bibnamefont {Blatt}},\ }\href {https://doi.org/10.1038/nature09801} {\bibfield  {journal} {\bibinfo  {journal} {Nature}\ }\textbf {\bibinfo {volume} {470}},\ \bibinfo {pages} {486} (\bibinfo {year} {2011})}\BibitemShut {NoStop}%
\bibitem [{\citenamefont {Harrington}\ \emph {et~al.}(2022)\citenamefont {Harrington}, \citenamefont {Monroe}, \citenamefont {Biercuk}, \citenamefont {Oliver}, \citenamefont {Kippenberg}, \citenamefont {Houck},\ and\ \citenamefont {Clerk}}]{harrington2022engineered}%
  \BibitemOpen
  \bibfield  {author} {\bibinfo {author} {\bibfnamefont {P.~M.}\ \bibnamefont {Harrington}}, \bibinfo {author} {\bibfnamefont {C.}~\bibnamefont {Monroe}}, \bibinfo {author} {\bibfnamefont {M.~J.}\ \bibnamefont {Biercuk}}, \bibinfo {author} {\bibfnamefont {W.~D.}\ \bibnamefont {Oliver}}, \bibinfo {author} {\bibfnamefont {T.~J.}\ \bibnamefont {Kippenberg}}, \bibinfo {author} {\bibfnamefont {A.~A.}\ \bibnamefont {Houck}},\ and\ \bibinfo {author} {\bibfnamefont {A.~A.}\ \bibnamefont {Clerk}},\ }\href {https://doi.org/10.1103/RevModPhys.94.045001} {\bibfield  {journal} {\bibinfo  {journal} {Rev. Mod. Phys.}\ }\textbf {\bibinfo {volume} {94}},\ \bibinfo {pages} {045001} (\bibinfo {year} {2022})}\BibitemShut {NoStop}%
\bibitem [{\citenamefont {Lodahl}\ \emph {et~al.}(2004)\citenamefont {Lodahl}, \citenamefont {Van~Driel}, \citenamefont {Nikolaev}, \citenamefont {Irman}, \citenamefont {Overgaag}, \citenamefont {Vanmaekelbergh},\ and\ \citenamefont {Vos}}]{lodahl2004control}%
  \BibitemOpen
  \bibfield  {author} {\bibinfo {author} {\bibfnamefont {P.}~\bibnamefont {Lodahl}}, \bibinfo {author} {\bibfnamefont {A.~F.}\ \bibnamefont {Van~Driel}}, \bibinfo {author} {\bibfnamefont {I.~S.}\ \bibnamefont {Nikolaev}}, \bibinfo {author} {\bibfnamefont {A.}~\bibnamefont {Irman}}, \bibinfo {author} {\bibfnamefont {K.}~\bibnamefont {Overgaag}}, \bibinfo {author} {\bibfnamefont {D.}~\bibnamefont {Vanmaekelbergh}},\ and\ \bibinfo {author} {\bibfnamefont {W.~L.}\ \bibnamefont {Vos}},\ }\href {https://doi.org/10.1038/nature02772} {\bibfield  {journal} {\bibinfo  {journal} {Nature}\ }\textbf {\bibinfo {volume} {430}},\ \bibinfo {pages} {654} (\bibinfo {year} {2004})}\BibitemShut {NoStop}%
\bibitem [{\citenamefont {Cywiński}\ \emph {et~al.}(2009)\citenamefont {Cywiński}, \citenamefont {Witzel},\ and\ \citenamefont {Das~Sarma}}]{cywinski2009prl}%
  \BibitemOpen
  \bibfield  {author} {\bibinfo {author} {\bibfnamefont {L.}~\bibnamefont {Cywiński}}, \bibinfo {author} {\bibfnamefont {W.~M.}\ \bibnamefont {Witzel}},\ and\ \bibinfo {author} {\bibfnamefont {S.}~\bibnamefont {Das~Sarma}},\ }\href {https://doi.org/10.1103/PhysRevLett.102.057601} {\bibfield  {journal} {\bibinfo  {journal} {Phys. Rev. Lett.}\ }\textbf {\bibinfo {volume} {102}},\ \bibinfo {pages} {057601} (\bibinfo {year} {2009})}\BibitemShut {NoStop}%
\bibitem [{\citenamefont {Eckel}\ \emph {et~al.}(2009)\citenamefont {Eckel}, \citenamefont {Reina},\ and\ \citenamefont {Thorwart}}]{JH_CoherentControl_TLS_nonMarkovian}%
  \BibitemOpen
  \bibfield  {author} {\bibinfo {author} {\bibfnamefont {J.}~\bibnamefont {Eckel}}, \bibinfo {author} {\bibfnamefont {J.~H.}\ \bibnamefont {Reina}},\ and\ \bibinfo {author} {\bibfnamefont {M.}~\bibnamefont {Thorwart}},\ }\href {https://doi.org/10.1088/1367-2630/11/8/085001} {\bibfield  {journal} {\bibinfo  {journal} {New Journal of Physics}\ }\textbf {\bibinfo {volume} {11}},\ \bibinfo {pages} {085001} (\bibinfo {year} {2009})}\BibitemShut {NoStop}%
\bibitem [{\citenamefont {Breuer}\ \emph {et~al.}(2016)\citenamefont {Breuer}, \citenamefont {Laine}, \citenamefont {Piilo},\ and\ \citenamefont {Vacchini}}]{RMP_BLP_2016}%
  \BibitemOpen
  \bibfield  {author} {\bibinfo {author} {\bibfnamefont {H.-P.}\ \bibnamefont {Breuer}}, \bibinfo {author} {\bibfnamefont {E.-M.}\ \bibnamefont {Laine}}, \bibinfo {author} {\bibfnamefont {J.}~\bibnamefont {Piilo}},\ and\ \bibinfo {author} {\bibfnamefont {B.}~\bibnamefont {Vacchini}},\ }\href {https://doi.org/10.1103/RevModPhys.88.021002} {\bibfield  {journal} {\bibinfo  {journal} {Rev. Mod. Phys.}\ }\textbf {\bibinfo {volume} {88}},\ \bibinfo {pages} {021002} (\bibinfo {year} {2016})}\BibitemShut {NoStop}%
\bibitem [{\citenamefont {de~Vega}\ and\ \citenamefont {Alonso}(2017)}]{Review_nonMark_Dynamics}%
  \BibitemOpen
  \bibfield  {author} {\bibinfo {author} {\bibfnamefont {I.}~\bibnamefont {de~Vega}}\ and\ \bibinfo {author} {\bibfnamefont {D.}~\bibnamefont {Alonso}},\ }\href {https://doi.org/10.1103/RevModPhys.89.015001} {\bibfield  {journal} {\bibinfo  {journal} {Rev. Mod. Phys.}\ }\textbf {\bibinfo {volume} {89}},\ \bibinfo {pages} {015001} (\bibinfo {year} {2017})}\BibitemShut {NoStop}%
\bibitem [{\citenamefont {Luchnikov}\ \emph {et~al.}(2020)\citenamefont {Luchnikov}, \citenamefont {Vintskevich}, \citenamefont {Grigoriev},\ and\ \citenamefont {Filippov}}]{luchnikov2020mlNM}%
  \BibitemOpen
  \bibfield  {author} {\bibinfo {author} {\bibfnamefont {I.~A.}\ \bibnamefont {Luchnikov}}, \bibinfo {author} {\bibfnamefont {S.~V.}\ \bibnamefont {Vintskevich}}, \bibinfo {author} {\bibfnamefont {D.~A.}\ \bibnamefont {Grigoriev}},\ and\ \bibinfo {author} {\bibfnamefont {S.~N.}\ \bibnamefont {Filippov}},\ }\href {https://doi.org/10.1103/PhysRevLett.124.140502} {\bibfield  {journal} {\bibinfo  {journal} {Phys. Rev. Lett.}\ }\textbf {\bibinfo {volume} {124}},\ \bibinfo {pages} {140502} (\bibinfo {year} {2020})}\BibitemShut {NoStop}%
\bibitem [{\citenamefont {Scarpetta}\ \emph {et~al.}(2025)\citenamefont {Scarpetta}, \citenamefont {Reina},\ and\ \citenamefont {Hjorth-Jensen}}]{scarpetta2025machinelearning}%
  \BibitemOpen
  \bibfield  {author} {\bibinfo {author} {\bibfnamefont {J.~M.}\ \bibnamefont {Scarpetta}}, \bibinfo {author} {\bibfnamefont {J.~H.}\ \bibnamefont {Reina}},\ and\ \bibinfo {author} {\bibfnamefont {M.}~\bibnamefont {Hjorth-Jensen}},\ }\href {https://arxiv.org/abs/2506.06555} {\bibinfo {title} {Machine learning non-markovian two-level quantum noise spectroscopy}} (\bibinfo {year} {2025}),\ \Eprint {https://arxiv.org/abs/2506.06555} {arXiv:2506.06555 [quant-ph]} \BibitemShut {NoStop}%
\bibitem [{\citenamefont {Altherr}\ and\ \citenamefont {Yang}(2021)}]{altherr2021nonMarkovMetrology}%
  \BibitemOpen
  \bibfield  {author} {\bibinfo {author} {\bibfnamefont {A.}~\bibnamefont {Altherr}}\ and\ \bibinfo {author} {\bibfnamefont {W.}~\bibnamefont {Yang}},\ }\href {https://doi.org/10.1103/PhysRevLett.127.273601} {\bibfield  {journal} {\bibinfo  {journal} {Phys. Rev. Lett.}\ }\textbf {\bibinfo {volume} {127}},\ \bibinfo {pages} {273601} (\bibinfo {year} {2021})}\BibitemShut {NoStop}%
\bibitem [{\citenamefont {Martina}\ \emph {et~al.}(2021)\citenamefont {Martina}, \citenamefont {Benedetti},\ and\ \citenamefont {Paris}}]{martina2021mlNM}%
  \BibitemOpen
  \bibfield  {author} {\bibinfo {author} {\bibfnamefont {S.}~\bibnamefont {Martina}}, \bibinfo {author} {\bibfnamefont {C.}~\bibnamefont {Benedetti}},\ and\ \bibinfo {author} {\bibfnamefont {M.~G.~A.}\ \bibnamefont {Paris}},\ }\href {https://doi.org/10.1103/PhysRevA.104.022429} {\bibfield  {journal} {\bibinfo  {journal} {Phys. Rev. A}\ }\textbf {\bibinfo {volume} {104}},\ \bibinfo {pages} {022429} (\bibinfo {year} {2021})}\BibitemShut {NoStop}%
\bibitem [{\citenamefont {Ronchetti}\ and\ \citenamefont {Huber}(2009)}]{ronchetti2009robust_statistics}%
  \BibitemOpen
  \bibfield  {author} {\bibinfo {author} {\bibfnamefont {E.~M.}\ \bibnamefont {Ronchetti}}\ and\ \bibinfo {author} {\bibfnamefont {P.~J.}\ \bibnamefont {Huber}},\ }\href@noop {} {\emph {\bibinfo {title} {Robust statistics}}}\ (\bibinfo  {publisher} {John Wiley \& Sons Hoboken, NJ, USA},\ \bibinfo {year} {2009})\BibitemShut {NoStop}%
\bibitem [{\citenamefont {Rousseeuw}\ and\ \citenamefont {Leroy}(2003)}]{rousseeuw2003robust_regression}%
  \BibitemOpen
  \bibfield  {author} {\bibinfo {author} {\bibfnamefont {P.~J.}\ \bibnamefont {Rousseeuw}}\ and\ \bibinfo {author} {\bibfnamefont {A.~M.}\ \bibnamefont {Leroy}},\ }\href@noop {} {\emph {\bibinfo {title} {Robust regression and outlier detection}}}\ (\bibinfo  {publisher} {John wiley \& sons},\ \bibinfo {year} {2003})\BibitemShut {NoStop}%
\end{thebibliography}
\end{document}